\documentstyle[preprint,tighten,aps]{revtex}

\begin{document}

\draft
\preprint{IFUP-TH 5/96}

\title{A strong-coupling analysis of two-dimensional
$\protect\bbox{{\rm O}(N)}$ $\protect\bbox{\sigma}$ models with
$\protect\bbox{N\geq 3}$ on square, triangular and honeycomb
lattices}

\author{Massimo Campostrini, Andrea Pelissetto, Paolo Rossi, 
and Ettore Vicari}
\address{Dipartimento di Fisica dell'Universit\`a and I.N.F.N.,
I-56126 Pisa, Italy}

\maketitle

\begin{abstract}
Recently-generated long strong-coupling series for the two-point
Green's functions of asymptotically free ${\rm O}(N)$ lattice $\sigma$
models are analyzed, focusing on the evaluation of dimensionless
renormalization-group invariant ratios of physical quantities and
applying resummation techniques to series in the inverse temperature
$\beta$ and in the energy $E$.  Square, triangular, and honeycomb
lattices are considered, as a test of universality and in order to
estimate systematic errors.  Large-$N$ solutions are carefully studied
in order to establish benchmarks for series coefficients and
resummations.  Scaling and universality are verified.  All invariant
ratios related to the large-distance properties of the two-point
functions vary monotonically with $N$, departing from their large-$N$
values only by a few per mille even down to $N=3$.
\end{abstract}



\section{Introduction}
\label{introduction}

The properties of physical systems in the vicinity of a critical
point, such as critical exponents and amplitude ratios, can be
extracted by a variety of methods, ranging from exact solutions to
Monte Carlo simulations.

In the absence of exact results, one of the most successful approaches
is based on the investigation of the strong-coupling series expansion,
which enjoys the property of a finite radius of convergence, often
(but not necessarily) coinciding with the extent of the
high-temperature phase.  More generally, when no singular points occur
on the real axis of the complex coupling plane, it is possible to
exploit strong-coupling results even beyond the convergence radius by
analytic continuations, which are based on appropriate resummation
methods.  Extending the length of the strong-coupling series and
improving the accuracy of the resummations are therefore the two most
compelling tasks within this approach to the study of the behavior of
systems in the critical region.

As part of an extended program of strong-coupling calculations we have
recently computed an extended series expansion of all nontrivial
two-point Green's functions
\begin{equation}
G(x) = \left<{\vec s}(0)\cdot{\vec s}(x)\right>
\end{equation}
for the nearest-neighbor lattice formulation of two-dimensional 
${\rm O}(N)$ $\sigma$ models on the square, triangular, and honeycomb
lattices, respectively up to 21st, 15th, and 30th order in the
strong-coupling expansion parameter $\beta$.  A complete presentation
of our strong-coupling computations for ${\rm O}(N)$ $\sigma$ models
in two and three dimensions will appear in a forthcoming paper.
A preliminary report of our calculations can be found in 
Ref.~\cite{lattice95}.

The relevance of a better understanding of 2-$d$ ${\rm O}(N)$ $\sigma$
models cannot be overestimated.  They appear in condensed matter
literature as prototype models for critical phenomena that are
essentially restricted to two-dimensional layers, including some
instances of high-$T_c$ superconductivity.  Moreover, they can be
employed as model field theories sharing some of the most peculiar
features of four-dimensional gauge theories, such as asymptotic
freedom and spontaneous mass generation.  This last statement must
however be qualified, since the above-mentioned properties, according
to common lore, are possessed only by those 2-$d$ ${\rm O}(N)$ models
such that $N>2$.

We focus here on these asymptotically free models, analyzing their
strong-coupling expansion in order to extract information that may be
relevant to the description of their continuum limit
($\beta\to\infty$), assuming $\beta_c=\infty$ to be the only
singularity on the real axis.  This hypothesis is favored by all
numerical evidence as well as by the successful application of the
extrapolation techniques that we shall discuss in the present paper.
The analysis of our strong-coupling series for 
models with $N\geq 2$ is presented in Ref.~\cite{Nm2}.

It is obviously quite hard to imagine that strong-coupling techniques
may be really accurate in describing the divergent behavior of such
quantities as the correlation length and the magnetic susceptibility.
Nevertheless, as our calculations will explicitly confirm, the
strong-coupling analysis may provide quite accurate continuum-limit
estimates when applied directly to dimensionless,
renormalization-group invariant ratios of physical quantities.  Two
basic ideas will make this statement more convincing.

(i) For any dimensionless, renormalization-group invariant ratio
$R(\beta)$, when $\beta$ is sufficiently large we may expect a
behavior 
\begin{equation}
R(\beta)-R^*\sim {1\over \xi^2(\beta)},
\label{scalR}
\end{equation}
where $R^*$ is the fixed point (continuum) value and $\xi$ is the
(diverging) correlation length.  Hence a reasonable estimate of $R^*$
may be obtained at the values of $\beta$ corresponding to large but
finite correlation lengths, where the function $R(\beta)$ flattens.
This is essentially the same idea underlying Monte Carlo studies of
asymptotically free theories, based on the identification of the
so-called scaling region.

(ii) On physical grounds, it is understandable that $\beta$ is not
necessarily the most convenient variable to parameterize phenomena
occuring around $\beta=\infty$.  An interesting alternative is based
on the observation that the strong-coupling series of the internal
energy
\begin{equation}
E = \beta + O(\beta^3)
\end{equation}
may be inverted to give $\beta$ as a series in $E$.  This series may
be substituted into other strong-coupling expansions, obtaining
expressions for physical quantities as power series in $E$.  It might
now be easier to reach the continuum limit, since it now occurs at a
finite value of the expansion variable, i.e., $E\to1$.

We hope to convince the reader that, by exploiting these ideas,
state-of-the-art strong-coupling calculations can be made at least as
accurate as the best Monte Carlo simulations presently available,
when applied to dimensionless renormalization-group invariant quantities.

We must stress that the analysis of the strong-coupling series
calculated on different lattices offers a possibility of testing
universality, and, on the other side, once universality is assumed, it
represents a further check for possible systematic errors and allows
their quantitative estimate; this estimate is usually a difficult task
in strong-coupling extrapolation methods such as those based on Pad\'e
approximants and their generalizations.

Our physical intuition of the behavior of ${\rm O}(N)$ models is
strongly guided by our knowledge of their large-$N$ behavior, and by
the evidence of a very weak dependence on $N$ of the dimensionless
ratios.  In order to extend our understanding to those lattices that
have not till now received a systematic treatment, and also in order
to establish a benchmark for the strong-coupling analysis, we decided
to start our presentation with a detailed comparative study of the
large-$N$ limit of various lattices, in the nearest-neighbor
formulation.  To the best of our knowledge, only the large-$N$
solution on the square lattice was already known explicitly
\cite{sqNi}.

The paper is organized as follows:

In Sec.~\ref{secNi} we present the large-$N$ limit solution of ${\rm
  O}(N)$ $\sigma$ models on the square, triangular and honeycomb
lattices, in the nearest-neighbor formulation, calculating several
physical quantities and showing explicitly the expected universality
properties.  The triangular- and honeycomb-lattice results are
original, and possess some intrinsic reasons of interest.  However,
readers willing to focus on square-lattice results are advised to jump
to Sec.~\ref{SCA} after reading Subs.~\ref{secse} and \ref{secsqNi},
where the notation is fixed.

Sec.~\ref{SCA} is devoted to a detailed analysis of the available
strong-coupling series of $G(x)$ and other physical quantities on the
square, triangular, and honeycomb lattices.  Most of the results we
shall show there concern the $N=3$ model.  The basic motivation for
this choice lies in the observation that all dependence in $N$ is
monotonic between 3 and $\infty$; hence the discussion of higher-$N$
results would be only a boring repetition of the considerations
presented here.  The reader not interested in the analysis of
triangular and honeycomb lattices may skip most of the discussion, by
focusing on Subs.~\ref{scsq}, where further definitions are introduced
and the square-lattice series are analyzed, and on Subs.~\ref{concl},
where all conclusions are drawn.

Apps.~\ref{apptr} and \ref{appex} provide the derivation and the
technical details of the large-$N$ calculations on the triangular and
honeycomb lattices respectively.  We present as well the calculation
of the $\Lambda$-parameters.

App.~\ref{singNinf} is a study of the complex temperature
singularities of the $N=\infty$ partition functions on the 
triangular and honeycomb lattices.

In Apps.~\ref{appscsq}, \ref{appsctr} and \ref{appscex} we present,
for selected values of $N$, the strong-coupling series of some
relevant quantities on the square, triangular, and honeycomb lattice
respectively.

\section{The large-$\protect\bbox{N}$ limit of lattice 
$\protect\bbox{{\rm O}(N)}$ $\protect\bbox{\sigma}$ models}
\label{secNi}

\subsection{The large-$\protect\bbox{N}$ saddle point equation}
\label{secse}

The nearest-neighbor lattice formulations on square, triangular and
honeycomb lattices are defined by the action
\begin{equation}
S_L= -N\beta\sum_{\rm links} {\vec s}_{x_l}\cdot {\vec s}_{x_r},
\qquad {\vec s}_x\cdot {\vec s}_x = 1,
\label{lattaction}
\end{equation}
where $\vec s$ is a $N$-component vector, the sum is performed over
all links of the lattice and $x_l,x_r$ indicate the sites at the ends
of each link.  The coordination number is $c=4,6,3$ respectively for
the square, triangular and honeycomb lattice.  The lattice spacing
$a$, which represents the length unit, is defined to be the length of
a link.  The volume per site is then $v_s=1,\sqrt{3}/2, 3\sqrt{3}/4$
(in unit of $a^2$) respectively for the square, triangular, and
honeycomb lattice.

Straightforward calculations show that the correct continuum
limit of ${\rm O}(N)$ $\sigma$ models, 
\begin{equation}
S= {N\over 2t} \int d^2x\, \partial_\mu {\vec s}(x)\cdot
\partial_\mu {\vec s}(x),
\qquad {\vec s}(x)\cdot {\vec s}(x) = 1,
\label{contaction}
\end{equation}
is obtained by identifying  
\begin{equation}
t={1\over \beta}, \  {1\over \sqrt{3}\beta},\ 
{\sqrt{3}\over \beta},
\label{temp}
\end{equation}
respectively for the square, triangular and honeycomb lattice.
Notice that 
\begin{equation}
\lambda\equiv t\beta = {4v_s\over c}
\label{tbeta}
\end{equation}
is the distance between nearest-neighbor
sites of the dual lattice in unit of the lattice spacing $a$.

When the number of field components $N$ per site goes to infinity,
one can use a saddle point equation to evaluate the partition 
function. Replacing the constraint $\vec s_x^{\,2}=1$
by a Fourier integral over 
a conjugate variable $\alpha_x$, we write the partition
function as
\begin{eqnarray}
Z&&\propto \int \prod_x d{\vec s}_x \,\delta( \vec s_x^{\,2}-1)\,
\exp N\beta \sum_{\rm links} {\vec s}_{x_l}\cdot {\vec s}_{x_r}
\nonumber \\
&&\propto\int \prod_x d\phi_x d\alpha_x \,\exp 
N\left[ \sum_x i{\alpha_x\over 2}\left( 1 - \phi_x^2\right)
-{\beta\over 2}\sum_{\rm links} \left( 
\phi_{x_l}-\phi_{x_r}\right)^2\right].
\label{fp}
\end{eqnarray}
Integrating out the $\phi$ variables we arrive at
the expression 
\begin{equation}
Z\propto \int d\alpha_x \,\exp {N\over 2}
\left( \sum_x i\alpha_x - {\rm Tr}\,\ln R\right),
\label{fp2}
\end{equation}
where 
\begin{equation}
R_{xy}= -{1\over t} \Delta_{xy} + i\alpha_x\delta_{xy},
\label{R}
\end{equation}
and $\Delta_{xy}$ is a generalized Laplacian operator, such
that 
\begin{equation}
\lambda\, \sum_{\rm links} \left(\phi_{x_l}-\phi_{x_r}\right)^2
= -\sum_{x,y} \phi_x \Delta_{xy} \phi_y.
\label{Q}
\end{equation}

The large-$N$ limit solution is obtained from the variational 
equation with respect to $\alpha_x$.
Looking for a translation invariant solution we set
\begin{equation}
i\alpha_x={v_s\over t}\,z.
\label{costalp}
\end{equation}
The matrix $R$ then becomes
\begin{equation}
R_{xy}= {1\over t} \left[ -\Delta_{xy}+zv_s\delta_{xy}\right],
\label{R2}
\end{equation}
and the saddle point equation is written as
\begin{equation}
1=\lim_{N_s\rightarrow\infty}
{1\over N_s} {\rm Tr}\,R^{-1},
\label{spe}
\end{equation}
where $N_s$ is the number of sites.

The large-$N$ fundamental two-point Green's function is
obtained by
\begin{equation}
G(x-y)= R^{-1}_{xy}.
\label{NiGx}
\end{equation}

In order to calculate the trace of $R^{-1}$, the easiest
procedure consists in Fourier transforming the operator
$R$. Such transformation is straightforward on lattices,
like square and triangular lattices, whose sites
are related by a translation group, and in these cases it
yields the diagonalization of the matrix $R_{xy}$.
The honeycomb lattice, not possessing a full translation
symmetry, presents some complications. 
In this case a partial diagonalization of $R_{xy}$ can be 
achieved following the procedure outlined in Ref.~\cite{SCUN2}.

\subsection{The square lattice}
\label{secsqNi}

Turning to the momentum space the variational equation
becomes
\begin{equation}
{1\over t}=\beta =\int_{-\pi}^\pi {d^2 k\over (2\pi)^2}
{1\over \widehat{k}^2+z}={1\over 2\pi} 
\rho_{\rm s}(z) K\left( \rho_{\rm s}(z) \right), 
\label{sesq}
\end{equation}
where 
\begin{equation}
\rho_{\rm s}(z)=\left(1 + {1\over 4}z\right)^{-1},
\label{rhos}
\end{equation}
and $K$ is the complete integral of the first kind.

Let's define the moments of $G(x)$
\begin{equation}
m_{2j}\equiv\sum_x (x^2)^j \,G(x).
\label{momgx}
\end{equation}
Straightforward calculations lead to the following
results
\begin{equation}
\chi\equiv m_0 = {t\over z},
\label{chisqin}
\end{equation}
\begin{equation}
\xi_{G}^2 \equiv M_{G}^{-2} \equiv 
{m_2\over 4\chi}= {1\over z},
\label{xigsqin}
\end{equation}
\begin{equation}
 u\equiv {m_2^2\over \chi m_4}=
{1\over 4}\left( 1 + {z\over 16}\right)^{-1}.
\label{omsqin}
\end{equation}
Notice that in the large-$N$ limit the renormalization constant of the
fundamental field is $Z=t$. $u$ is a renormalization-group invariant
quantity.

The mass-gap should be extracted from the long distance
behavior of the two-point Green's function, which is also
related to the imaginary momentum singularity of the 
Fourier transform of $G(x)$.
In the absence of a strict rotation invariance, one actually
may define different estimators of the mass-gap having 
the same continuum limit.
On the square lattice one may consider $\mu_{\rm s}$ and 
$\mu_{\rm d}$ obtained respectively by the equations
\begin{eqnarray}
&&\tilde{G}^{-1}(p_1=i\mu_{\rm s},p_2=0)=0,\nonumber \\
&&\tilde{G}^{-1}\left(p_1=i{\mu_{\rm d}\over\sqrt{2}},
p_2=i{\mu_{\rm d}\over\sqrt{2}}\right) = 0.
\label{msmd}
\end{eqnarray}
$\mu_{\rm s}$ and $\mu_{\rm d}$ 
determine respectively the long
distance behavior of the side and diagonal wall-wall
correlations constructed with $G(x)$. 
In generalized Gaussian models, such as the large-$N$ limit
of ${\rm O}(N)$ models, it turns out convenient to define
the following quantities 
\begin{eqnarray}
&&M_{\rm s}^2= 2\left( {\rm cosh} 
\mu_{\rm s} - 1\right),\nonumber \\
&&M_{\rm d}^2= 4\left( {\rm cosh} 
{\mu_{\rm d}\over \sqrt{2}} -1\right).
\label{MsMd}
\end{eqnarray}
In the continuum limit 
\begin{equation}
{M_{\rm s}\over \mu_{\rm s}}\,,{M_{\rm d}
\over \mu_{\rm d}}\rightarrow 1,
\label{msmd2}
\end{equation}
therefore $M_{\rm s}$ and $M_{\rm d}$ may be also used
as estimators of the mass-gap. 

In the large-$N$ limit 
\begin{equation}
M_{\rm s}^2 = M_{\rm d}^2 = z=M_{G}^2.
\label{msmdmg}
\end{equation}

The rotational invariance of $G(x)$ at large distance,
$d\gg\xi$, is checked by the ratios $\mu_{\rm s}/\mu_{\rm d}$.
Using the above results one can evaluate the scaling violation terms:
\begin{equation}
{\mu_{\rm s}\over \mu_{\rm d}}=
{ 
\ln\left( {1\over 2}\sqrt{z} + \sqrt{ 1 + 
{1\over 4}z}\right)\over
\sqrt{2}\ln\left( {1\over 2\sqrt{2}}\sqrt{z} + \sqrt{ 1 + 
{1\over 8}z}\right)}
 = 1 - {1\over 48}z + {71\over 23040}z^2+
O\left(z^3\right).
\label{rotviol}
\end{equation}
 
Another test of scaling is provided by the ratio
\begin{equation}
{\mu_{\rm s}\over M_{G}}
= {2\over \sqrt{z}}
\ln\left( {\sqrt{z}\over 2} + \sqrt{ 1 + 
{z\over 4}}\right)=
1 - {1\over 24}z + {3\over 640}z^2+
O\left(z^3\right).
\label{scalviol}
\end{equation}

The internal energy can be easily calculated obtaining
\begin{equation}
E \equiv \langle {\vec s}_x\cdot {\vec s}_{x+\mu} \rangle =
R^{-1}_{x,x+\mu}=
1\,-\,{1\over 4\beta}\,+\,{z\over 4}.
\label{energysq}
\end{equation}
Therefore
\begin{equation}
{1\over 2}\sum_\mu
\langle ({\vec s}_{x+\mu}-{\vec s}_x)^2 \rangle =
{1\over 2\beta}\,-\,{z\over 2},
\label{condlatt}
\end{equation}
where the term proportional to $z$ is related to the condensate $T$ of
the trace of the energy-momentum tensor~\cite{CRcond}
\begin{equation}
{\beta(t)\over 2t^2}
\partial_\mu {\vec s}(x)\cdot\partial_\mu {\vec s}(x).
\label{temtr}
\end{equation}
In the large-$N$ limit
\begin{equation}
\beta(t)=-{1\over 2\pi}t^2,
\label{beta}
\end{equation}
therefore from the expression (\ref{energysq}) we deduce
\begin{equation}
{T\over M_{G}^2}={1\over 4\pi}.
\label{condsq}
\end{equation}

Another interesting quantity which can be evaluated in the large-$N$
limit is the zero-momentum four-point renormalized coupling constant,
defined by
\begin{equation}
g_r = -{\chi_4\over \chi^2\xi_{G}^2}
\label{gr0}
\end{equation}
where
\begin{equation}
\chi_4 = \sum_{x,y,z} \langle  {\vec s}_0\cdot {\vec s}_x 
\, {\vec s}_y\cdot {\vec s}_z \rangle_c.
\label{chi4}
\end{equation}
$g_r$ is zero in the large-$N$ limit, where the theory is Gaussian-like
and thus $\chi_4=0$. 
Its value in the continuum limit
\begin{equation}
g_r^* = {8\pi\over N}+ O\left({1\over N^2}\right)
\label{gr1}
\end{equation}
can be also evaluated in the large-$N$ expansion of
the continuum formulation of the ${\rm O}(N)$ models~\cite{gr}.
On the square lattice, by using the saddle point equation we find
\begin{equation}
Ng_r = -2 {\partial \ln z\over \partial \beta},
\label{gr2}
\end{equation}
which can be made more explicit by writing
\begin{equation}
Ng_r = 4\pi{1+\rho_{\rm s}\over \rho_{\rm s} E(\rho_{\rm s})} =
8\pi\left[1 + {z\over 8}\left(\ln {z\over 32}\,+\,2\right) 
+ O(z^2)\right],
\label{gr3}
\end{equation}
where $E$ is an elliptic function.

All the above results can be expressed as functions of $\beta$
by solving the saddle point equation.
Concerning asymptotic scaling, and therefore 
solving the saddle point equation at large $\beta$, one finds
\begin{equation}
M_{G}\simeq 4\sqrt{2}\,\exp \left(-{2\pi\over t}\right).
\label{asysq}
\end{equation}

The analytic structure of the various observables
has been investigated in Ref.~\cite{BCMO}.
The complex $\beta$-singularities are square-root branch points,
indeed quantities like $\chi$ and $\xi_G^2$
behave as 
\begin{equation}
A(\beta)+B(\beta) \sqrt{\beta-\beta_s}
\end{equation}
around a singular point $\beta_s$, where $A(\beta)$ and $B(\beta)$ are
regular in the neighborhood of $\beta_s$.  The singularities closest
to the origin are located at $\bar{\beta}=0.32162\,(\pm 1\pm i)$.
Such singularities determine the convergence radius of the
strong-coupling expansion, which is therefore $\beta_r=0.45484$,
corresponding to a correlation length $\xi_{G}=3.17160$.

\subsection{The triangular lattice}
\label{sectrNi}

On the triangular lattice, using the results of App.~\ref{apptr}, 
the saddle point equation can be written as
\begin{equation}
{1\over t}=\sqrt{3}\beta = \int^\pi_{-\pi} {dk_1\over 2\pi}
\int^{2\pi/\sqrt{3}}_{-2\pi/\sqrt{3}}
{dk_2\over 2\pi} {1\over \Delta(k)+z}
\label{setr}
\end{equation}
where 
\begin{equation}
\Delta(k)=4\left[ 1 - {1\over 3}\left(
\cos k_1+2\cos {k_1\over 2}\cos {\sqrt{3}k_2\over 2}\right)\right]
\label{deltatr}
\end{equation}
and the momentum integration is performed over the Brillouin
zone corresponding to a triangular lattice.
By rather straightforward calculations (making also use
of some of the formulas of Ref.~\cite{Gradshteyn})
the saddle point equation can be written as
\begin{equation}
{1\over t}=\sqrt{3}\beta =
{1\over 2\pi} \left( 1 + {z\over 6}\right)^{-1/4}\,\rho_{\rm t}(z)
\,K(\rho_{\rm t}(z)),
\label{setr2}
\end{equation}
where
\begin{equation}
 \rho_{\rm t}(z)= \left( 1+{z\over 6}\right)^{1/4}\,
\left[ {1\over 2} + {z\over 8} + {1\over 2}
\left( 1 + {z\over 6}\right)^{1/2} \right]^{-1/2}\, 
\left[ {5\over 2} + {3z\over 8} - {3\over 2}
\left( 1 + {z\over 6}\right)^{1/2} \right]^{-1/2}. 
\label{setr3}
\end{equation}

Using the results of App.~\ref{apptr} one can find
\begin{equation}
\chi={t\over v_s z}={2\over 3\beta z},
\label{chitrin}
\end{equation}
\begin{equation}
 \xi_{G}^2 \equiv M_{G}^{-2} =  {1\over z}\, ,
\label{xitrin}
\end{equation}
\begin{equation}
u\equiv {m_2^2\over \chi m_4}
={1\over 4}\left( 1 + {z\over 16}\right)^{-1} .
\label{omtrin}
\end{equation}

An estimator of the mass-gap $\mu_{\rm t}$ can be extracted from the
long distance behavior of the 
wall-wall correlation function defined in
Eq.~(\ref{walldeftr}), indeed for $x\gg 1$
\begin{equation}
G_{\rm t}^{(\rm w)}(x)\propto e^{-\mu_{\rm t} x}.
\label{ldgtr}
\end{equation}
In the large-$N$ limit one finds
\begin{equation}
M^2_{\rm t}\equiv{8\over 3}\left( {\rm cosh} {\sqrt{3}\over 2}
\mu_{\rm t} -1\right)=z=M^2_{G}.
\label{Mtr}
\end{equation}
A test of scaling is provided by the ratio 
\begin{equation}
{\mu_{\rm t}\over M_{G}}=
{2\over \sqrt{3z}}
{\rm Arccosh}\,\left[ 1 + {3\over 8}z \right]
 =  1-{1\over 32}z + {9\over 10240} z^2
+O\left( z^3\right),
\label{scalvioltr}
\end{equation}
where scaling violations are of the same order as those 
found on the square lattice for the corresponding
quantity, cfr.\ Eq.~(\ref{scalviol}).

The internal energy is given by the following
expression
\begin{equation}
E=\langle {\vec s}_{x_l}\cdot {\vec s}_{x_r}\rangle =
1 - {1\over 6\beta} + {z\over 4} ,
\label{etr}
\end{equation}
leading again to the result (\ref{condsq}) 
for the condensate of the trace of the
energy-momentum tensor, in agreement with universality. 

We calculated $g_r$ on the triangular lattice, finding 
the following expression (in the derivation we made use of the
saddle point equation (\ref{setr}))
\begin{equation}
Ng_r = - {2\over \sqrt{3}} {\partial \ln z\over \partial \beta},
\label{gr2t}
\end{equation}
which can be written in a more explicit form using 
Eq.~(\ref{setr2}): 
\begin{eqnarray}
 Ng_r&=&4\pi\left( 1 + {1\over 6}z\right)^{1/4}
{1\over z}\left[ {E(\rho_{\rm t})\over 1-\rho_{\rm t}^2} 
{\partial \rho_{\rm t}\over \partial z}- 
{1\over 24}\left(1+{1\over 6}z\right)^{-1}\rho_{\rm t} 
  K(\rho_{\rm t})\right]^{-1}
\nonumber \\
&=&
8\pi\left[1 + {z\over 8}\left(\ln {z\over 48}\,+\,{11\over 6}\right) 
+ O(z^2)\right],
\label{gr3t}
\end{eqnarray}
where the continuum value of $Ng_r$, obtained for $z\rightarrow 0$, is
in agreement with the results (\ref{gr1}) and (\ref{gr3}).

In the weak coupling region $t\rightarrow 0$ the
saddle point equation leads to the asymptotic
scaling formula
\begin{equation}
M_{G}\simeq 4\sqrt{3} \exp \left( 
-{ 2\pi\over t}\right).
\label{asytr}
\end{equation}
The equations (\ref{asysq}) and
(\ref{asytr}) are in agreement with the large-$N$ limit of the  
ratio of the $\Lambda$-parameters of the square 
and triangular lattice formulations
calculated in App.~\ref{apptr}, cfr.\ Eq.~(\ref{ratioltr2}),
using perturbation theory.

We have investigated the analytic structure in the complex
$\beta$-plane. Details of such study are presented in
App.~\ref{singNinf}.  As on the square lattice, the singularities are
square-root branch points. Those closest to the origin are placed at
$\bar{\beta}= 0.206711\pm \,i\,0.181627$, leading to a convergence
radius for the strong-coupling expansion $\beta_r=0.275169$, which
corresponds to a correlation length $\xi_{G}=2.98925$.

\subsection{The honeycomb lattice}
\label{iNhl}

The analysis of models defined on the honeycomb lattice presents a few
subtleties caused by fact that, unlike square and triangular lattices,
not all sites are related by a translation, not allowing a
straightforward definition of a Fourier transform. Nevertheless,
observing that sites at even distance in the number of lattice links
form triangular lattices, one can define a Fourier-like transformation
that partially diagonalizes the Gaussian propagator (up to $2\times 2$
matrices)~\cite{SCUN2}.  In this section we present the relevant
results, some details of their derivation are reported in
App.~\ref{appex}.

Using the expression of $R^{-1}$ of Eq.~(\ref{greengex}) 
we write the saddle point equation in the following form
\begin{equation}
{1\over t}={\beta\over\sqrt{3}}=
\int^{{2\over3}\pi}_{-{2\over3}\pi} {dk_1\over 2\pi}
\int^{\pi/\sqrt{3}}_{-\pi/\sqrt{3}}
{dk_2\over 2\pi} {1+{1\over 4}z\over 
\Delta(k)+z\left(1+{1\over 8}z\right)}
\label{seex}
\end{equation}
where 
\begin{equation}
\Delta(k)={8\over 9}\left[ 2 - \cos {\sqrt{3}\over 2}k_2
\left( \cos {3\over 2}k_1  + \cos {\sqrt{3}\over 2}k_2\right)  
\right],
\label{deltaex}
\end{equation}
and integrating over the momentum  we arrive at
\begin{equation}
{1\over t}={\beta\over\sqrt{3}}=
{1\over 2\pi} \left( 1 + {z\over 4}\right)^{1/2}
\rho_{\rm h}(z) \,K(\rho_{\rm h}(z)),
\label{seex2}
\end{equation}
where
\begin{equation}
\rho_{\rm h}(z)= \left( 1 + {z\over 4}\right)^{1/2} 
\left( 1 + {3z\over 8}\right)^{-3/2} 
\left( 1 + {z\over 8}\right)^{-1/2}. 
\label{seex3}
\end{equation}

From Eq.~(\ref{greengex}) we also derive
\begin{equation}
\chi ={t\over v_s z}={4\over 3\beta z},
\label{chihoin}
\end{equation}
\begin{equation}
\xi_{G}^2 \equiv M^{-2}_{G}=
{1\over z},
\label{xihoin}
\end{equation}
\begin{equation}
u ={1\over 4}\left( 1 + {z\over 16}\right)^{-1}.
\label{omhoin}
\end{equation}

The two orthogonal wall-wall correlation functions 
$G^{(\rm w)}_{\rm v}(x)$ and $G^{(\rm w)}_{\rm h}(x)$ defined in
Eqs.~(\ref{g1}) and (\ref{g2}) allow one to define two estimators of
the mass-gap from their long distance behavior
\begin{eqnarray}
G^{(\rm w)}_{\rm v}(x)\propto e^{-\mu_{\rm v} x},\nonumber\\ 
G^{(\rm w)}_{\rm h}(x)\propto e^{-\mu_{\rm h} x},
\label{ldbg}
\end{eqnarray}
where $x$ is the distance between the two walls in 
unit of the lattice spacing.
In the continuum limit $\mu_{\rm v}=\mu_{\rm h}$ 
and they both reproduce
the physical mass propagating in the fundamental channel.
As on the square and triangular lattices,   it is convenient
to define the quantities
\begin{eqnarray}
&&M_{\rm v}^2 = {8\over 9}\left( {\rm cosh} {3
\mu_{\rm v}\over 2} - 1\right),\nonumber \\
&&M_{\rm h}^2 = {8\over 3}\left( {\rm cosh} 
{\sqrt{3}\mu_{\rm h}\over 2} -1\right),
\label{M1M2}
\end{eqnarray}
which, in the continuum limit, 
are also estimators of the mass gap. 
In the large-$N$ limit one finds
\begin{eqnarray}
&&M_{\rm v}^2 = z\left( 1+ {z\over 8}\right),\nonumber \\
&&M_{\rm h}^2=z.
\label{M1M22}
\end{eqnarray}
Notice that in the continuum large-$N$ limit the result  
\begin{equation}
{M\over M_{G}}=1,
\label{momg}
\end{equation}
where $M$ is any mass-gap
estimator, is found for all lattice formulations considered.

On the honeycomb lattice the maximal violation of full rotational
symmetry occurs for directions differing by a $\pi/6$ angle, and
therefore, taking into account its discrete rotational symmetry, also
by a $\pi/2$ angle. So a good test of rotation invariance of $G(x)$ at
large distance is provided by the ratio $\mu_{\rm v}/\mu_{\rm h}$:
\begin{equation}
{\mu_{\rm v}\over \mu_{\rm h}}=
{
{\rm Arccosh}\,\left[ 1 + {9\over 8}z \left( 1 + {1\over 8}
z\right)\right]
\over
\sqrt{3}{\rm Arccosh}\,\left[ 1 + {3\over 8}z \right]}
= 1+{1\over 640} z^2
+O\left( z^3\right).
\label{rotscalho}
\end{equation}
As expected from the better rotational symmetry of the honeycomb
lattice, rotation invariance is set earlier than for the square
lattice, indeed the $O(z)$ scaling violation is absent.

A test of scaling is provided by the ratio 
\begin{equation}
{\mu_{\rm h}\over M_{G}}=
{2\over \sqrt{3z}}
{\rm Arccosh}\,\left[ 1 + {3\over 8}z \right]
= 1-{1\over 32}z + {9\over 10240} z^2
+O\left( z^3\right),
\label{scalviolho}
\end{equation}
where scaling violations are of the same order of those 
found on the square lattice for the corresponding
quantity, cfr.\ Eq.~(\ref{scalviol}).

The internal energy is given by
\begin{equation}
E=1 - {1\over 3\beta} + {z\over 4}
=1 - {1\over 3\beta} + {M_{G}^2\over 4},
\label{eneex}
\end{equation}
where the term proportional to $M_{G}^2$
verifies again universality.

In the weak coupling region $t\rightarrow 0$ the
saddle point equation leads to the asymptotic
scaling formula
\begin{equation}
M_{G}\simeq 4 \exp \left( 
-{ 2\pi\over t}\right).
\label{asyex}
\end{equation}
The equations (\ref{asysq}) and
(\ref{asyex}) are in agreement with the large-$N$ limit of the  
ratio of the $\Lambda$-parameters of the square 
and triangular lattice formulations
calculated in App.~\ref{appex}, cfr.\ Eq.~(\ref{ratiolho2}),
using perturbation theory.

In Fig.~\ref{asyiN} we compare asymptotic scaling from
the various lattices considered, plotting the ratio
between $M_{G}$ and the corresponding
asymptotic formula (cfr.\ Eqs.~(\ref{asysq}),
(\ref{asytr}) and (\ref{asyex})).
Notice that in the large-$N$ limit corrections to 
asymptotic scaling are $O(M^2_{G})$, in that corrections
$O(1/\ln M_{G})$ are suppressed by a factor $1/N$.

We have investigated the analytic structure in the complex 
temperature-plane of the $N=\infty$ model on the honeycomb lattice
(details are reported in App.~\ref{singNinf}).
As on the square and triangular lattices,
singularities are square-root branch points, and those closest to the
origin are placed on the imaginary axis
at $\bar{\beta}=\pm i 0.362095$. 
The convergence radius for the strong-coupling expansion
is associated to a quite small correlation length:
$\xi_{G}=1.00002$.

\section{Continuum results from strong coupling}
\label{SCA}

\subsection{Analysis of the series}
\label{analysis}

In this section we analyze the strong-coupling series of some of the
physical quantities which can be extracted from the two-point
fundamental Green's function.  We especially consider dimensionless
renormalization-group invariant ratios, whose value in the scaling
region, i.e., their asymptotic value for $\beta\rightarrow \infty$,
concerns the continuum physics.  Some strong-coupling series for
selected values of $N$ are reported in the Apps.~\ref{appscsq},
\ref{appsctr} and \ref{appscex} respectively for the square,
triangular and honeycomb lattices.  The series in the energy are
obtained by inverting the strong-coupling series of the energy
$E=\beta+O(\beta^3)$ and substituting into the original series in
$\beta$.

Our analysis of the series of dimensionless renormalization 
group invariant ratios of physical quantities,
such as those defined in the previous section, is 
based on Pad\'e approximant (PA) techniques.
For a review on the resummation techniques cfr.\ 
Ref.~\cite{Guttmann}.

PA's are expected to converge well to meromorphic
analytic functions. More flexibility is achieved by applying
the PA analysis to the logarithmic derivative  
(Dlog-PA analysis), and therefore enlarging the class
of functions which can be reproduced to those having
branch-point singularities.
The accuracy and the convergence of the PA's  
depend on how well the function considered, 
or its logarithmic derivative, can be reproduced by a 
meromorphic analytic function, and may change when considering
different representations of the same quantity.
By comparing the results from different series representations
of the same quantity one may check for possible
systematic errors in the resummation procedure employed.

In our analysis we constructed $[l/m]$ PA's and Dlog-PA's
of both the series in $\beta$ and in the energy.
$l,m$ are the orders of the polynomials
respectively at the numerator and at the denominator
of the ratio forming the $[l/m]$ PA of the series at hand, 
or of its logarithmic derivative (Dlog-PA).
While $[l/m]$ PA's provide directly 
the quantity at hand, in a Dlog-PA analysis one gets 
a $[l/m]$ approximant by reconstructing the original quantity
from the $[l/m]$ PA of its logarithmic derivative,
i.e., a $[l/m]$ Dlog-PA of the series $A(x)=\sum_{i=0}^\infty a_i x^i$
is obtained by
\begin{equation}
A_{l/m}(x) =  a_0\exp \int_0^x 
dx' \, {\rm Dlog}_{l/m} A(x').
\label{appA}
\end{equation}
where ${\rm Dlog}_{l/m} A(x)$ indicates the $[l/m]$ PA
of the logarithmic derivative of $A(x)$.

We recall that a $[l/m]$ PA uses $n=l+m$ terms of the series,
while a $[l/m]$ Dlog-PA requires $n=l+m+1$ terms.
Continuum estimates are then obtained by evaluating the approximants
of the energy series at $E=1$, and those of the $\beta$
series at a value of $\beta$ corresponding to a reasonably
large correlation length. 

As final estimates we take the average of the results from the
quasi-diagonal (i.e., with $l\simeq m$) PA's using all available terms
of the series. The errors we will display are just indicative, and
give an idea of the spread of the results coming from different PA's.
They are the square root of the variance around the estimate of the
results, using also quasidiagonal PA's constructed from shorter
series.  Such errors do not always provide a reliable estimate of the
uncertainty, which may be underestimated especially when the structure
of the function (or of its logarithmic derivative) is not well
approximated by a meromorphic analytic function. In such cases a more
reliable estimate of the real uncertainty should come from the
comparison of results from the analysis of different series
representing the same quantity, which in general are not expected to
have the same analytic structure.

In the following of this section we present 
the main results obtained from our strong-coupling analysis.
Most of them will concern the $N=3$ case.

\subsection{The square lattice}
\label{scsq}

On the square lattice we have calculated the two-point Green's
function up to $O(\beta^{21})$, from which we have extracted
strong-coupling series of the quantities $E$, $\chi$, $\xi_{G}^2$,
$u$, $M_{\rm s}^2$, $M_{\rm d}^2$, already introduced in
Sec.~\ref{secsqNi}, and of the ratios $r\equiv M_{\rm s}^2/M_{\rm
  d}^2$, $s\equiv M_{\rm s}^2/M_{G}^2$.  Some of the above series for
selected values of $N$ are reported in App.~\ref{appscsq}.  Our
strong-coupling series represent a considerable extension of the 14th
order calculations of Ref.~\cite{Luscher}, performed by means of a
linked cluster expansion, which have been rielaborated and analyzed in
Ref.~\cite{Butera}.  We also mention recent works where the linked
cluster expansion technique has been further developed and
calculations of series up to 18th order~\cite{Reisz} and 19th
order~\cite{Butera2} for d=2,3,4 have been announced.

In order to investigate the analytic structure in the complex
$\beta$-plane we have performed a study of the singularities of the
Dlog-PA's of the strong-coupling series of $\chi$ and $\xi_{G}^2$.  As
expected by asymptotic freedom, no indication of the presence of a
critical point at a finite real value of $\beta$ emerges from the
strong-coupling analysis of $N\geq 3$ models, confirming earlier
strong-coupling studies~\cite{Butera}.  The singularities closest to
the origin, emerging from the Dlog-PA analysis of $\chi$ and
$\xi_{G}^2$, are located at a pair of complex conjugate points, rather
close to the real axis in the $N=3$ case (where $\bar{\beta}\simeq
0.59\pm i 0.16$) and moving, when increasing $N$, toward the
$N=\infty$ limiting points $\bar{\beta}=0.32162\,(1\pm i)$.  In Table
\ref{zeroes} such singularities are reported for some values of $N$.
The singularity closest to the origin determines the convergence
radius of the corresponding strong-coupling series. For example for
$N=3$ the strong-coupling convergence radius turns out to be
$\beta_r\simeq 0.61$, which corresponds to a quite large correlation
length $\xi\simeq 65$.  We recall that the partition function on the
square lattice has the symmetry $\beta \rightarrow -\beta$, which must
be also realized in the locations of its complex singularities.

By rotation invariance the ratio 
$r\equiv M_{\rm s}^2/M_{\rm d}^2$ should go to one
in the continuum limit. Therefore the analysis of such ratio
should be considered as a test of the procedure employed 
to estimate continuum physical quantities. 
In the large-$N$ limit $r=1$ at all values of $\beta$.
This is not anymore true 
at finite $N$, where the strong-coupling series of 
$M_{\rm s}^2$ and $M_{\rm d}^2$ differ from each other, 
as shown in App.~\ref{appscsq}.
From $G(x)$ up to $O(\beta^{21})$ we could calculate
the ratio $r$ up to $O(\beta^{14})$. 
The results of our analysis of the series of $r$ for $N=3$
are summarized in Table \ref{sqr}. There we report
the values of the PA's and Dlog-PA's of the $E$-series at $E=1$, 
and of those of the $\beta$-series at $\beta=0.55$,
which corresponds to a reasonably large 
correlation length $\xi\simeq 25$.
We considered PA's and Dlog-PA's with 
$l+m\geq 11$ and $m\geq l\geq 5$.
The most precise determinations of $r^*$,
the value of $r$ at the continuum limit,  come from Dlog-PA's,
whose final estimates are $r^*=1.0000(12)$ from the $E$-approximants,
and $r^*=1.0002(6)$ from the $\beta$-approximants (at $\beta=0.55$).
The precision of these results is remarkable.

For all $N\geq 3$ the violation of rotation invariance in the large
distance behavior of $G(x)$, monitored by the ratio 
$\mu_{\rm s}/\mu_{\rm d}$, turns out quantitatively very close to that
at $N=\infty$ when considered as function of $\xi_G$ (in a plot the
$N=3$ curve of $\mu_{\rm s}/\mu_{\rm d}$ versus $\xi_{G}$ as obtained
from the strong-coupling analysis would be hardly distinguishable from
the exact $N=\infty$ one).  $\mu_{\rm s}/\mu_{\rm d}$ is one within
about one per mille already at $\xi\simeq 4$.

Calculating a few more components of $G(x)$ at larger orders
(i.e., those involved by the wall-wall correlation
function at distance 6 and 7 respectively up to $O(\beta^{22})$
and $O(\beta^{23})$),
we computed the ratio 
\begin{equation}
s\equiv {M_{\rm s}^2\over M_{G}^2}
\label{sdef}
\end{equation}
up to $O(\beta^{16})$, by applying the technique described in
Refs.~\cite{SCUN1,RV}.  We recall that at $N=\infty$ we found $s=1$
independently of $\beta$.  No exact results are known about the
continuum limit $s^*$ of the ratio $s$, except for its large-$N$
limit: $s^*=1$.  Both large-$N$ and Monte Carlo estimates indicate a
value very close to one.  From a $1/N$ expansion~\cite{Flyv,CR}:
\begin{equation}
s^*= 1 - {0.006450\over N} + O\left( {1\over N^2}\right).
\label{largeNs}
\end{equation}
Monte Carlo simulations at $N=3$~\cite{Meyer}
gave $s=0.9988(16)$ at $\beta={1.7/3}=0.5666...$ ($\xi\simeq 35$), and
$s=0.9982(18)$ at $\beta=0.6$ ($\xi\simeq 65$),
leading to the estimate $s^*=0.9985(12)$.

In Table \ref{sqs} we report, for $N=3$, the values of PA's and
Dlog-PA's of the energy and $\beta$ series of $s$ respectively at
$E=1$ and at $\beta=0.55$.  We considered PA's and Dlog-PA's with
$l+m\geq 13$ and $m\geq l\geq 5$.  Combining PA and Dlog-PA results,
our final estimates are $s^*=0.998(3)$ from the $E$-approximants, and
$s^*=0.998(1)$ from the $\beta$ approximants evaluated at
$\beta=0.55$, in full agreement with the estimates from the $1/N$
expansion and Monte Carlo simulations.  With increasing $N$, the
central estimate of $s^*$ tends to be closer to one.

The scaling-violation pattern of the quantity $\mu_{\rm s}/M_{G}$ for
$N=3$ is similar to the pattern for $N=\infty$ (cfr.\ 
Eq.~(\ref{scalviol})), i.e., it is stable within a few per
mille for $\xi\gtrsim 5$.

Another dimensionless renormalization-group invariant quantity we have
considered is $u\equiv m_2^2/(\chi m_4)$, whose large-$N$ limit has
been calculated in the previous section, cfr.\ Eq.~(\ref{omsqin}). 
At finite $N$ its continuum limit $u^*$ is not known. 
From the expression of the self-energy calculated up to
$O(1/N)$~\cite{Flyv,CR,CRselfenergy}, one can obtain
\begin{equation}
u^*= {1\over 4}\left[ 1 - {0.006198\over N} 
    + O\left( {1\over N^2}\right) \right].
\label{largeNu}
\end{equation}
It is interesting to notice that the $O(1/N)$ correction in
Eqs.~(\ref{largeNs}) and (\ref{largeNu}) is very small.

At $N=3$ the analysis of the $O(\beta^{21})$ strong-coupling series of
$u$ detected a simple pole close to the origin at $\beta_0=-0.085545$
for the $\beta$-series, and at $E_0=-0.086418$ for the energy series,
corresponding to $M^2_{G}=-16.000$, which, within the precision of our
strong-coupling estimate, is also the location of the pole in the
corresponding $N=\infty$ expression (\ref{omsqin}).  Being a simple
pole, this singularity can be perfectly reproduced by a standard PA
analysis, and indeed we found PA's to be slightly more stable than
Dlog-PA's in the analysis of $u$.  The results concerning $N=3$,
reported in Table \ref{sqom} (for PA's with $l+m\geq 16$ and $m\geq
l\geq 8$), lead to the estimates $u^*=0.2498(6)$ from the energy
analysis, and $u^*=0.2499(5)$ form the $\beta$ analysis (at
$\beta=0.55$).  The agreement with the large-$N$ formula (\ref{largeNu})
is satisfactory.
In Fig.~\ref{figomsq} the curve $u(E)$ as obtained
from the $[10/10]$ PA and the exact curve $u(E)$ at $N=\infty$ (cfr.\ 
Eq.~\ref{omsqin}) are plotted, showing almost no differences.

In Table \ref{sum} we give a
summary of the determinations of $r^*$, $s^*$, and $u^*$ 
from PA's and Dlog-PA's of the energy and $\beta$-series.

We mention that we also tried to analyze series in the variable
\begin{equation}
z={I_{N/2}(N\beta)\over I_{N/2-1}(N\beta)},
\label{chcoeff}
\end{equation}
which is the character coefficient of the fundamental representation.
As for the $E$-series, the continuum limit should be reached at a
finite value $z\rightarrow 1$, and estimates of $r^*$,$s^*$ and $u^*$
may be obtained evaluating the approximants of the corresponding
$z$-series at $z=1$.  We obtained results much less precise than those
from the analysis of the $E$-series.  Maybe because of the
thermodynamical meaning of the internal energy, resummations by PA's
and Dlog-PA's of the $E$-series turn out much more effective,
providing rather precise results even at the continuum limit $E=1$.

The strong-coupling approach turns out to be less effective to the
purpose of checking asymptotic scaling.  In Table \ref{mc}, we
compare, for $N=3,4,8$, $\xi_{G}$ as obtained from the plain 
21st order series of $\xi_{G}^2$ and from its Dlog-PA's with
some Monte Carlo results available in the literature.  Resummation
by integral approximants~\cite{IA} provides results substantially
equivalent to those of Dlog-PA's.  For $N=3$ Dlog-PA's follow
 Monte Carlo data reasonably well up to about the convergence radius
$\beta_r\simeq 0.6$ of the strong-coupling expansion, but they fail
beyond $\beta_r$. On the other hand it is well known that for $N=3$
the asymptotic scaling regime is set at larger
$\beta$-values~\cite{CEPS}.  More sophisticated analysis can be found
in Refs.~\cite{Butera,Bonnier}, but they do not seem to lead to a
conclusive result about the asymptotic freedom prediction in the
$O(3)$ $\sigma$ model.  At larger $N$, the convergence radius
decreases, but on the other hand the asymptotic scaling regime should
be reached earlier.  At $N=4$ and $N=8$ the 21st order plain
series of $\xi_{G}^2$ provides already quite good estimates of $\xi_G$
within the convergence radius when compared with Monte Carlo results.
Again Pad\'e-type resummation fails for $\beta>\beta_r$.  We mention
that at $N=4$ the convergence radius $\beta_r\simeq 0.60$ corresponds
to $\xi_G\simeq 25$, and at $N=8$ $\beta_r\simeq 0.55$ corresponds to
$\xi_G\simeq 8$.

In order to check asymptotic scaling we consider the ratio
$\Lambda_{\rm s}/\Lambda_{2l}$, where
$\Lambda_{\rm s}$ is the effective $\Lambda$-parameter which
can be extracted by
\begin{equation}
\Lambda_{\rm s}\equiv\left( {\Lambda_{\rm s}\over M}\right)
M={M\over R_{\rm s}},
\label{effla}
\end{equation}
where $M$ is 
an estimator of the mass-gap,
$R_{\rm s}$ is the mass-$\Lambda$ parameter ratio
in the square lattice nearest-neighbor formulation~\cite{Hasenfratz}
\begin{equation}
R_{\rm s}= R_{\overline{\rm MS}}\times
\left( {\Lambda_{\overline{\rm MS}}\over 
\Lambda_{\rm s}}\right)=\left( {8\over e}\right)^{1\over N-2}
{1\over \Gamma\left( 1 + {1\over N-2}\right)}\times
\sqrt{32} \exp\left[ {\pi\over 2(N-2)}\right],
\label{RL}
\end{equation}
and $\Lambda_{2l}$ is the corresponding two-loop formula
\begin{equation}
\Lambda_{2l}= \left( {2\pi N \over N-2}\beta\right)^{1\over N-2}
\exp \left( -{2\pi N\over N-2}\beta\right).
\label{twoloopla}
\end{equation}
The ratio $\Lambda_{\rm s}/\Lambda_{2l}$ should go to one in the
continuum limit, according to asymptotic scaling.  The available
series of $M_G^2$ are longer than any series of the mass-gap
estimators; therefore, neglecting the very small difference between
$M_{G}$ and $M$ (we have seen that for $N\geq 3$ $(M_{G}-M)/M\lesssim
10^{-3}$ in the continuum limit), for which formula (\ref{RL})
holds, we use $M_{G}$ as estimator of $M$.  In Fig.~\ref{asysc} we
plot $\Lambda_{\rm s}/ \Lambda_{2l}$ for various values of $N$,
$N=3,4,8$, and for comparison the exact curve for $N=\infty$.  As
already noted in Ref.~\cite{Wolff} by a Monte Carlo study, for
$N=3,4$ at $\xi\simeq 10$ the asymptotic scaling regime is still far
(about 50\% off at $N=3$ and $15\%$ at $N=4$), while for $N=8$ it is
verified for $\xi\gtrsim 4$ within a few per cent, and notice that the
convergence radius $\beta_r\simeq 0.55$ corresponds to $\xi\simeq 8$.
Anyway with increasing $N$ curves of $\Lambda_{\rm s}/ \Lambda_{2l}$
clearly approach the exact $N=\infty$ limit.

\subsection{The triangular lattice}
\label{sctr}

On the triangular lattice we have calculated the two-point Green's
function up to $O(\beta^{15})$, from which we have extracted
strong-coupling series of the quantities $E$, $\chi$, $\xi_{G}^2$,
$u$, $M_{\rm t}^2$, already introduced in Sec.~\ref{sectrNi}, and of
the ratios $s\equiv M_{\rm t}^2/M_{G}^2$.  Some of the above series
for $N=3$ are reported in App.~\ref{appsctr}.

Like ${\rm O}(N)$ $\sigma$ models on the square lattice, 
no indication of the presence
of a critical point at a finite real value of $\beta$ emerges
from the strong-coupling analysis for $N\geq 3$. 
By a Dlog-PA analysis of the $O(\beta^{15})$
strong-coupling series of $\chi$ and $\xi_{G}^2$ at $N=3$, 
we found that the singularities closest to the origin is 
$\bar{\beta}\simeq 0.358\pm i 0.085$, giving rise to a convergence
radius $\beta_r\simeq 0.37$ which should correspond to a rather
large correlation length: $\xi_{G}\simeq 70$.
Increasing $N$ such singularities move toward their
$N=\infty$ limit $\bar{\beta}=0.206711\pm\,i\,0.181627$.
Some details of this analysis are given in Table \ref{zeroes}.

In our analysis of dimensionless quantities we considered, as on the
square lattice, both the series in the energy and in $\beta$.  The
estimates concerning the continuum limit are obtained by evaluating
the approximants of the energy series at $E=1$, and those of the
$\beta$-series at a $\beta$ associated to a reasonably large
correlation length.  For $N=3$ we chose $\beta=0.33$, whose
corresponding correlation length should be $\xi\simeq 22$, according
to a strong-coupling estimate.

Calculating a few more components of $G(x)$ at larger orders (i.e.,
those involved by the wall-wall correlation function at distance
$\sqrt{3}/2 \times 5$ up to $O(\beta^{16})$), we computed the
ratio $s\equiv M_{\rm t}^2/M_{G}^2$ up to $O(\beta^{11})$
\cite{SCUN1,RV}.  For $N=3$ the analysis of the strong-coupling series
of $s$ (some details are given in Table \ref{trs}) leads to the
estimate $s^*=0.998(3)$ from the energy approach, and $s^*=0.998(1)$
evaluating the approximants at $\beta=0.33$ (we considered PA's and
Dlog-PA's with $l+m\geq 8$ and $m\geq l\geq 4$).  Such results are in
perfect agreement with those found for the square lattice.

PA's and Dlog-PA's (with 
$l+m\geq 11$ and  $m\geq l\geq 5$)
of the strong-coupling series of $u$ expressed 
in terms of the energy, evaluated at $E=1$, lead to the estimate
$u^*=0.249(1)$ at $N=3$. The analysis of the series in $\beta$
gives $u^*=0.2502(4)$.
Again universality is satisfied. 

A summary of the results on the triangular lattice can be
found in Table \ref{sum}.

As on the square lattice we checked asymptotic scaling by looking at
the ratio $\Lambda_{\rm t}/ \Lambda_{2l}$, where $\Lambda_{\rm t}$ is
the effective $\Lambda$-parameter on the triangular lattice, defined
in analogy with Eq.~(\ref{effla}). Beside
the formulas concerning asymptotic scaling given for the square
lattice case, cfr.\ Eqs.~(\ref{effla}-\ref{twoloopla}), we need here
the $\Lambda$-parameter ratio $\Lambda_{\rm t}/\Lambda_{\rm s}$
calculated in App.~\ref{apptr}, cfr.\ Eq.~(\ref{ratioltr2}).  We again
used $M_G$ as approximate estimator of the mass-gap $M$.
Fig.~\ref{asysctr} shows curves of $\Lambda_{\rm t}/ \Lambda_{2l}$ for
various values of $N$, $N=3,4,8$, and for comparison the exact curve
for $N=\infty$.  Such results are similar to those found on the square
lattice: for $N=3,4$ asymptotic scaling regime is still far at
$\xi_G\simeq 10$, but it is verified within a few per cent at $N=8$,
where the correlation length corresponding to the strong-coupling
convergence radius is $\xi\simeq 8$.

\subsection{The honeycomb lattice}
\label{scho}

On the honeycomb lattice we have calculated 
the two-point Green's function 
up to $O(\beta^{30})$, from which we extracted 
strong-coupling series of the quantities $E$, $\chi$, 
$\xi_{G}^2$, $u$, $M_{\rm v}^2$, $M_{\rm h}^2$, 
already introduced in  Sec.~\ref{iNhl},
and of the ratios $r\equiv M_{\rm v}^2/M_{\rm h}^2$,
$s\equiv M_{\rm h}^2/M_{G}^2$.
Some of the above series for $N=3$ are reported 
in App.~\ref{appscex}.

At $N=3$ a Dlog-PA analysis of the $O(\beta^{30})$
strong-coupling series of $\chi$ and $\xi_{G}^2$ 
detected two couples of complex conjugate singularities,
one on the imaginary axis at $\bar{\beta}\simeq\pm i0.460$, 
quite close to the origin, and the other 
at $\bar{\beta}\simeq 0.93\pm i 0.29$.
The singularity on the imaginary axis
leads to a rather small convergence radius in terms
of correlation length, indeed  at $\beta\simeq 0.46$ 
we estimate $\xi\simeq 2.6$. 
At $N=4$ we found $\bar{\beta}\simeq\pm i0.444$, 
and $\bar{\beta}\simeq 0.88\pm i 0.41$.
At larger $N$ the singularities closest
to the origin converge toward the $N=\infty$ value
$\bar{\beta}=\pm i \,0.362095$.
Notice that, as on the square lattice, the partition function on the
honeycomb lattice enjoys the symmetry $\beta\rightarrow -\beta$.

Again we analyzed both the series in the energy and in $\beta$.
The estimates concerning the continuum limit are obtained by evaluating
the approximants of the energy series at $E=1$, and 
those of the $\beta$-series at $\beta=0.85$ for the $N=3$ case,
which should correspond to $\xi\simeq 22$.

By rotation invariance the ratio 
$r\equiv M_{\rm h}^2/M_{\rm v}^2$ should go to one
in the continuum limit. 
From $G(x)$ up to $O(\beta^{30})$ we extracted
the ratio $r$ up to $O(\beta^{20})$. 
Again PA's and Dlog-PA's 
of the energy series evaluated at $E=1$
and of the $\beta$-series evaluated at $\beta=0.85$
(some details are given in Table \ref{hor})
give the correct result in the continuum limit: 
respectively $r^*=1.00(1)$ and $r^*=1.001(1)$ at $N=3$
(we considered PA's and Dlog-PA's with 
$l+m\geq 16$ and $m\geq l\geq 7$).

Calculating a few more components of $G(x)$ at larger orders (i.e.,
those involved by $G^{({\rm w})}_{\rm h}(x)$, defined in
Eq.~(\ref{g2}), at distances $x=\sqrt{3}/2 \times 9$ and
$x=\sqrt{3}/2 \times 10$ respectively at $O(\beta^{34})$ and
$O(\beta^{35})$), we computed the ratio $s\equiv M_{\rm h}^2/M_{G}^2$
up to $O(\beta^{25})$ \cite{SCUN1,RV}.  For $N=3$ the analysis of the
strong-coupling series of $s$ gives $s^*=0.999(3)$ from the
$E$-approximants and $s^*=0.9987(5)$ from the $\beta$-approximants
evaluated at $\beta=0.85$ (some details are given in Table \ref{hos}),
in agreement with the result found on the other lattices.  We
considered PA's and Dlog-Pa's with $l+m\geq 22$, $m\geq l\geq 10$.

The analysis of the energy series of $u$ confirms universality:
PA's and Dlog-PA's (with $l\leq m$,
$l+m\geq 26$, $l\geq 12$) of the energy series
evaluated at $E=1$ give $u^*=0.249(3)$,
and those of the $\beta$-series at $\beta=0.85$ $u^*=0.2491(3)$.
As for the square lattice, the curve $u(E)$ obtained 
from the PA's at $N=3$ and the exact curve $u(E)$ 
at $N=\infty$, cfr.\ Eq.~(\ref{omhoin}), 
would be hardly distinguishable if plotted together.

As noted above, the convergence radius $\beta_r$ is small in terms of
correlation length for all values of $N$: it goes from $\xi\simeq 1.0$
at $N=\infty$ to $\xi\simeq 2.6$ at $N=3$. Nevertheless in this case
Dlog-PA resummations seem to give reasonable estimates of $\xi_G$ even
beyond $\beta_r$ (apparently up to about the next singularity closest
to the origin).  In Fig.~\ref{asyscho} we show curves of 
$\Lambda_{\rm h}/ \Lambda_{2l}$, where $\Lambda_{\rm h}$ is the
effective $\Lambda$-parameter on the honeycomb lattice, for various
values of $N$, $N=3,4,8$, and for comparison the exact curve for
$N=\infty$.  The necessary ratio of $\Lambda$-parameters has been
calculated in App.~\ref{appex}, cfr.\ Eqs.~(\ref{ratiolho}) and
(\ref{ratiolho2}).

\subsection{Conclusions}
\label{concl}

We have shown that quite accurate continuum limit estimates of
dimensionless renormalization-group invariant quantities, such as $s$
and $u$ (cfr.\ Eqs.~(\ref{sdef}) and (\ref{omsqin})), can be obtained
by analyzing their strong-coupling series and applying resummation
techniques both in the inverse temperature variable $\beta$ and in the
energy variable $E$.  In particular, in order to get continuum
estimates from the analysis of the energy series, we evaluated the
corresponding PA's and Dlog-PA's at $E=1$, i.e., at the continuum
limit.  This idea was already applied to the calculation of the
continuum limit of the zero-momentum four-point coupling $g_r$,
obtaining accurate results~\cite{gr}.  These results look very
promising in view of a possible application of such strong-coupling
analysis to four-dimensional gauge theories.

The summary in Table \ref{sum} of our $N=3$ strong-coupling results
for the continuum values $r^*$, $s^*$ and $u^*$,
for all lattices we have considered, shows that
universality is verified within a precision of few per mille, leading
 to the final estimates $s^*\simeq 0.9985$ and $u^*\simeq 0.2495$
with an uncertainty of about one per mille.
The comparison with the exact $N=\infty$ results, $s^*=1$ and
$u^*=1/4$, shows that quantities like $s^*$ and $u^*$, which describe the
small momentum universal behavior of $\widetilde{G}(p)$ in the
continuum limit, change very little and apparently monotonically from
$N=3$ to $N=\infty$, suggesting that
at $N=3$ $\widetilde{G}(p)$ is essentially Gaussian at small momentum.

Let us make this statement more precise.
In the critical region
one can expand the dimensionless 
renormalization-group invariant function
\begin{equation}
L(p^2/M_G^2)\equiv {\widetilde{G}(0)\over \widetilde{G}(p)}
\label{elle}
\end{equation}
around $y\equiv p^2/M_G^2=0$, writing
\begin{eqnarray}
L(y)&=&1 + y + l(y)\nonumber \\
l(y)&=&\sum_{i=2}^\infty c_i y^i.
\label{lexp}
\end{eqnarray}
$l(y)$ parameterizes the difference from a generalized Gaussian
propagator.  One can easily relate the coefficients $c_i$ of the
expansion (\ref{lexp}) to dimensionless renormalization-group
invariant ratios involving the moments $m_{2j}$ of $G(x)$.

It is worth observing that
\begin{equation}
u^* = {1\over 4 (1 - c_2)}.
\label{uc2}
\end{equation} 
In the large-$N$ limit the function $l(y)$ is
depressed by a factor $1/N$.
Moreover the coefficients of its low-momentum expansion are very small.
They can be derived from the $1/N$ expansion of the 
self-energy~\cite{Flyv,CR,CRselfenergy}. 
In the leading order in the $1/N$ expansion one finds  
\begin{eqnarray}
c_{2}&\simeq&-{0.00619816...\over N},\nonumber \\ 
c_{3}&\simeq&{0.00023845...\over N},\nonumber \\
c_{4}&\simeq&-{0.00001344...\over N},\nonumber \\ 
c_{5}&\simeq&{0.00000090...\over N},
\label{c1N}
\end{eqnarray}
etc..  For sufficiently large $N$ we then expect
\begin{equation}
c_i\ll c_2\ll 1 \;\;\;\;\;\;\;\;\;{\rm for}\;\;\;\; i\geq 3.
\label{crel}
\end{equation}  
As a consequence, since 
the zero of $L(y)$ closest 
to the origin is $y_0=-s^*$, the value of $s^*$ 
is substantially fixed  by the term proportional to
$(p^2)^2$ in the inverse propagator, through the approximate relation 
\begin{equation}
s^*-1\simeq c_2 \simeq 4 u^* - 1 .
\label{s4u}
\end{equation}
Indeed in the large-$N$ limit one finds from Eqs.~(\ref{largeNs}) and 
(\ref{largeNu})
\begin{equation}
s^*-4u^*={-0.000252\over N}+O\left( {1\over N^2}\right),
\end{equation}
where the coefficient of the $1/N$ term is much
smaller than those of $s^*$ and $u^*$.

From this large-$N$ analysis one expects 
that even at $N=3$ the function $l(y)$ be small in a relatively large
region around $y=0$. 
This is confirmed by the strong-coupling estimate of $u^*$,
which, using Eq.~(\ref{uc2}), leads to  $c_2 \simeq -0.002$.
Furthermore, the comparison of the estimates of $s^*$ and $u^*$
shows that $s^*-4u^*\simeq 0$ within the precision of our analysis,
consistently with Eq.~(\ref{crel}).
It is interesting to note that similar results have been
obtained for the models with $N\leq 2$,
and in particular for the Ising Model, i.e., for $N=1$, where the
strong-coupling analysis turns out to be very precise~\cite{Nm2}.

We can conclude that the two-point Green function for all $N\geq 3$ is
almost Gaussian in a large region around $p^2=0$, i.e.,
$|p^2/M_G^2|\lesssim 1$, and the small corrections to Gaussian
behavior are essentially determined by the $(p^2)^2$ term in the
expansion of the inverse propagator.

Differences from Gaussian behavior will become important at
sufficiently large momenta, as predicted by simple weak coupling
calculations supplemented by a renormalization group resummation.
Indeed the asymptotic behavior of $G(x)$ for $x\ll1/M$ (where $M$ is
the mass gap) turns out to be
\begin{equation}
G(x) \sim \left(\ln{1\over xM}\right)^{\gamma_1/b_0}, \qquad
{\gamma_1\over b_0} = {N-1\over N-2} \,;
\end{equation}
$b_0$ and $\gamma_1$ are the first coefficients of the
$\beta$-function and of the anomalous dimension of the fundamental
field $\vec s$ respectively.  Let us remind that a free Gaussian
Green's function behaves like $\ln (1/x)$.
Important differences are
present in other Green's functions even at small momentum, as shown in
the analysis of the four-point zero-momentum renormalized coupling,
whose definition involves the zero-momentum four-point correlation
function (\ref{chi4})~\cite{gr}.
However monotonicity in $N$ seems to be a persistent feature.

Our strong-coupling calculations allow also a check of asymptotic
scaling for a relatively large range of correlation lengths. For all
lattices considered the ratio between the effective
$\Lambda$-parameter extracted from the mass-gap and its two-loop
approximation, $\Lambda/\Lambda_{2l}$, when considered as function of
$\xi_G$, shows similar patterns with changing $N$.  Confirming earlier
Monte Carlo studies, large discrepancies from asymptotic scaling are
observed for $N=3$ in the range of correlation lengths we could
reliably investigate, i.e.,  $\xi\lesssim 50$. At $N=8$ and for all
lattices considered, asymptotic scaling within a few per cent is
verified for $\xi\gtrsim 4$, and increasing $N$ the ratio
$\Lambda/\Lambda_{2l}$ approaches smoothly its $N=\infty$ limit.

\acknowledgments

It is a pleasure to thank B.~Alles
for useful and stimulating discussions.

\appendix

\section{The triangular lattice}
\label{apptr}

The sites $\vec{x}$ of a finite periodic triangular 
lattice can be represented in Cartesian coordinates by
\begin{eqnarray}
&&\vec{x}(l_1,l_2)= l_1\vec{\eta}_1 
+ l_2\vec{\eta}_2,\nonumber \\
&& l_1=1,...L_1, \ l_2=1,...L_2,\nonumber \\
&& \vec{\eta}_1=\left( 1,0\right),\quad
\vec{\eta}_2= \left( {1\over 2},{\sqrt{3}\over 2}\right).
\label{trcoord}
\end{eqnarray}
We set $a=1$, where the lattice space $a$ is the length of a link.
The total number of sites, links, and triangles is respectively 
$N_s=L_1 L_2$, $N_l=3N_s$, $N_{tr}=2N_s$.
Taking into account periodic boundary conditions,
a finite lattice Fourier transform can be defined by
\begin{eqnarray}
&&\phi(\vec{k}) =\sum_{\vec{x}} e^{i\vec{k}\cdot \vec{x}}
\phi(\vec{x})\nonumber \\
&&\phi(\vec{x}) ={1\over v_s N_s}
\sum_{\vec{k}} e^{i\vec{k}\cdot \vec{x}}\phi(\vec{k}),
\label{ft}
\end{eqnarray}
where $v_s=\sqrt{3}/2$ is the volume per site, and the set 
of momenta is
\begin{eqnarray}
&&\vec{k}(m_1,m_2)= {2\pi\over L_1}m_1\vec{\rho}_1 
+ {2\pi\over L_2}m_2\vec{\rho}_2,\nonumber \\
&& m_1=1,...L_1, \ m_2=1,...L_2,\nonumber \\
&& \vec{\rho}_1=\left( 1,-{1\over \sqrt{3}}\right), \quad
\vec{\rho}_2= \left( 0,{2\over \sqrt{3}}\right).
\label{trmom}
\end{eqnarray}
Notice that
\begin{equation}
\vec{k}\cdot \vec{x}={2\pi\over L_1}m_1l_1 
+{2\pi\over L_2}m_2l_2.
\label{ooo}
\end{equation}

To begin with, let us
discuss the Gaussian model on the triangular lattice, 
which is defined by the action
\begin{equation}
S_{G}={\kappa\over 2}\sum_{\rm links}
\left( \phi_{x_l}-\phi_{x_r}\right)^2
\label{gaussian}
\end{equation}
where $x_l$, $x_r$ indicate the sites at the ends of 
each link. Performing the Fourier transform (\ref{ft}) 
one derives the propagator
\begin{eqnarray}
&&\langle \phi(k)\phi(q)\rangle=
{v_s\over \sqrt{3}\kappa} 
{1\over \Delta(k)}\delta_{k+q,0},\nonumber \\
&&\Delta(k)=4\left[ 1 - {1\over 3}\left(
\cos k_1+2\cos {k_1\over 2}
\cos {\sqrt{3}k_2\over 2}\right)\right].
\label{prop}
\end{eqnarray}
From these formulae one can easily obtain the large-$N$ limit, since
the model becomes Gaussian for $N\to\infty$.

In the massive Gaussian model  
one may define an exponentiated wall-wall correlation
function by
\begin{equation}
G_{\rm t}^{(\rm w)}\left({\sqrt{3}\over 2}l_2\right)=
\sum_{l_1} G(l_1\vec{\eta}_1+l_2\vec{\eta}_2).
\label{walldeftr}
\end{equation}

In the ${\rm O}(N)$ $\sigma$ models,
in order to evaluate the ratio between the $\Lambda$-parameter
of the ${\overline{\rm MS}}$ renormalization scheme
and the triangular nearest-neighbor lattice regularization,
we calculated the correlation function $G(x)$ in perturbation
theory.
In the $x$-space we obtained (neglecting $O(a^2)$ terms)
\begin{eqnarray}
&&G(x)= 1 + {N-1\over N} t F(a/x) + O(t^2),\nonumber \\
&&F(a/x)= {1\over 2\pi}\left(
\ln {a\over x} - \gamma_{_E} - \ln 2
-{1\over 2}\ln 3\right).
\label{gxtr}
\end{eqnarray}
In the $p$-space
\begin{eqnarray}
&&\widetilde{G}(k)={N-1\over N} {t\over k^2} 
\left[ 1 + {t\over N}\left( D(ak)+{1\over 2\sqrt{3}}
\right)+ O(t^2)\right] ,\nonumber \\
&&D(ak)= {1\over 2\pi}\left(
\ln {ak} - 2\ln 2
-{1\over 2}\ln 3\right).
\label{gptr}
\end{eqnarray}
The above results required the calculation of the following
integral
\begin{eqnarray}
&&\int^\pi_{-\pi} {dk_1\over 2\pi}
\int^{2\pi/\sqrt{3}}_{-2\pi/\sqrt{3}}
{dk_2\over 2\pi} {e^{ikx}-1\over \Delta(k)}=
F(a/x)+O(a/x),\nonumber \\
&&\int^\pi_{-\pi} {dk_1\over 2\pi}
\int^{2\pi/\sqrt{3}}_{-2\pi/\sqrt{3}}
{dk_2\over 2\pi} {\Delta(q)-\Delta(k)-\Delta(k+q)
\over \Delta(k)\Delta(k+q)}=2D(aq)+O(aq),
\label{inttr}
\end{eqnarray}
where the extremes of integration are chosen to cover
the appropriate Brillouin zone, which can be determined
from the finite lattice momenta (\ref{trmom}).

Comparing the above results with the two-point 
Green's function renormalized in the ${\overline{\rm MS}}$ scheme
\begin{eqnarray}
&&G_{\overline{\rm MS}}\left(x\mu=2e^{-\gamma_{_E}}\right)=
1 + O(t_r^2),\nonumber\\
&&\widetilde{G}_{\overline{\rm MS}}\left({k\over \mu}=1\right)=
{N-1\over N}{t_r\over k^2}\left[ 1  + O(t_r^2)\right],
\label{grms}
\end{eqnarray}
and following a standard procedure, one can determine
the ratio of the $\Lambda$-parameters 
\begin{equation}
{\Lambda_{\overline{\rm MS}}\over \Lambda_{\rm t}}
=4\sqrt{3}\,\exp \left( {\pi\over(N-2)}{1\over \sqrt{3}}\right),
\label{ratioltr}
\end{equation}
where $\Lambda_{\rm t}$ is the $\Lambda$-parameter
of the ${\rm O}(N)$ $\sigma$ models on the triangular lattice.
Furthermore comparing with the Green function calculated on the
square lattice~\cite{Falcioni} one can also derive
\begin{equation}
{\Lambda_{\rm s}\over \Lambda_{\rm t}}
= \sqrt{3\over 2}\,\exp\left[ {\pi\over N-2}
\left( {1\over\sqrt{3}}-{1\over 2}\right)\right],
\label{ratioltr2}
\end{equation}
where $\Lambda_{\rm s}$ is the $\Lambda$-parameter on the
square lattice.

\section{The honeycomb lattice}
\label{appex}

The sites $\vec{x}$ of a finite periodic honeycomb 
lattice can be represented in Cartesian coordinates by
\begin{eqnarray}
&&\vec{x}= \vec{x}\,'+p\,\vec{\eta}_p
\nonumber \\
&&\vec{x}\,'= l_1\,\vec{\eta}_1 
+ l_2\,\vec{\eta}_2,\nonumber \\
&& l_1=1,...L_1, \ l_2=1,...L_2, \ 
p=0,1,\nonumber \\
&& \vec{\eta}_1=\left( {3\over 2},{\sqrt{3}\over 2}
\right),\quad
\vec{\eta}_2= \left( 0,\sqrt{3}\right),
\quad \vec{\eta}_p=\left(1,0\right).
\label{hocoord}
\end{eqnarray}
We set $a=1$, where the lattice space $a$ is the length of a link.
The total number of sites, links and hexagons is respectively
$N_s=2L_1 L_2$, $N_{\rm l}=3L_1L_2$, $N_{\rm h}=L_1L_2$.  The
coordinate $p$ can be interpreted as the parity of the corresponding
lattice site: sites with the same parity are connected by an even
number of links.

The two sublattices identified by
$\vec{x}_+(l_1,l_2)\equiv\vec{x}(l_1,l_2,0)$ and
$\vec{x}_-(l_1,l_2)\equiv \vec{x}(l_1,l_2,1)$ forms a triangular
lattice.  Each link of the honeycomb lattice connects sites belonging
to different sublattices.  Triangular lattices have a more symmetric
structure, in that their sites are characterized by a group of
translations.  It is then convenient to rewrite a field
$\phi(\vec{x})\equiv\phi(l_1,l_2,p)$ in terms of two new fields
$\phi_+(\vec{x}_+)\equiv \phi(\vec{x}_+)$ and $\phi_-(\vec{x}_-)\equiv
\phi(\vec{x}_-)$ defined respectively on the sublattices $\vec{x}_+$
and $\vec{x}_-$.  Taking into account periodic boundary conditions, a
finite lattice Fourier transform can be consistently
defined~\cite{SCUN2}
\begin{eqnarray}
&&\phi_\pm(\vec{k})=v_{\rm h}\,\sum_{\vec{x}_\pm} e^{i\vec{k}\cdot
\vec{x}_\pm}\,\phi_\pm (\vec{x}_\pm),\nonumber \\
&&\phi_\pm(\vec{x}_\pm)={1\over v_{\rm h}L_1L_2}\,
\sum_{\vec{k}} e^{-i\vec{k}\cdot
\vec{x}_\pm}\,\phi_\pm (\vec{k}),
\label{a3}
\end{eqnarray}
where $v_{\rm h}=3\sqrt{3}/2$ is the volume of an hexagon, and
the set of momenta is
\begin{eqnarray}
&&\vec{k}={2\pi\over L_1}m_1\vec{\rho}_1
+ {2\pi\over L_2}m_2\vec{\rho}_2
\nonumber \\
&&m_1=1,...L_1,\ m_2=1,...L_2,\nonumber \\
&&\vec{\rho}_1=\left( {2\over 3},0\right),\quad
\vec{\rho}_2=\left( -{1\over 3},{1\over \sqrt{3}}\right).
\label{a4}
\end{eqnarray}

Using the results reported in Ref.~\cite{SCUN2}
one can find the following expression for
the matrix $R^{-1}$ (cfr.\ Eq.~(\ref{R2})):
\begin{eqnarray}
R^{-1}(\vec{x};\vec{y})=&&R^{-1}(\vec{x}\,',p_x;\vec{y}\,',p_y)
= \nonumber \\
=&&{t\over v_sN_s}\sum_{\vec{k}} 
e^{i\vec{k}\cdot\left(\vec{x}\,'-\vec{y}\,'\right)}
{1\over \Delta(k)+z\left(1 + {1\over 8}z\right)} 
\left( \matrix{ 1 + {1\over 4}z& e^{-ik_1}H(k)^*\cr
 e^{ik_1}H(k)& 1+{1\over 4}z\cr}\right)
\label{greengex}
\end{eqnarray}
where
\begin{eqnarray}
&&\Delta(k)={8\over 9}\left[ 2 - \cos {\sqrt{3}\over 2}k_2
\left( \cos {3\over 2}k_1  + \cos {\sqrt{3}\over 2}k_2\right)  
\right],\nonumber \\
&&H(k)= e^{-ik_1}{1\over 3}
\left( 1 + 2 e^{i{3\over 2}k_1} \cos {\sqrt{3}\over 2}k_2
\right).
\label{deltaex2}
\end{eqnarray}
In the large-$N$ limit $R^{-1}(x;y)$ represents 
the two-point Green's function. 

In Ref.~\cite{SCUN1}, guided by the analysis of the Gaussian
model on the honeycomb lattice, two wall-wall correlation
functions were defined:
\begin{equation}
G^{(\rm w)}_{\rm v}(\case{3}{2}l_1)=\sum_{l_2} 
G(l_1\vec{\eta}_1+l_2\vec{\eta}_2),
\label{g1}
\end{equation}
with the sum running over sites of positive parity forming a vertical
line;
\begin{equation}
G^{(\rm w)}_{\rm h}(\case{1}{2} \sqrt{3} l) =
\sum_{l_2,p} 
G\bigl((l-2l_2)\vec{\eta}_1+l_2\vec{\eta}_2+p\vec{\eta}_p\bigr), 
\label{g2}
\end{equation}
where the sum is performed over all sites having the same coordinate
$x_2$.  $G^{(\rm w)}_{\rm v}(x)$ and $G^{(\rm w)}_{\rm h}(x)$ allows
the definition of two estimators of the mass gap, $\mu_{\rm v}$ and
$\mu_{\rm h}$, whose ratio must go to one in the continuum limit by
rotation invariance.  On the honeycomb lattice the maximal violation
of full rotational symmetry occurs for directions differing by a
$\pi/6$ angle, and therefore, taking into account its discrete
rotational symmetry, also by a $\pi/2$ angle. So a good test of
rotation invariance is provided by the ratio between masses extracted
from the long distance behaviors of a couple of orthogonal wall-wall
correlation functions constructed with $G(x)$ such as $\mu_{\rm
  v}/\mu_{\rm h}$.

In the ${\rm O}(N)$ $\sigma$ models,
in order to evaluate the ratio between the 
$\Lambda$-parameter of the ${\overline{\rm MS}}$ renormalization
scheme and the honeycomb nearest-neighbor lattice regularization.
we calculated, in perturbation theory, 
the correlation function 
\begin{equation}
G(x_+-y_+)=\langle {\vec s}_{x_+}\cdot {\vec s}_{y_+}\rangle.
\label{ge}
\end{equation}
In the $x$-space we obtained (neglecting $O(a^2)$ terms)
\begin{eqnarray}
&&G(x)= 1+{N-1\over N} t F(a/x)+O(t^2),\nonumber \\
&&F(a/x)= {1\over 2\pi}\left(
\ln {a\over x} - \gamma_{_E} - \ln 2\right).
\label{gex}
\end{eqnarray}
In the $p$-space
\begin{eqnarray}
&&\widetilde{G}(k)={N-1\over N} {t\over k^2} 
\left[ 1 + {t\over N}\left( D(ak)+{1\over 3\sqrt{3}}
\right)+ O(t^2)\right] ,\nonumber \\
&&D(ak)= {1\over 2\pi}\left(
\ln {ak} - 2\ln 2\right).
\label{gep}
\end{eqnarray}
The relevant formulas required by the 
above calculations are can be found in Ref.~\cite{SCUN2}.

Comparing the above results with the two-point 
Green's function renormalized in the ${\overline{\rm MS}}$,
cfr.\ Eq.~(\ref{grms}), one can obtain 
the ratio of $\Lambda$-parameters 
\begin{equation}
{\Lambda_{\overline{\rm MS}}\over \Lambda_{\rm h}}
=4\,\exp\left( {\pi\over N-2}\,{2\over 3\sqrt{3}}\right)
\label{ratiolho}
\end{equation}
where $\Lambda_{\rm h}$ is the $\Lambda$-parameter
of the honeycomb lattice,
and also
\begin{equation}
{\Lambda_{\rm h}\over \Lambda_{\rm s}}
= \sqrt{2} \exp\left[ {\pi\over N-2}\left(
{1\over 2} - {2\over 3\sqrt{3}}\right)\right].
\label{ratiolho2}
\end{equation}

We also give few orders of the perturbative expansion
of the internal energy
\begin{equation}
E=1 - {N-1\over N}{t\over 3\sqrt{3}}
-  {N-1\over N^2}{t^2\over 54}+O(t^3).
\label{enepert}
\end{equation}

\section{Complex-Temperature Singularities at
$\protect\bbox{N=\infty}$ on the triangular and honeycomb lattices} 
\label{singNinf}

In this Appendix we will compute the complex-temperature singularities
for the $N=\infty$ model on the various lattices we considered. We
will follow closely the analysis of Ref.~\cite{BCMO} for the square
lattice.  The $N=\infty$ solution is written in all cases in terms of
a variable $w$ related to the inverse temperature $\beta$ by the gap
equation which has the generic form
\begin{equation}
\beta=\, h(w) \rho(w) K(\rho(w)) \equiv b(w)
\label{gap-generic}
\end{equation}
for suitable analytic functions $h(w)$ and $\rho(w)$ and variable $w$,
which will be defined below.  Here $K(w)$ is the complete elliptic
integral of the first kind.  In general $b(w)$ will be defined on the
complex plane with suitable cuts. To the purpose of computing the
singularities of the inverse function $w(\beta)$ we will be forced to
consider $b(w)$ on its Riemann surface $\cal R$. Moreover our
discussion will be only local and thus we will determine all
singularities which appear in the Riemann surface of $w(\beta)$.
Notice that not all of them will necessarily appear in the principal
sheet of $w(\beta)$.

Let us now consider a point $w_0\in\cal R$ and let $\beta_0 = b(w_0)$. 
To study $w(\beta)$ in a 
neighborhood of $\beta_0$ let us expand $b(w)$ around $w_0$.
If $b'(w_0)$ is different from zero, then $w(\beta)$
is obviously analytic in a neighborhood of $\beta_0$ and admits an
expansion of the form 
\begin{equation}
w =\, w_0 + {1\over b'(w_0)} (\beta-\beta_0) + O((\beta-\beta_0)^2) .
\end{equation}
Instead if $b'(w_0)=0$ $\beta_0$ is a 
singular point. Indeed let $k$ be the smallest integer such that 
$b^{(k)}(w_0)\not=0$. Then in a neighborhood of $\beta_0$ we have
\begin{equation}
w =\, w_0 + \left({k!\over b^{(k)}(w_0)}\right)^{1/k} 
     (\beta-\beta_0)^{1/k} + O((\beta-\beta_0)^{2/k}),
\end{equation}
and therefore $\beta_0$ is a $k$th-root singular point of $w(\beta)$.
Thus in order to determine the singularities of $w(b)$ we must
determine the zeroes of $b'(w)$ on the Riemann surface of the function
$b(w)$.

Beside the expansion in $\beta$ we will be interested in expanding our
observables in terms of the energy $E$. We want thus to study the
singularities of the various observables when expressed in terms of
$E$.  In practice we must study the functions $w(E)$ and $\beta(E)$.

For all lattices we consider we have
\begin{equation}
E =  {\alpha\over \beta}+\, e(w)\, =\, {\alpha\over b(w)} + \, e(w)
\end{equation}
where $\alpha$ is a constant and $e(w)$ an analytic function.  We will
verify in each specific case that the zeroes of $b(w)$ do not give
rise to any singularity. Then by an argument completely analogous to
the one given for $w(\beta)$, the singularities of $w(E)$ are
determined by the zeroes of ${dE\over dw}$ over the Riemann surface of
the function $E(w)$, which, because of the fact that $e(w)$ is a
simple rational function of $w$, coincides with the Riemann surface of
$b(w)$.

Finally let us discuss $\beta(E)$. Of course locally we can rewrite it
as $\beta(w(E))$. We shall show that $d\beta/dw$ and $dE/dw$ never
vanish at the same point.  Then it is simple to convince oneself that
the singularities of $\beta(E)$ coincide with those of $w(E)$.

\subsection{Analytic structure of the complete elliptic integral 
$\protect\bbox{K(w)}$}

The analytic properties of the function $K(w)$ are well known
(cfr.\ e.g.\ Refs.~\cite{Chandra,Akhiezer}).
First of all $K(w)$ is really a function of $w^2$; indeed it 
has a representation in terms of hypergeometric functions 
as \cite{Gradshteyn,Akhiezer}
\begin{equation}
K(w) = {\pi\over2} \, {}_2 F_1 \left({1\over2},{1\over2};1;w^2\right).
\end{equation}
From this representation one may see that $K(w)$ is analytic in the
$w^2$-plane cut along the real axis from one to infinity. We want now
to discuss the extension of $K(w)$ to its Riemann surface.  First of
all let us introduce the {\em theta function}
\begin{equation}
\theta_3 (v|\tau)\,=\, \sum_{m=-\infty}^\infty 
e^{\pi i (m^2\tau + 2 m v)} .
\end{equation}
$\theta_3(v|\tau)$ is an entire function
of $\tau$ and $v$ for ${\rm Im}\tau > 0$. Moreover it satisfies the 
following properties (cfr.\ Ref.~\cite{Chandra}, Chapter 5)
\begin{eqnarray}
\theta_3(v|\tau+2) &=& \theta_3(v|\tau), \label{thetapiu2}\\
\theta_3(v/\tau|-1/\tau) &=& \sqrt{-i\tau}
    e^{\pi i v^2/\tau}\, \theta_3(v|\tau). \label{thetainvtau}
\end{eqnarray}
In terms of $\theta_3(v|\tau)$ we define the modular function 
\begin{equation}
  \lambda(\tau) \, =\, e^{\pi i \tau} 
      {\theta^4_3 (-\tau/2|\tau) \over \theta^4_3 (0|\tau)}
\end{equation}
which is also an entire function of $\tau$ for ${\rm Im}\, \tau > 0$.
The function $\lambda(\tau)$ has several important properties
(cfr.\ Ref.~\cite{Chandra}, Chapter 7):

(i) Consider the group $\Gamma_2$ of transformations
\begin{equation}
\tau' = {a \tau + b\over c \tau + d}
\label{modular-transf}
\end{equation}
with 
\begin{equation}
   M\equiv \, \left(\matrix{a & b\cr c & d\cr}\right) \in
   {\rm SL}(2,Z) ,\ \left(\matrix{a & b\cr c & d\cr}\right) \equiv
    \left(\matrix{1 & 0\cr 0 & 1}\right)\; \hbox{\rm mod 2} .
\label{Gamma2-def}
\end{equation}
The function $\lambda(\tau)$ is invariant under $\Gamma_2$, i.e.,  
$\lambda(\tau)=\lambda(\tau')$. Let us notice that
$\Gamma_2$ is generated by the 
transformations
\begin{eqnarray}
    \tau' &=& \tau + 2 \label{gen1}\\
    \tau' &=& {\tau\over 2\tau + 1} \,. \label{gen2}
\end{eqnarray}

(ii) Let $D$ be the domain of the complex plane bounded by the lines
${\rm Re}\tau=\pm1$ and the circles $|\tau\pm 1/2| = 1/2$, including
the boundaries with ${\rm Re}\tau<0$. Then for every $\tau'$ with
${\rm Im} \tau'>0$ there exists a unique $\tau\in D$ and a
transformation (\ref{modular-transf}) connecting $\tau$ and $\tau'$.
Thus $D$ is the {\em fundamental domain} of the group $\Gamma_2$.

(iii) If $c$ is a complex number different from 0 and 1, the equation 
$\lambda(\tau) = c $ has one and only one solution in $D$. Moreover 
$\lambda(x + i \infty) = 0$ for real $x$.

Using the function $\lambda(\tau)$ we obtain a complete
parameterization of the Riemann surface of $K(w)$. Indeed, because of
the last property, we can perform the change of variables $w^2 =
\lambda(\tau)$, $\tau\in D$; then it can be shown that~\cite{Akhiezer}
\begin{equation}
K(\sqrt{\lambda(\tau)}) = {\pi\over2} \theta_3^2 (0|\tau) .
\end{equation}
The complete Riemann surface is then obtained by letting $\tau$ vary
in the whole upper complex plane. Notice that since $\theta_3(0|\tau)$
never vanishes for $\hbox{\rm Im}\, \tau > 0$, the analytic extension
of $K(w)$ is always non-zero.

We want now to express the analytic extension of $K(w)$ on every
Riemann sheet in terms of functions defined on the principal sheet.
This will allow us to go back using $w^2$ as fundamental variable.
This is easily accomplished using the second property of
$\lambda(\tau)$.  Indeed if $\tau'$ is a generic point with positive
imaginary part, there exists a unique $\tau\in D$ such that
(\ref{modular-transf}) holds for suitable integers $a$, $b$, $c$, $d$
satisfying (\ref{Gamma2-def}).  Thus
\begin{equation}
  \theta_3^2 (0|\tau') \, =\, \theta_3^2 
  \left( 0| {a \tau + b \over c \tau + d}\right) .
\end{equation}
To compute the r.h.s., let us notice that the transformations
(\ref{modular-transf}) are generated by (\ref{gen1}) and (\ref{gen2}).
Then, using (\ref{thetapiu2}) and
\begin{equation}
    \theta_3^2\left(0| {\tau\over \pm 2 \tau + 1}\right) \, 
   = \theta_3^2(0|\tau) \pm \, 2 i \theta_3^2(0|-1/\tau),
\end{equation}
which can be derived from (\ref{thetapiu2}) and (\ref{thetainvtau}),
we get easily, by induction,
\begin{equation}
  \theta_3^2 (0|\tau') \, =\, s \left(
  d \theta_3^2 (0|\tau) + i c \theta_3^2(0|-1/\tau) \right)
\end{equation}
where $s$ assumes the values $\pm 1$. The presence of this sign is due
to the fact that the matrix in (\ref{Gamma2-def}) which corresponds to
the transformation (\ref{modular-transf}) is defined up to a sign.  If
we define
\begin{equation}
T_1 \equiv \,\left(\matrix{1 & 2\cr 0 & 1\cr}\right), \quad
T_2 \equiv \,\left(\matrix{1 & 0\cr 2 & 1\cr}\right) \nonumber
\end{equation}
and fix the signs in $M$ so that $M$ is a product of $T_1$ and $T_2$ 
and their inverses only, then $s=1$.
Finally, using $\lambda(-1/\tau) = 1 - \lambda(\tau)$
which easily follows from (\ref{thetainvtau}), we get
\begin{equation}
{\pi\over2} \theta_3^2 (0|\tau') \, =\, s 
   \left( d K(w) + i c K(\sqrt{1 - w^2}) \right)
\end{equation}
with $w^2 = \lambda(\tau')$. As we shall see the sign $s$ plays no role
in the subsequent discussion.

We have thus reached the following result~\cite{BCMO}: the Riemann
surface of $K(w)$ is obtained by considering matrices $M$ generating 
transformations belonging to $\Gamma_2$: 
as $M$ is defined modulo a sign we can always assume $d>0$. 
Then each sheet of the Riemann surface is labeled by the pair 
$(d,c)$ of one row. (We write $(d,c)$ instead of $(c,d)$
to use the same notation of Ref.~\cite{BCMO}.) The extension 
$K_{(d,c)}(w)$ on this sheet is given by
\begin{equation}
 s (d K(w) + i c K(\sqrt{1 - w^2}) )
\end{equation}
where $s$ is a sign depending on $M$.

Finally let us notice that $K(\sqrt{z^*}) = K(\sqrt{z})^*$
so that $K_{(d,c)}(\sqrt{z^*}) = K_{(d,-c)}(\sqrt{z})^*$.
This observation will allow us to consider only the case $d,c>0$.

\subsection{Location of the singularities}

The singularities of $w(\beta)$ on the square lattice have been
studied in Ref.~\cite{BCMO}. Here we will restrict our attention to
the case of the triangular and honeycomb lattice. Let us first
simplify the expressions (\ref{setr2}) and (\ref{seex2}) introducing
new variables
\begin{eqnarray}
w = \left( 1 + {z\over6} \right)^{1/2}  
&& \  \hbox{\rm on the triangular lattice} \\
w = \left( 1 + {z\over4} \right)^{\hphantom{1/2}}
&& \  \hbox{\rm on the honeycomb lattice} .
\end{eqnarray}
Then 
\begin{eqnarray}
\beta = \, {1\over 2\pi} \, {1\over \sqrt{3 w}} \, \rho K(\rho) 
&& \;\; \hbox{\rm on the triangular lattice} \\
\beta = \, {1\over 2\pi} \, {\sqrt{3 w}} \, \rho K(\rho) 
&& \;\; \hbox{\rm on the honeycomb lattice} 
\end{eqnarray}
where for both lattices
\begin{equation}
\rho = \, {4 \sqrt{w}\over (3w-1)^{3/2} (w+1)^{1/2} } \;\; .
\end{equation}
The gap equation has an important property: as 
\begin{equation}
   \rho(-w)^2 = {\rho(w)^2\over \rho(w)^2 - 1}
\end{equation}
and the elliptic integral satisfies the property $K(i z/\sqrt{1-z^2})
= \sqrt{1 - z^2} K(z)$ we can immediately derive that $\beta(-w) = \pm
\beta(w)$ where the upper (resp.\ lower) sign refers to the triangular
(resp.\ honeycomb) lattice.

Let us now discuss the Riemann surface of $b(w)$. As $\rho^2$ is a
meromorphic function with two poles in the $w$-plane for $w=1/2$ and
$w=-1$, we can apply the discussion of the previous paragraph. We must
then consider the prefactor in front, which contains a square-root
with two branching points and which has thus a double-sheeted Riemann
surface. The Riemann surface is then labeled by two integers $(d,c)$
as we discussed in the previous paragraph, and a sign $s$ which
specifies the sheet of the Riemann surface of the prefactor.  Thus we
have
\begin{eqnarray}
\beta_{s,d,c} = s {1\over 2\pi} {1\over \sqrt{3 w}} 
\rho K_{(d,c)}(\rho) && \  \hbox{\rm on the tr.\ latt.} \\
\beta_{s,d,c} = s {1\over 2\pi} {\sqrt{3 w}} \rho K_{(d,c)}(\rho) 
&& \  \hbox{\rm on the honeycomb lattice} 
\end{eqnarray}
To identify the singularities we must then find the zeroes in the
complex plane of
\begin{equation}
{d \beta_{s,d,c}\over dw} =\, - {s\rho \over 4 \pi w} \,
   {1\over (3 w)^{\pm 1/2}} \left[ \pm K_{(d,c)}(\rho) +\,
    E_{(d,c)} (\rho) {(3 w - 1)^2 \over (3 w+1) (w-1)} \right]
\label{equazione}
\end{equation}
where the upper (resp.\ lower) sign refers to the triangular 
(resp.\ honeycomb) lattice and
\begin{equation}
E_{(d,c)}(w) =\, d E(w) 
+ i c \left[ K(\sqrt{1-w^2}) - E(\sqrt{1 - w^2}) \right] .
\end{equation}
The zeroes of (\ref{equazione}) have been studied numerically 
as in Ref.~\cite{BCMO}. In Table \ref{tabellazeribeta} we report the 
solutions with positive real and imaginary part we have found 
for the lowest values of $(d,c)$. We have verified that in all cases 
$d^2\beta_{s,d,c}/dw^2 \not=0$ at the singularity: thus all points
are square-root branch points. Notice that if $\beta$ is a 
singularity also $-\beta$ and $\pm \beta^{*}$ are singularities.
Our results are somewhat different from those of Ref.~\cite{BCMO} on
the square lattice. Indeed in our case we have found more than one
zero with ${\rm Re}\, \beta>0$ and ${\rm Im}\, \beta>0$
on each sheet. The principal sheet is an exception, as it 
is free of singularity for the triangular lattice, while it contains
a pair of purely imaginary singularities for the honeycomb lattice.
We stress that our search for zeroes of (\ref{equazione}) has been
done numerically and thus we cannot exclude that some zeroes have
been overlooked. However we are confident that at least in the region
$|\beta|<2$ our list is exhaustive for the values of $(d,c)$ we have
examined.

To check the value of the zero with lowest $|\beta|$ we can compare
with a direct determination of the singularity from an analysis of the
high-temperature series of $\chi$ by Dlog-PA's or first-order
inhomogeneous integral approximants (IA's).  In Table \ref{HTzeroes}
we report the results of such an analysis.  These numbers are in very
good agreement with the exact results, although the spread of the
Dlog-PA's largely underestimates the true error.  This is probably due
to the fact that Dlog-PA's are unable to reconstruct the exact
singularity. Indeed for $\beta\to\beta_{sing}$ we have $\chi = \chi_0
+ \chi_1 (\beta-\beta_{sing})^{1/2} + ...$: as $\chi_0\not =0 $ this
behavior can be reproduced by IA's but not by Dlog-PA's.

Let us also notice that for the triangular lattice 
only the singularities with positive real part appear in our analysis.

With series with 30 terms or more it is also possible to get an 
estimate of a second singularity. For the triangular lattice 
the series with 30 terms has a second singularity at 
$\beta = - 0.698(6)\pm i 0.776(5)$ which corresponds to the second
point in the sheet with $d=1$ and $c=2$. For the honeycomb lattice
we get $\pm 0.449(13)\pm i 0.610(17)$ (30 terms) and 
$\pm 0.4504(7)\pm i 0.5909(25)$ (60 terms). The IA's are less stable
and the 30-term series do not yield any result. With 60 terms
on the honeycomb lattice we get $\pm 0.449(3)\pm i 0.585(2)$ which
is in perfect agreement with the exact result.

Beside considering the series in $\beta$ we have also considered
series with the energy $E$ as variable. In this case we must consider
the zeroes of $dE/dw$. Explicitly for the three lattices we have (for
the square lattice we take $w=\rho_{\rm s}$):
\begin{eqnarray}
{d E\over dw} &=& {1\over 4\beta^2} {d\beta\over dw} - {1\over w^2} 
 \  \hbox{\rm square latt.}  \\
{d E\over dw} &=& {1\over 6\beta^2} {d\beta\over dw} + 3 w
 \  \hbox{\rm triangular latt.}  \\
{d E\over dw} &=& {1\over 3\beta^2} {d\beta\over dw} + 1
 \  \hbox{\rm honeycomb latt.}  
\end{eqnarray}
It is evident from this formulae that $dE/dw\not=0$ where 
$d\beta/dw=0$. Thus, as we said at the beginning of this Appendix,
the analysis of $dE/dw$ provides all singularities of 
$w(E)$ and $\beta(E)$.
We get for the nearest singularities:
\begin{eqnarray} 
E &=& \pm 0.330261131671 \, i \qquad
     \hbox{\rm square latt.} \nonumber \\
E &=& -0.290013856190 \pm 0.138180553789\, i \qquad
     \hbox{\rm triangular latt.} \nonumber \\
E &=&\pm 0.303078379027 \pm 0.402035415796\, i \qquad 
     \hbox{\rm honeycomb latt.} \nonumber 
\end{eqnarray}
For the square and honeycomb lattice the singularity appears on the 
principal sheet of $E(w)$ while for the triangular lattice it belongs 
the the sheet with $(d,c)=(1,2)$.

From the position of the singularities we can now compute the
convergence radius of the high-temperature series on the real axis.
In terms of the correlation length, we find that the series converge
up to $\xi_{conv}$ where:
\begin{enumerate}
\item square lattice: $\beta$-series: $\xi_{conv} = 3.17160$;
                   $E$-series: $\xi_{conv} = 1.38403$.
\item triangular lattice: $\beta$-series: $\xi_{conv} = 2.98925$;
                   $E$-series: $\xi_{conv} = 1.66706$.
\item honeycomb lattice: $\beta$-series: $\xi_{conv} = 1.00002$;
                   $E$-series: $\xi_{conv} = 2.43450$.
\end{enumerate}
The series converge thus in a very small $\beta$-disk: however 
PA's and IA's are quite successful in providing good estimates 
in a larger domain of the $\beta$-plane: indeed for the 
square (triangular, honeycomb respectively) lattice, 
series with 21 (15, 30 resp.) terms give estimates which differ
from the exact result less than 1\% till $\xi\approx 10$ 
(5,15 resp.).

\subsection{Conformal transformations}

Once the singularities are known one can use a conformal 
transformation to get rid of the nearest ones (cfr.\ e.g.\ 
\cite{Guttmann}) and thus accelerate the convergence of the 
approximants.

Let us first consider the triangular lattice which has two
singularities located at $\beta=\rho e^{\pm i\theta}$ with $\rho =
0.27516911105$, $\theta=0.720896055$. As in Ref.~\cite{BCMO} we
consider a transformation of the form
\begin{equation}
    \beta = \rho w\left[ 1 + P(w) Q(w) +\, \mu \, Q(w)^2\right]
\label{trconf1}
\end{equation}
where $Q(w) = 1 - 2w \cos\theta + w^2$ is a polynomial which vanishes 
for $w=e^{\pm i\theta}$. $P(w)$ is determined requiring that
$d\beta/dw = 0$ for $w=e^{\pm i\theta}$. The simplest polynomial with
this property is given by 
\begin{equation}
     P(w) = - {1\over 2 \sin^2\theta} 
         \left(\cos 2 \theta -\, w \cos \theta\right) \;\; .
\label{trconf1P}
\end{equation}
We will also consider a second possibility given by
\begin{equation}
\beta = \rho w \left[1 
    - {1\over4}\left( {\cos 4 \theta\over \sin^2 2 \theta}
    - {\cot 2 \theta\over \sin 2 \theta} w^2\right) 
     \left(1 - 2 w^2 \cos 2\theta + w^4\right) +\, \mu
     \left(1 - 2 w^2 \cos 2\theta + w^4\right)^2 \right]
\end{equation}
This transformation would also work if we had four singularities at 
$\beta = \pm \rho e^{\pm i \theta}$. It is easy to see that for 
$\theta=\pi/4$ it reduces to the transformation used in \cite{BCMO}
for the square lattice.
For the honeycomb lattice the nearest singularities are 
$\beta = \pm i\rho$ with $\rho = 0.3620955333$. 
We can thus use (\ref{trconf1}) with 
$\theta = \pi/2$, i.e.,
\begin{equation}
\beta = \rho w \left[1 + {1\over2} (1 + w^2) + \mu (1 + w^2)^2\right]
\;\; .
\end{equation}
In all cases $\mu$ is a free parameter which can be used to optimize 
the transformation. 

In order to compare the series with and without conformal
transformation we have compared the results for the magnetic
susceptibility for series with $15$ and $30$ terms on the triangular
lattice and honeycomb lattice respectively.

For the triangular lattice we have considered the $[6/7]$, $[7/6]$,
$[6/8]$, $[7/7]$ and $[8/6]$ DLog-PA's. For the series without
conformal transformation we have found that all PA's have
singularities on the real axis or with a small imaginary part 
(${\rm Im}\, \beta\lesssim 0.2$) with $0.4\lesssim {\rm Re}\, \beta
\lesssim 0.6$. Excluding only the PA with the nearest singularity
($[6/7]$ for which $\beta_{sing} \simeq 0.414$) we find that the
estimates agree within 1\% till $\beta\approx 0.33$ ($\xi = 5.31$)
where we get $\chi = 57.69(56)$ (this value is the average of the
estimates of the various PA's while the error is the maximum
difference between two different PA's) which must be compared with the
exact value $\chi = 56.9837$.  Let us now consider the series obtained
from the conformal transformation (\ref{trconf1}). The results are now
much more stable: at $\beta=0.33$ the series with $\mu=0$ gives
$\chi=56.978(25)$ while for $\mu=0.5$ we get $\chi=56.9804(24)$. The
estimates of the series with $\mu=0$ agree within 1\% till
$\beta=0.46$ ($\xi=21.58$) where we get $\chi = 672(6)$ (exact value
$\chi = 674.86$) , while for the case $\mu=0.5$ the same is true till
$\beta=0.61$ ($\xi=110.27$). In this last case however the
fluctuations of the approximants are not a good estimate of the error:
indeed we get $\chi = 12861(126)$ to be compared with the exact value
$\chi=13290$. The estimates agree within 1\% with the exact result
only till $\beta=0.56$ ($\xi=64.00$).

Let us now consider the second transformation.  The Dlog-PA's for $\mu
= 0$ do not have any singularity on the real axis for $\beta<3$ and
their estimates are extremely stable. For the value we have considered
before, $\beta = 0.33$, they give $\chi = 56.9808 (5)$ which is in
excellent agreement with the exact value although the error bar is
clearly underestimated. The five different PA's agree within 1\% till
$\beta=0.72$ ($\xi = 365.05$) where the estimate is $\chi =
110200(1100)$ to be compared with the exact value $\chi=123387$. Again
the error bar is underestimated and the true error is 11\%. Indeed the
estimate from the PA's agrees with the exact value within 1\% only
till $\beta = 0.53$ ($\xi = 46.18$).

If we use the series with $\mu \not=0$ the stability of the PA's
decreases and singularities on the real axis begin to appear near the
origin.  For $\mu=0.1$ the PA's are still reasonably stable and we get
for $\beta=0.33$, $\chi = 57.03(6)$. PA's agree within 1\% till $\beta
= 0.41$ ($\xi = 12.55$) where we get $\chi = 257.9(2.4)$ to be
compared with the estimate with $\mu=0$, $\chi = 255.83(2)$ and the
exact value $\chi = 256.02$.

An analogous analysis can be done for the honeycomb lattice.
In this case we have considered the $[14/14]$, $[13/16]$, $[14/15]$,
$[15/14]$, $[16/13]$ Dlog-PA's excluding in each case 
those having a singularity on the positive real axis with 
$\beta<3$. We find that PA's of the 
standard series agree within 1\% till $\beta=1.15$ ($\xi = 16.23$)
where $\chi = 303.2(2.8)$ to be compared with the exact result
$\chi = 305.53$. The conformally transformed PA's give instead
$\chi = 305.14(69)$ for $\mu = 0$ and $\chi = 305.84(63)$ for 
$\mu = 0.3$. The conformally transformed PA's agree within 1\%
till $\beta = 1.35$ ($\xi = 33.50$) 
where we get $\chi = 1100(15)$ for $\mu = 0$
and $1114(12)$ for $\mu = 0.3$. This should be compared 
with the exact value $\chi = 1108.13$ and with the estimate from the 
standard series $\chi = 1073(47)$.

From this analysis it emerges that the use of the conformal
transformation is extremely useful, especially for the triangular
lattice where one gets results which are better by a factor of 10 --
100. Of course the interesting problem would be to generalize the
method to finite values of $N$. There are two problems here: first of
all the exact location of the singularities is not known. Moreover
also the exact nature of the singularity must be determined from the
series.  The first problem is probably not a very serious one: indeed
if we redo the analysis we have presented using the values of the
zeroes obtained from a Dlog-PA's analysis of the series itself the
results are essentially unchanged. The really serious problem (at
least for low values of $N$) is the nature of the singularity: indeed
the transformations we have considered apply only to square-root
branch points. If the singularity is different new transformations
must be used.

\section{Strong-coupling series on the square lattice}
\label{appscsq}

A complete presentation of our strong-coupling series is beyond the
scope of the present paper.  A forthcoming publication will include
all the relevant ``raw'' series.  In order to enable the interested
readers to perform their own analysis, we present here just the series
for the internal energy $E$, the magnetic susceptibility $\chi$, the
three mass scales $M_G^2$, $M_{\rm s}^2$, and $M_{\rm d}^2$, and the
renormalization-group invariant quantity $u$, on the square lattice,
for the most interesting values of $N$, i.e., $N=3,4,8$.  The
following Appendices will be devoted to the triangular and honeycomb
lattice.

\subsection{$\protect\bbox{N=3}$}

\begin{mathletters}
\begin{eqnarray}
E &=&
   \beta + \case{7}{5}\beta^{3} - \case{24}{35}\beta^{5} - 
   \case{3439}{875}\beta^{7} - \case{21872}{1925}\beta^{9} - 
   \case{2876163636}{153278125}\beta^{11} + 
   \case{236181936}{21896875}\beta^{13} + 
   \case{9960909191551}{65143203125}\beta^{15} 
\nonumber \\ &+& 
   \case{2128364407641312}{4665255546875}\beta^{17} + 
   \case{34746325087320066964}{36692234876171875}\beta^{19} + 
   \case{42007349682504569392}{54800091048828125}\beta^{21} +
   O(\beta^{23}), \\
\chi &=&
   1 + 4\beta + 12\beta^{2} + \case{168}{5}\beta^{3} + 
   \case{428}{5}\beta^{4} + \case{1448}{7}\beta^{5} + 
   \case{84144}{175}\beta^{6} + \case{942864}{875}\beta^{7} + 
   \case{2055588}{875}\beta^{8} 
\nonumber \\ &+& 
   \case{6845144}{1375}\beta^{9} + 
   \case{3478216992}{336875}\beta^{10} + 
   \case{643017322016}{30655625}\beta^{11} + 
   \case{915294455744}{21896875}\beta^{12} + 
   \case{12550612712128}{153278125}\beta^{13}  
\nonumber \\ &+& 
   \case{120892276630256}{766390625}\beta^{14} + 
   \case{3896992088570128}{13028640625}\beta^{15} + 
   \case{50948877169965252}{91200484375}\beta^{16} + 
   \case{5670666438003413208}{5513483828125}\beta^{17} 
\nonumber \\ &+& 
   \case{6224368647227625667744}{3335657716015625}\beta^{18} + 
   \case{612580053518389456455744}{183461174380859375}\beta^{19} + 
   \case{1080648046932437417271696}{183461174380859375}\beta^{20} 
\nonumber \\ &+& 
   \case{43219200109596671558015312}{4219607010759765625}\beta^{21} + 
   O(\beta^{22}), \\
M_G^2 &=&
   \beta^{-1} - 4 + \case{18}{5}\beta + \case{638}{175}\beta^{3} - 
   \case{32}{25}\beta^{4} - \case{212}{875}\beta^{5} - 
   \case{8256}{875}\beta^{6} + \case{2392562}{336875}\beta^{7} - 
   \case{893248}{30625}\beta^{8} + \case{67614504}{13934375}\beta^{9}  
\nonumber \\ &+& 
   \case{1315441408}{11790625}\beta^{10} - 
   \case{305969023608}{766390625}\beta^{11} + 
   \case{515730116144}{766390625}\beta^{12} - 
   \case{129453356216124}{456002421875}\beta^{13} - 
   \case{38248494356608}{26823671875}\beta^{14} 
\nonumber \\ &+& 
   \case{15923896875925326898}{3335657716015625}\beta^{15} - 
   \case{102727516154055776}{25080133203125}\beta^{16} - 
   \case{11977717369812123408}{1467689395046875}\beta^{17} 
\nonumber \\ &+& 
   \case{872009509667914065856}{26208739197265625}\beta^{18} - 
   \case{12129265223066172243071804}{274274455699384765625}\beta^{19} + 
   O(\beta^{20}), \\
M_{\rm s}^2 &=&
   \beta^{-1} - 4 + \case{18}{5}\beta + \case{638}{175}\beta^{3} - 
   \case{32}{25}\beta^{4} - \case{772}{875}\beta^{5} - 
   \case{912}{175}\beta^{6} - \case{1730018}{336875}\beta^{7} - 
   \case{790736}{153125}\beta^{8} - \case{226737372}{13934375}\beta^{9}  
\nonumber \\ &+& 
   \case{1892118016}{58953125}\beta^{10} - 
   \case{280950072668}{3831953125}\beta^{11} + 
   \case{642095551792}{3831953125}\beta^{12} - 
   \case{451334756421336}{2280012109375}\beta^{13} + 
   \case{1890027233038144}{4694142578125}\beta^{14}  
\nonumber \\ &-& 
   \case{269818891521799734}{1282945275390625}\beta^{15}
   + O(\beta^{16}), \\
M_{\rm d}^2 &=&
   \beta^{-1} - 4 + \case{18}{5}\beta + \case{122}{35}\beta^{3} - 
   \case{4086}{875}\beta^{5} - \case{16737807}{1684375}\beta^{7} - 
   \case{18994322761}{1532781250}\beta^{9} + 
   \case{21507562283}{3831953125}\beta^{11}  
\nonumber \\ &+& 
   \case{7336555515481743}{79800423828125}\beta^{13}
   + O(\beta^{15}), \\
u &=&
   4\beta - 48\beta^{2} + \case{2808}{5}\beta^{3} - \case{32832}{5}\beta^{4} + 
   \case{2686616}{35}\beta^{5} - \case{157029504}{175}\beta^{6} + 
   \case{9178202064}{875}\beta^{7} - \case{107291190016}{875}\beta^{8} 
 \nonumber \\ &+& 
   \case{13796318675224}{9625}\beta^{9} - 
   \case{5644659282845824}{336875}\beta^{10} + 
   \case{30023104630997669536}{153278125}\beta^{11} - 
   \case{31905781361056472448}{13934375}\beta^{12}  
\nonumber \\ &+& 
   \case{315591420176616436544}{11790625}\beta^{13} - 
   \case{239797700666526161503744}{766390625}\beta^{14} + 
   \case{238270573588813562802881872}{65143203125}\beta^{15} 
\nonumber \\ &-& 
   \case{19497320035839102967394487936}{456002421875}\beta^{16} + 
   \case{333113044056059312586697767112}{666465078125}\beta^{17} 
\nonumber \\ &-& 
   \case{19489567025124572887718011335312384}{3335657716015625}\beta^{18} + 
   \case{2506117021969764882798368740582197376}{36692234876171875}\beta^{19} 
\nonumber \\ &-& 
   \case{146479827799664519145353950407208628096}{183461174380859375}
        \beta^{20} + 
   \case{39383302102402718602571807022847349864016}{4219607010759765625}
        \beta^{21}
\nonumber \\ &+& 
   O(\beta^{22}).
\end{eqnarray}
\end{mathletters}

\subsection{$\protect\bbox{N=4}$}

\begin{mathletters}
\begin{eqnarray}
E &=&
   \beta  + \case{4}{3}\beta^{3} - \case{4}{3}\beta^{5} - 
   \case{28}{5}\beta^{7} - \case{472}{45}\beta^{9} - 
   \case{344}{945}\beta^{11} + \case{256324}{2835}\beta^{13} + 
   \case{12150256}{42525}\beta^{15} + \case{18845248}{91125}\beta^{17}  
\nonumber \\ &-& 
   \case{68510312996}{63149625}\beta^{19} - 
   \case{339976954532}{63149625}\beta^{21} +
   O(\beta^{23}), \\
\chi &=&
   1 + 4\beta + 12\beta^{2} + \case{100}{3}\beta^{3} + 84\beta^{4} + 
   \case{596}{3}\beta^{5} + \case{1348}{3}\beta^{6} + 
   \case{14564}{15}\beta^{7} + \case{91132}{45}\beta^{8} + 
   \case{549332}{135}\beta^{9} 
\nonumber \\ &+& 
   \case{213868}{27}\beta^{10} + 
   \case{8483180}{567}\beta^{11} + \case{77643788}{2835}\beta^{12} + 
   \case{138079108}{2835}\beta^{13} + \case{3569641036}{42525}\beta^{14} + 
   \case{17867938876}{127575}\beta^{15} 
\nonumber \\ &+& 
   \case{1066237156}{4725}\beta^{16} + 
   \case{74408983028}{212625}\beta^{17} + 
   \case{992250411932}{1913625}\beta^{18} + 
   \case{45911479386812}{63149625}\beta^{19} + 
   \case{11971698881708}{12629925}\beta^{20}  
\nonumber \\ &+& 
   \case{69463206903148}{63149625}\beta^{21} + 
   O(\beta^{22}), \\
M_G^2 &=&
   \beta^{-1} - 4 + \case{11}{3}\beta + \case{34}{9}\beta^{3} - 
   \case{4}{3}\beta^{4} - \case{224}{135}\beta^{5} - \case{80}{9}\beta^{6} + 
   \case{4}{81}\beta^{7} - \case{544}{45}\beta^{8} - 
   \case{58150}{1701}\beta^{9} + \case{75032}{405}\beta^{10} 
\nonumber \\ &-& 
   \case{7465004}{18225}\beta^{11} + \case{20575936}{42525}\beta^{12} + 
   \case{204232808}{382725}\beta^{13} - \case{359904856}{127575}\beta^{14} + 
   \case{34575314246}{5740875}\beta^{15} - 
   \case{938720152}{382725}\beta^{16} 
\nonumber \\ &-& 
   \case{3116875126738}{189448875}\beta^{17} + 
   \case{47791394648}{1148175}\beta^{18} - 
   \case{128064328183586}{3978426375}\beta^{19} + 
   O(\beta^{20}), \\
M_{\rm s}^2 &=&
   \beta^{-1} - 4 + \case{11}{3}\beta + \case{34}{9}\beta^{3} - 
   \case{4}{3}\beta^{4} - \case{314}{135}\beta^{5} - \case{14}{3}\beta^{6} - 
   \case{836}{81}\beta^{7} + \case{404}{405}\beta^{8} - 
   \case{215318}{8505}\beta^{9} + \case{4700}{81}\beta^{10} 
\nonumber \\ &-& 
   \case{443038}{6075}\beta^{11} + \case{24410738}{127575}\beta^{12} - 
   \case{24732128}{382725}\beta^{13} + \case{139715614}{1148175}\beta^{14} + 
   \case{3172203256}{5740875}\beta^{15}
   + O(\beta^{16}), \\
M_{\rm d}^2 &=&
   \beta^{-1} - 4 + \case{11}{3}\beta + \case{65}{18}\beta^{3} - 
   \case{3341}{540}\beta^{5} - \case{6781}{540}\beta^{7} - 
   \case{1026007}{136080}\beta^{9} + \case{76323223}{2041200}\beta^{11} + 
   \case{6716460083}{36741600}\beta^{13}
\nonumber \\ &+& 
   O(\beta^{15}), \\
u &=&
   4\beta - 48\beta^{2} + \case{1684}{3}\beta^{3} - 6560\beta^{4} + 
   \case{229940}{3}\beta^{5} - \case{2686576}{3}\beta^{6} + 
   \case{52315868}{5}\beta^{7} - \case{1833750976}{15}\beta^{8} 
\nonumber \\ &+& 
   \case{192827317268}{135}\beta^{9} - \case{2252963987728}{135}\beta^{10} + 
   \case{61420977655724}{315}\beta^{11} - 
   \case{2152899064881952}{945}\beta^{12} 
\nonumber \\ &+& 
   \case{3593447617956212}{135}\beta^{13} - 
   \case{1469484644425188368}{4725}\beta^{14} + 
   \case{7358240125587547636}{2025}\beta^{15}  
\nonumber \\ &-& 
   \case{1805422873668489252736}{42525}\beta^{16} + 
   \case{316414144175527075263484}{637875}\beta^{17} - 
   \case{11090798958461896523514928}{1913625}\beta^{18}  
\nonumber \\ &+& 
   \case{475138205142448133881786892}{7016625}\beta^{19} - 
   \case{49962959062246660123963410208}{63149625}\beta^{20}  
\nonumber \\ &+& 
   \case{583759338332900106172953943052}{63149625}\beta^{21}
   + O(\beta^{22}).
\end{eqnarray}
\end{mathletters}

\subsection{$\protect\bbox{N=8}$}

\begin{mathletters}
\begin{eqnarray}
E &=&
   \beta + \case{6}{5}\beta^{3} - \case{38}{15}\beta^{5} - 
   \case{19924}{2625}\beta^{7} + \case{908}{525}\beta^{9} + 
   \case{103032376}{1771875}\beta^{11} + 
   \case{284005868}{1771875}\beta^{13} - 
   \case{801343307824}{3410859375}\beta^{15}
\nonumber \\ &-& 
   \case{74654425999556}{30697734375}\beta^{17} - 
   \case{5113078463377748}{1995352734375}\beta^{19} + 
   \case{1298678765126787388}{69837345703125}\beta^{21} +
   O(\beta^{22}), \\
\chi &=&
   1 + 4\beta + 12\beta^{2} + \case{164}{5}\beta^{3} + 
   \case{404}{5}\beta^{4} + \case{548}{3}\beta^{5} + 
   \case{9708}{25}\beta^{6} + \case{2023244}{2625}\beta^{7} + 
   \case{1258836}{875}\beta^{8}
\nonumber \\ &+& 
   \case{6597628}{2625}\beta^{9} + 
   \case{160511644}{39375}\beta^{10} + \case{433554844}{70875}\beta^{11} + 
   \case{4869604484}{590625}\beta^{12} + 
   \case{17075166692}{1771875}\beta^{13} + 
   \case{180200884868}{20671875}\beta^{14}  
\nonumber \\ &+& 
   \case{407707485308}{136434375}\beta^{15} - 
   \case{18904178053196}{2046515625}\beta^{16} - 
   \case{861976694991428}{30697734375}\beta^{17} - 
   \case{2351633289032404}{51162890625}\beta^{18}  
\nonumber \\ &-& 
   \case{421417265877650188}{9976763671875}\beta^{19} + 
   O(\beta^{20}), \\
M_G^2 &=&
   \beta^{-1} - 4 + \case{19}{5}\beta + \case{298}{75}\beta^{3} - 
   \case{28}{25}\beta^{4} - \case{5254}{875}\beta^{5} - 
   \case{2128}{375}\beta^{6} - \case{654656}{39375}\beta^{7} + 
   \case{98032}{5625}\beta^{8} - \case{95888554}{1771875}\beta^{9}  
\nonumber \\ &+& 
   \case{5421752}{28125}\beta^{10} - \case{1465956704}{20671875}\beta^{11} - 
   \case{2887857152}{8859375}\beta^{12} + 
   \case{21729605024744}{10232578125}\beta^{13} - 
   \case{3782033297144}{930234375}\beta^{14}  
\nonumber \\ &+& 
   \case{354224260413706}{153488671875}\beta^{15} + 
   \case{72788687225944}{7308984375}\beta^{16} - 
   \case{204193924410409426}{4655823046875}\beta^{17} + 
   \case{40434518800599992}{767443359375}\beta^{18}  
\nonumber \\ &+& 
   \case{7755263425304938826}{241744658203125}\beta^{19} + 
   O(\beta^{20}), \\
M_{\rm s}^2 &=&
   \beta^{-1} - 4 + \case{19}{5}\beta + \case{298}{75}\beta^{3} - 
   \case{28}{25}\beta^{4} - \case{5744}{875}\beta^{5} - 
   \case{182}{75}\beta^{6} - \case{853316}{39375}\beta^{7} + 
   \case{343772}{28125}\beta^{8} - \case{25597438}{1771875}\beta^{9} 
\nonumber \\ &+& 
   \case{10063732}{140625}\beta^{10} + 
   \case{7958412578}{103359375}\beta^{11} + 
   \case{637810442}{44296875}\beta^{12} + 
   \case{24367739089468}{51162890625}\beta^{13} - 
   \case{22983973787546}{23255859375}\beta^{14}  
\nonumber \\ &+& 
   \case{828277050985928}{767443359375}\beta^{15}
   + O(\beta^{16}), \\
M_{\rm d}^2 &+&
   \beta^{-1} - 4 + \case{19}{5}\beta + \case{23}{6}\beta^{3} - 
   \case{101219}{10500}\beta^{5} - \case{1598337}{87500}\beta^{7} + 
   \case{3042153439}{141750000}\beta^{9} + 
   \case{716944244341}{4961250000}\beta^{11} 
\nonumber \\ &+& 
   \case{668912806018181}{2728687500000}\beta^{13}
   + O(\beta^{15}), \\
u &=&
   4\beta - 48\beta^{2} + \case{2804}{5}\beta^{3} - \case{32736}{5}\beta^{4} + 
   \case{1146292}{15}\beta^{5} - \case{22299088}{25}\beta^{6} + 
   \case{27328827644}{2625}\beta^{7} - \case{106327225472}{875}\beta^{8} 
\nonumber \\ &+& 
   \case{1241049893108}{875}\beta^{9} - 
   \case{72427584678352}{4375}\beta^{10} + 
   \case{48910909763139412}{253125}\beta^{11} - 
   \case{190295816674721312}{84375}\beta^{12}  
\nonumber \\ &+& 
   \case{46643731943589538628}{1771875}\beta^{13} - 
   \case{2117208265068678482704}{6890625}\beta^{14} + 
   \case{83213961927741336432836}{23203125}\beta^{15}  
\nonumber \\ &-& 
   \case{142777012206408305834115008}{3410859375}\beta^{16} + 
   \case{14998421484199502791600778428}{30697734375}\beta^{17}  
\nonumber \\ &-& 
   \case{291768940616590173820050993776}{51162890625}\beta^{18} + 
   \case{132815395692009950946195947516884}{1995352734375}\beta^{19}  
\nonumber \\ &-& 
   \case{861233015742221409021216372394912}{1108529296875}\beta^{20} + 
   \case{211098292286114752398400491787924004}{23279115234375}\beta^{21}
   + O(\beta^{22}).
\end{eqnarray}
\end{mathletters}

\section{Strong-coupling series on the triangular lattice}
\label{appsctr}

We present here the series for the internal energy $E$, the magnetic
susceptibility $\chi$, the two mass scales $M_G^2$ and $M_{\rm t}^2$,
and the renormalization-group invariant quantity $u$, on the
triangular lattice, for $N=3,4,8$.

\subsection{$\protect\bbox{N=3}$}

\begin{mathletters}
\begin{eqnarray}
E &=&
   \beta  + 2\beta^{2} + \case{17}{5}\beta^{3} + 4\beta^{4} + 
   \case{132}{175}\beta^{5} - \case{352}{25}\beta^{6} - 
   \case{53833}{875}\beta^{7} - \case{246072}{1225}\beta^{8} - 
   \case{38879794}{67375}\beta^{9} - \case{6397316}{4375}\beta^{10}  
\nonumber \\ &-& 
   \case{8815166536}{2786875}\beta^{11} - 
   \case{63896967072}{11790625}\beta^{12} - 
   \case{81729013664}{15640625}\beta^{13} + 
   \case{23710274924992}{2360483125}\beta^{14} + 
   \case{415243711166548533}{5016026640625}\beta^{15} 
\nonumber \\ &+& 
   O(\beta^{16}) \\
\chi &=& 
   1 + 6\beta + 30\beta^{2} + \case{672}{5}\beta^{3} + 
   \case{2802}{5}\beta^{4} + \case{388452}{175}\beta^{5} + 
   \case{1478784}{175}\beta^{6} + \case{3891432}{125}\beta^{7} + 
   \case{683113506}{6125}\beta^{8}
\nonumber \\ &+& 
   \case{131380789212}{336875}\beta^{9} + 
   \case{449739783516}{336875}\beta^{10} + 
   \case{12494938177056}{2786875}\beta^{11} + 
   \case{90716274919896}{6131125}\beta^{12}  
\nonumber \\ &+& 
   \case{36820597127739144}{766390625}\beta^{13} + 
   \case{9067857431725940688}{59012078125}\beta^{14} + 
   \case{2430225926454897621936}{5016026640625}\beta^{15} +
   O(\beta^{16}) \\
M_G^2 &=&
   \case{2}{3}\beta^{-1} -4 +  \case{56}{15}\beta + \case{32}{5}\beta^{2} + 
   \case{4552}{525}\beta^{3} + \case{2176}{525}\beta^{4} - 
   \case{40312}{2625}\beta^{5} - \case{288944}{3675}\beta^{6} - 
   \case{13545688}{67375}\beta^{7}
\nonumber \\ &-& 
   \case{372621008}{1010625}\beta^{8} - 
   \case{338341556368}{459834375}\beta^{9} - 
   \case{33053110624}{18393375}\beta^{10} - 
   \case{5766016722184}{2299171875}\beta^{11} + 
   \case{279009107673392}{59012078125}\beta^{12}  
\nonumber \\ &+& 
   \case{72472337922926768}{2149725703125}\beta^{13} +
   O(\beta^{14}) \\
M_{\rm t}^2 &=&
   \case{2}{3}\beta^{-1}-4 + \case{56}{15}\beta + \case{32}{5}\beta^{2} + 
   \case{4496}{525}\beta^{3} + \case{11944}{2625}\beta^{4} - 
   \case{13836}{875}\beta^{5} - \case{6627032}{91875}\beta^{6} - 
   \case{337042602}{1684375}\beta^{7}
\nonumber \\ &-& 
   \case{1079483686}{2358125}\beta^{8} - 
   \case{408531412013}{459834375}\beta^{9} - 
   \case{3031149364472}{2299171875}\beta^{10} +
   O(\beta^{11}) \\
u &=&
   6\beta - 108\beta^{2} + \case{9552}{5}\beta^{3} - 33840\beta^{4} + 
   \case{104891592}{175}\beta^{5} - \case{371571936}{35}\beta^{6} + 
   \case{164533949256}{875}\beta^{7}
\nonumber \\ &-& 
   \case{20399809562064}{6125}\beta^{8} + 
   \case{2838986498807112}{48125}\beta^{9} - 
   \case{50284702987803792}{48125}\beta^{10} + 
   \case{405247036983420442416}{21896875}\beta^{11}  
\nonumber \\ &-& 
   \case{50244722315696297988096}{153278125}\beta^{12} + 
   \case{4449723414146572825129368}{766390625}\beta^{13} - 
   \case{866958410946002558379625488}{8430296875}\beta^{14}  
\nonumber \\ &+& 
   \case{1827333381636135987619199322672}{1003205328125}\beta^{15} +
   O(\beta^{16})
\end{eqnarray}
\end{mathletters}

\subsection{$\protect\bbox{N=4}$}

\begin{mathletters}
\begin{eqnarray}
E &=&
   \beta  + 2\beta^{2} + \case{10}{3}\beta^{3} + \case{10}{3}\beta^{4} - 
   \case{8}{3}\beta^{5} - \case{238}{9}\beta^{6} - \case{4312}{45}\beta^{7} - 
   \case{12316}{45}\beta^{8} - \case{30212}{45}\beta^{9} - 
   \case{60434}{45}\beta^{10} 
\nonumber \\ &-& 
   \case{1579292}{945}\beta^{11} + 
   \case{24363404}{14175}\beta^{12} + \case{869521126}{42525}\beta^{13} + 
   \case{3674593642}{42525}\beta^{14} + \case{34840110848}{127575}\beta^{15} +
   O(\beta^{16}) \\
\chi &=& 
   1 + 6\beta + 30\beta^{2} + 134\beta^{3} + 554\beta^{4} + 2162\beta^{5} + 
   \case{24166}{3}\beta^{6} + \case{144242}{5}\beta^{7} + 
   \case{1496254}{15}\beta^{8}
\nonumber \\ &+& 
   \case{15035566}{45}\beta^{9} + 
   \case{3260594}{3}\beta^{10} + \case{361103026}{105}\beta^{11} + 
   \case{50064022798}{4725}\beta^{12} + 
   \case{450734165906}{14175}\beta^{13}  
\nonumber \\ &+& 
   \case{439170936682}{4725}\beta^{14} + 
   \case{11246618825102}{42525}\beta^{15} +
   O(\beta^{16}) \\
M_G^2 &=&
   \case{2}{3}\beta^{-1}-4 + \case{34}{9}\beta + \case{20}{3}\beta^{2} + 
   \case{248}{27}\beta^{3} + \case{32}{9}\beta^{4} - 
   \case{9268}{405}\beta^{5} - \case{14024}{135}\beta^{6} - 
   \case{323204}{1215}\beta^{7}
\nonumber \\ &-& 
   \case{198104}{405}\beta^{8} - 
   \case{19703848}{25515}\beta^{9} - \case{45897388}{42525}\beta^{10} + 
   \case{357801536}{382725}\beta^{11} + \case{2167022524}{127575}\beta^{12} + 
   \case{83057677936}{1148175}\beta^{13} 
\nonumber \\ &+& 
   O(\beta^{14}) \\
M_{\rm t}^2 &=&
   \case{2}{3}\beta^{-1}-4 + \case{34}{9}\beta + \case{20}{3}\beta^{2} + 
   \case{245}{27}\beta^{3} + \case{107}{27}\beta^{4} - 
   \case{18581}{810}\beta^{5} - \case{40037}{405}\beta^{6} - 
   \case{324218}{1215}\beta^{7}
\nonumber \\ &-& 
   \case{2772953}{4860}\beta^{8} - 
   \case{189804947}{204120}\beta^{9} - \case{493593379}{765450}\beta^{10} +
   O(\beta^{11}) \\
u &=&
   6\beta - 108\beta^{2} + 1910\beta^{3} - 33828\beta^{4} + 599078\beta^{5} - 
   10609380\beta^{6} + \case{2818307806}{15}\beta^{7}  
\nonumber \\ &-& 
   \case{49910910532}{15}\beta^{8} + \case{2651696611738}{45}\beta^{9} - 
   \case{46960305901996}{45}\beta^{10} + 
   \case{17464545810431714}{945}\beta^{11}  
\nonumber \\ &-& 
   \case{220920665969769296}{675}\beta^{12} + 
   \case{82160433528738212702}{14175}\beta^{13} - 
   \case{1455022824164792634328}{14175}\beta^{14}  
\nonumber \\ &+& 
   \case{5153554644036578704006}{2835}\beta^{15} +
   O(\beta^{16})
\end{eqnarray}
\end{mathletters}

\subsection{$\protect\bbox{N=8}$}

\begin{mathletters}
\begin{eqnarray}
E &=&
   \beta  + 2\beta^{2} + \case{10}{3}\beta^{3} + \case{10}{3}\beta^{4} - 
   \case{8}{3}\beta^{5} - \case{238}{9}\beta^{6} - \case{4312}{45}\beta^{7} - 
   \case{12316}{45}\beta^{8} - \case{30212}{45}\beta^{9} - 
   \case{60434}{45}\beta^{10} 
\nonumber \\ &-& 
   \case{1579292}{945}\beta^{11} + 
   \case{24363404}{14175}\beta^{12} + \case{869521126}{42525}\beta^{13} + 
   \case{3674593642}{42525}\beta^{14} + \case{34840110848}{127575}\beta^{15} +
   O(\beta^{16}) \\
\chi &=& 
   1 + 6\beta + 30\beta^{2} + 134\beta^{3} + 554\beta^{4} + 2162\beta^{5} + 
   \case{24166}{3}\beta^{6} + \case{144242}{5}\beta^{7} + 
   \case{1496254}{15}\beta^{8} 
\nonumber \\ &+& 
   \case{15035566}{45}\beta^{9} + 
   \case{3260594}{3}\beta^{10} + \case{361103026}{105}\beta^{11} + 
   \case{50064022798}{4725}\beta^{12} + 
   \case{450734165906}{14175}\beta^{13}  
\nonumber \\ &+& 
   \case{439170936682}{4725}\beta^{14} + 
   \case{11246618825102}{42525}\beta^{15} +
   O(\beta^{16}) \\
M_G^2 &=&
   \case{2}{3}\beta^{-1} -4 + \case{34}{9}\beta + \case{20}{3}\beta^{2} + 
   \case{248}{27}\beta^{3} + \case{32}{9}\beta^{4} - 
   \case{9268}{405}\beta^{5} - \case{14024}{135}\beta^{6} - 
   \case{323204}{1215}\beta^{7} - \case{198104}{405}\beta^{8}  
\nonumber \\ &-& 
   \case{19703848}{25515}\beta^{9} - \case{45897388}{42525}\beta^{10} + 
   \case{357801536}{382725}\beta^{11} + \case{2167022524}{127575}\beta^{12} + 
   \case{83057677936}{1148175}\beta^{13} +
   O(\beta^{14}) \\
M_{\rm t}^2 &=&
   \case{2}{3}\beta^{-1}-4 + \case{34}{9}\beta + \case{20}{3}\beta^{2} + 
   \case{245}{27}\beta^{3} + \case{107}{27}\beta^{4} - 
   \case{18581}{810}\beta^{5} - \case{40037}{405}\beta^{6} - 
   \case{324218}{1215}\beta^{7} - \case{2772953}{4860}\beta^{8}  
\nonumber \\ &-& 
   \case{189804947}{204120}\beta^{9} - \case{493593379}{765450}\beta^{10} +
   O(\beta^{11}) \\
u &=&
   6\beta - 108\beta^{2} + 1910\beta^{3} - 33828\beta^{4} + 599078\beta^{5} - 
   10609380\beta^{6} + \case{2818307806}{15}\beta^{7}  
\nonumber \\ &-& 
   \case{49910910532}{15}\beta^{8} + \case{2651696611738}{45}\beta^{9} - 
   \case{46960305901996}{45}\beta^{10} + 
   \case{17464545810431714}{945}\beta^{11}  
\nonumber \\ &-& 
   \case{220920665969769296}{675}\beta^{12} + 
   \case{82160433528738212702}{14175}\beta^{13} - 
   \case{1455022824164792634328}{14175}\beta^{14}  
\nonumber \\ &+& 
   \case{5153554644036578704006}{2835}\beta^{15} +
   O(\beta^{16})
\end{eqnarray}
\end{mathletters}

\section{Strong-coupling series on the honeycomb lattice}
\label{appscex}

We present here the series for the internal energy $E$, the magnetic
susceptibility $\chi$, the three mass scales $M_G^2$, $M_{\rm v}^2$, and
$M_{\rm h}^2$, and the renormalization-group invariant quantity $u$, on the
honeycomb lattice, for $N=3,4,8$.

\subsection{$\protect\bbox{N=3}$}

\begin{mathletters}
\begin{eqnarray}
E &=&
   \beta  - \case{3}{5}\beta^{3} + \case{88}{35}\beta^{5} - 
   \case{1761}{175}\beta^{7} + \case{14902}{385}\beta^{9} - 
   \case{3439769596}{21896875}\beta^{11} + 
   \case{2045078768}{3128125}\beta^{13} - 
   \case{5158924941321}{1861234375}\beta^{15}  
\nonumber \\ &+& 
   \case{28899507590512836}{2425932884375}\beta^{17} - 
   \case{172552140904013898254}{3335657716015625}\beta^{19} + 
   \case{415753493847105514488}{1835409748046875}\beta^{21}  
\nonumber \\ &-& 
   \case{871596813626516225704857972}{872691449952587890625}\beta^{23} + 
   \case{297323210393498586435793176}{67130111534814453125}\beta^{25}  
\nonumber \\ &-& 
   \case{93896946096105935045256448288}{4757152640718994140625}\beta^{27} +
   \case{6327983006511909061486897651308648928}%
      {71634557740672624383232421875}\beta^{29} +
    O(\beta^{31}) \\
\chi &=&
   1 + 3\beta + 6\beta^{2} + \case{51}{5}\beta^{3} + \case{84}{5}\beta^{4} + 
   \case{978}{35}\beta^{5} + \case{7128}{175}\beta^{6} + 
   \case{345}{7}\beta^{7} + \case{12018}{175}\beta^{8} + 
   \case{239238}{1925}\beta^{9}
\nonumber \\ &+& 
   \case{52831068}{336875}\beta^{10} + 
   \case{1370148342}{21896875}\beta^{11} + 
   \case{36519432}{398125}\beta^{12} + \case{2120228676}{3128125}\beta^{13} + 
   \case{81696113448}{109484375}\beta^{14}  
\nonumber \\ &-& 
   \case{4005762392397}{2605728125}\beta^{15} -
   \case{5554706676084}{3648019375}\beta^{16} + 
   \case{522250407043193142}{60648322109375}\beta^{17} + 
   \case{28557036836995675104}{3335657716015625}\beta^{18} - 
\nonumber \\ &-& 
   \case{84308531983195740042}{2382612654296875}\beta^{19} -
   \case{568341711070335425448}{16678288580078125}\beta^{20} + 
   \case{60997222972929327707388}{383600637341796875}\beta^{21} + 
\nonumber \\ &+& 
   \case{26213984562109070694215784}{174538289990517578125}\beta^{22} - 
   \case{32254495435296445139757114}{45931128944873046875}\beta^{23} - 
   \case{43715635286204172906375312}{67130111534814453125}\beta^{24}  
\nonumber \\ &+& 
   \case{546580409405561136704024484}{174538289990517578125}\beta^{25} +
   \case{2495378933349261313782223032}{872691449952587890625}\beta^{26}  
\nonumber \\ &-& 
   \case{682220109367777429732601484621372}{48718000193603218994140625}
      \beta^{27} -
   \case{728104326241248730928651925895608}{57712092537037659423828125}
      \beta^{28}  
\nonumber \\ &+& 
   \case{562809908853468932757594137982353385288}%
      {8954319717584078047904052734375}\beta^{29} + 
   \case{8519985565279875808402224146574969010656}%
      {152223435198929326814368896484375}\beta^{30} 
\nonumber \\ &+& 
    O(\beta^{31}) \\
M_G^2 &=&
   \case{4}{3}\beta^{-1}-4 + \case{52}{15}\beta - 
   \case{316}{175}\beta^{3} + \case{5472}{875}\beta^{5} - 
   \case{64}{25}\beta^{6} - \case{19122724}{1010625}\beta^{7} + 
   \case{47808}{4375}\beta^{8} + \case{1316732616}{21896875}\beta^{9}  
\nonumber \\ &-& 
   \case{7823616}{153125}\beta^{10} - 
   \case{472105313248}{2299171875}\beta^{11} + 
   \case{1973349088}{8421875}\beta^{12} + 
   \case{46661184606992}{65143203125}\beta^{13} - 
   \case{26076544910176}{26823671875}\beta^{14}  
\nonumber \\ &-& 
   \case{1783696991213198452}{667131543203125}\beta^{15} + 
   \case{3724145632458848}{938828515625}\beta^{16} + 
   \case{5630519133419621984}{526682797265625}\beta^{17} - 
   \case{14678226956585492416}{877804662109375}\beta^{18}  
\nonumber \\ &-& 
   \case{23307722455628565130807096}{523614869971552734375}\beta^{19} + 
   \case{131635204541138381824}{1796123385546875}\beta^{20} + 
   \case{14931213920583922857376688}{79335586359326171875}\beta^{21}  
\nonumber \\ &-& 
   \case{12490315953001109355570208}{37943106519677734375}\beta^{22} - 
   \case{268173518152038112779584900384}{335986208231746337890625}\beta^{23}  
\nonumber \\ &+& 
   \case{1303640079018735894344620352}{872691449952587890625}\beta^{24} + 
   \case{38043748942433272601810917870624}{11242615429293050537109375}
      \beta^{25}  
\nonumber \\ &-& 
   \case{35163452856807151015094953488864}{5174187606768893603515625}
      \beta^{26} - 
   \case{32741212467941707889297214938863984773664}%
      {2283351527983939902215533447265625}\beta^{27} 
\nonumber \\ &+& 
   \case{1538444172762671916055304195595321312}%
      {49801555715150600933837890625}\beta^{28} +
    O(\beta^{29}) \\
M_{\rm v}^2 &=&
   \case{2}{9}\beta^{-2} -\case{38}{45} +  \case{1418}{1575}\beta^{2} + 
   \case{2804}{7875}\beta^{4} - \case{4877702}{3031875}\beta^{6} + 
   \case{83484008}{21896875}\beta^{8} - 
   \case{510322072076}{34487578125}\beta^{10}  
\nonumber \\ &+& 
   \case{241521726513358}{4104021796875}\beta^{12} - 
   \case{35557059082336258372}{150104597220703125}\beta^{14} + 
   \case{3656759742100547860823}{3752614930517578125}\beta^{16} +
    O(\beta^{18}) \\
M_{\rm h}^2 &=&
   \case{4}{3}\beta^{-1} -4 + \case{52}{15}\beta - 
   \case{316}{175}\beta^{3} + \case{15296}{2625}\beta^{5} - 
   \case{64}{125}\beta^{6} - \case{6651852}{336875}\beta^{7} + 
   \case{1088}{4375}\beta^{8} + \case{4502031416}{65690625}\beta^{9}  
\nonumber \\ &-& 
   \case{203936}{459375}\beta^{10} - 
   \case{192671527632}{766390625}\beta^{11} + 
   \case{1013665088}{294765625}\beta^{12} + 
   \case{1306699165544048}{1368007265625}\beta^{13} - 
   \case{1577414046048}{134118359375}\beta^{14}  
\nonumber \\ &-& 
   \case{187249702135322630044}{50034865740234375}\beta^{15} + 
   \case{27983550800928}{670591796875}\beta^{16} + 
   \case{197406026102200399712}{13167069931640625}\beta^{17} - 
   \case{561057571773872512}{4389023310546875}\beta^{18}  
\nonumber \\ &-& 
   \case{9420235976547662005883368}{154004373521044921875}\beta^{19} + 
   \case{17183199427949180266496}{43780507522705078125}\beta^{20} + 
   \case{3314631986901973406398391696}{13090371749288818359375}\beta^{21}  
\nonumber \\ &-& 
   \case{1688528041491767277609504}{1328008728188720703125}\beta^{22} - 
   \case{5344697433752690539376157188912}{5039793123476195068359375}
      \beta^{23} 
\nonumber \\ &+& 
   \case{710002489923384145123709548576}{176392759321666827392578125}
      \beta^{24} +
    O(\beta^{25}) \\
u &=&
   3\beta - 27\beta^{2} + \case{1176}{5}\beta^{3} - \case{10233}{5}\beta^{4} + 
   \case{623418}{35}\beta^{5} - \case{27129006}{175}\beta^{6} + 
   \case{9444396}{7}\beta^{7} - \case{82196703}{7}\beta^{8}  
\nonumber \\ &+& 
   \case{983643153048}{9625}\beta^{9} - 
   \case{299630551162554}{336875}\beta^{10} + 
   \case{169503983781937692}{21896875}\beta^{11} - 
   \case{1475232154782748902}{21896875}\beta^{12}  
\nonumber \\ &+& 
   \case{987637288715664984}{1684375}\beta^{13} - 
   \case{50792385588434618148}{9953125}\beta^{14} + 
   \case{82664811252077369258388}{1861234375}\beta^{15}  
\nonumber \\ &-& 
   \case{35253092409175923838220943}{91200484375}\beta^{16} +  
   \case{15694806775998694225868699994}{4665255546875}\beta^{17}  
\nonumber \\ &-& 
   \case{1993179427702306936700632573158}{68074647265625}\beta^{18} + 
   \case{170001570169775679767297711329632}{667131543203125}\beta^{19}  
\nonumber \\ &-& 
   \case{7397813830453458326408000754198846}{3335657716015625}\beta^{20} + 
   \case{7404260658028457836903051615667215212}{383600637341796875}
      \beta^{21}  
\nonumber \\ &-& 
   \case{1172825899888916586985509963759764004972}{6981531599620703125}
      \beta^{22} + 
   \case{1275921695053691221896293444849460211245084}%
      {872691449952587890625}\beta^{23}  
\nonumber \\ &-& 
   \case{1586377139730021695207597717642888690409438}%
      {124670207136083984375}\beta^{24} + 
   \case{96646235832092490263933098748637598311624744}%
      {872691449952587890625}\beta^{25}  
\nonumber \\ &-& 
   \case{221351162997126072386711604146938113185602308}%
      {229655644724365234375}\beta^{26}  
\nonumber \\ &+& 
   \case{408671731065098636806177681425625771659960813158664}%
      {48718000193603218994140625}\beta^{27}  
\nonumber \\ &-& 
   \case{273870833487071174807420156540338331788208519364045012}%
      {3751286014907447862548828125}\beta^{28}  
\nonumber \\ &+& 
   \case{5689559760891835362935463361457661566503603613628700393824}%
      {8954319717584078047904052734375}\beta^{29}  
\nonumber \\ &-& 
   \case{4208991513353764866716270905427277208225624532617948200865512}%
      {761117175994646634071844482421875}\beta^{30} +
    O(\beta^{31})
\end{eqnarray}
\end{mathletters}

\subsection{$\protect\bbox{N=4}$}

\begin{mathletters}
\begin{eqnarray}
E &=&
   \beta  - \case{2}{3}\beta^{3} + \case{8}{3}\beta^{5} - 
   \case{512}{45}\beta^{7} + \case{6254}{135}\beta^{9} - 
   \case{112352}{567}\beta^{11} + \case{495496}{567}\beta^{13} - 
   \case{501611216}{127575}\beta^{15} + 
   \case{11453817058}{637875}\beta^{17}  
\nonumber \\ &-& 
   \case{748041961864}{9021375}\beta^{19} + 
   \case{4879081930372}{12629925}\beta^{21} - 
   \case{468805978966751344}{258597714375}\beta^{23} + 
   \case{5164375543360356758}{603394666875}\beta^{25}  
\nonumber \\ &-& 
   \case{3308377214206458689408}{81458280028125}\beta^{27} + 
   \case{5256059401974658161694}{27152760009375}\beta^{29} +
    O(\beta^{31}) \\
\chi &=&
   1 + 3\beta + 6\beta^{2} + 10\beta^{3} + 16\beta^{4} + 26\beta^{5} + 
   \case{110}{3}\beta^{6} + \case{598}{15}\beta^{7} + 
   \case{752}{15}\beta^{8} + \case{4492}{45}\beta^{9} + 
   \case{1114}{9}\beta^{10} 
\nonumber \\ &-& 
   \case{21976}{945}\beta^{11} - 
   \case{18344}{315}\beta^{12} + \case{622724}{945}\beta^{13} + 
   \case{11892428}{14175}\beta^{14} - \case{102416012}{42525}\beta^{15} - 
   \case{19192244}{6075}\beta^{16} + \case{10289936}{875}\beta^{17}  
\nonumber \\ &+& 
   \case{9473001266}{637875}\beta^{18} - 
   \case{1147872224624}{21049875}\beta^{19} - 
   \case{285404814608}{4209975}\beta^{20} + 
   \case{5401807541912}{21049875}\beta^{21} + 
   \case{138663144738428}{442047375}\beta^{22}  
\nonumber \\ &-& 
   \case{14938117796303512}{12314176875}\beta^{23} - 
   \case{126027986916733444}{86199238125}\beta^{24} + 
   \case{386190762389548268}{67043851875}\beta^{25} + 
   \case{1379265161509279622}{201131555625}\beta^{26}  
\nonumber \\ &-& 
   \case{745692645593229011368}{27152760009375}\beta^{27} - 
   \case{878237978318189566412}{27152760009375}\beta^{28} + 
   \case{3567754536639184909436}{27152760009375}\beta^{29}  
\nonumber \\ &+& 
   \case{1783902695490692741638}{11636897146875}\beta^{30} +
    O(\beta^{31}) \\
M_G^2 &=&
   \case{4}{3}\beta^{-1}-4 + \case{32}{9}\beta - \case{56}{27}\beta^{3} + 
   \case{2944}{405}\beta^{5} - \case{8}{3}\beta^{6} - 
   \case{30136}{1215}\beta^{7} + \case{376}{27}\beta^{8} + 
   \case{2169952}{25515}\beta^{9} - \case{27872}{405}\beta^{10}  
\nonumber \\ &-& 
   \case{17364176}{54675}\beta^{11} + \case{139568}{405}\beta^{12} + 
   \case{1403817608}{1148175}\beta^{13} - 
   \case{206305088}{127575}\beta^{14} - 
   \case{12041211784}{2460375}\beta^{15} + \case{44783968}{6075}\beta^{16}  
\nonumber \\ &+& 
   \case{2327367655528}{113669325}\beta^{17} - 
   \case{5538622408}{164025}\beta^{18} - 
   \case{1056805727795792}{11935279125}\beta^{19} + 
   \case{2704092978728}{17222625}\beta^{20}  
\nonumber \\ &+& 
   \case{909455600757890632}{2327379429375}\beta^{21} - 
   \case{4903884524065528}{6630710625}\beta^{22} - 
   \case{85272481476107112904}{48874968016875}\beta^{23} + 
   \case{19055775959948008}{5425126875}\beta^{24}  
\nonumber \\ &+& 
   \case{821836927190253245168}{104732074321875}\beta^{25} - 
   \case{7802995733962068568}{465475885875}\beta^{26} - 
   \case{15607345431297974660104}{439874712151875}\beta^{27}  
\nonumber \\ &+& 
   \case{19608530438150678273776}{244374840084375}\beta^{28} +
    O(\beta^{29}) \\
M_{\rm v}^2 &=&
   \case{2}{9}\beta^{-2} -\case{22}{27} +  \case{8}{9}\beta^{2} + 
   \case{494}{1215}\beta^{4} - \case{8284}{3645}\beta^{6} + 
   \case{161257}{25515}\beta^{8} - \case{3197867}{127575}\beta^{10} + 
   \case{40478257}{382725}\beta^{12}  
\nonumber \\ &-& 
   \case{23351747878}{51667875}\beta^{14} +
   \case{2016622630922}{1023023925}\beta^{16} +
    O(\beta^{18}) \\
M_{\rm h}^2 &=&
   \case{4}{3}\beta^{-1}-4 + \case{32}{9}\beta - \case{56}{27}\beta^{3} + 
   \case{2764}{405}\beta^{5} - \case{16}{27}\beta^{6} - 
   \case{30376}{1215}\beta^{7} + \case{64}{81}\beta^{8} + 
   \case{2357776}{25515}\beta^{9} - \case{184}{135}\beta^{10}  
\nonumber \\ &-& 
   \case{19782236}{54675}\beta^{11} + \case{76984}{10935}\beta^{12} +
   \case{1686885896}{1148175}\beta^{13} - \case{2132936}{76545}\beta^{14} - 
   \case{5039762128}{820125}\beta^{15} + \case{76496944}{688905}\beta^{16}  
\nonumber \\ &+& 
   \case{2987093529676}{113669325}\beta^{17} - 
   \case{22231013972}{51667875}\beta^{18} - 
   \case{50559997058744}{442047375}\beta^{19}  
\nonumber \\ &+& 
   \case{775676060228}{465010875}\beta^{20} + 
   \case{391764598151298668}{775793143125}\beta^{21} -
   \case{512747827960696}{76726794375}\beta^{22} - 
   \case{110242538271813248108}{48874968016875}\beta^{23}  
\nonumber \\ &+& 
   \case{43888147443715136}{1611262681875}\beta^{24} +
    O(\beta^{25}) \\
u &=&
   3\beta - 27\beta^{2} + 235\beta^{3} - 2043\beta^{4} + 17765\beta^{5} - 
   154479\beta^{6} + \case{20149303}{15}\beta^{7} - 
   \case{58403373}{5}\beta^{8}
\nonumber \\ &+& 
   \case{4570670527}{45}\beta^{9} - 
   \case{13248238489}{15}\beta^{10} + \case{7257685461779}{945}\beta^{11} - 
   \case{4207327698821}{63}\beta^{12} + 
   \case{548778458844239}{945}\beta^{13}  
\nonumber \\ &-& 
   \case{3408540645268981}{675}\beta^{14} + 
   \case{1867275863842718053}{42525}\beta^{15} - 
   \case{51546286924640369}{135}\beta^{16} + 
   \case{235318541138701614541}{70875}\beta^{17}  
\nonumber \\ &-& 
   \case{6138706350282887902919}{212625}\beta^{18} + 
   \case{5284592580750019875772771}{21049875}\beta^{19} - 
   \case{5105854777719427773522979}{2338875}\beta^{20}  
\nonumber \\ &+& 
   \case{399586148153799097820463863}{21049875}\beta^{21} - 
   \case{24322483746219174453342441067}{147349125}\beta^{22} + 
\nonumber \\ &+& 
   \case{123726647366907086076547668505151}{86199238125}\beta^{23} - 
   \case{71725146330946050495871958406491}{5746615875}\beta^{24}  
\nonumber \\ &+& 
   \case{21829256518148380793466834521759833}{201131555625}\beta^{25} - 
   \case{569455321970865221978936851386596153}{603394666875}\beta^{26}  
\nonumber \\ &+& 
   \case{222828956715504343457272488503713477607}{27152760009375}
      \beta^{27} - 
   \case{13181163564610119027733292785978467713}{184712653125}\beta^{28}  
\nonumber \\ &+& 
   \case{16848860749534433639413131084425510515541}{27152760009375}
      \beta^{29} -
   \case{775190057499031799755350461405515599727}{143665396875}\beta^{30} 
\nonumber \\ &+& 
    O(\beta^{31})
\end{eqnarray}
\end{mathletters}

\subsection{$\protect\bbox{N=8}$}

\begin{mathletters}
\begin{eqnarray}
E &=&
   \beta  - \case{4}{5}\beta^{3} + \case{46}{15}\beta^{5} - 
   \case{7552}{525}\beta^{7} + \case{2302}{35}\beta^{9} - 
   \case{555317288}{1771875}\beta^{11} + 
   \case{2746333288}{1771875}\beta^{13} - 
   \case{1785780786208}{227390625}\beta^{15}  
\nonumber \\ &+& 
   \case{49727658484666}{1227909375}\beta^{17} - 
   \case{422283776631137752}{1995352734375}\beta^{19} + 
   \case{26014686098048746528}{23279115234375}\beta^{21}  
\nonumber \\ &-& 
   \case{151047237869123839244528}{25383189111328125}\beta^{23} + 
   \case{10528747447001620271543026}{329981458447265625}\beta^{25} - 
   \case{54722318647021454789447096}{318009342041015625}\beta^{27}  
\nonumber \\ &+& 
   \case{141254571303908072786177902882}{151461489427294921875}\beta^{29} +
    O(\beta^{31}) \\
\chi &=&
   1 + 3\beta + 6\beta^{2} + \case{48}{5}\beta^{3} + \case{72}{5}\beta^{4} + 
   \case{112}{5}\beta^{5} + \case{746}{25}\beta^{6} + 
   \case{170}{7}\beta^{7} + \case{3392}{175}\beta^{8} + 
   \case{2392}{35}\beta^{9} + \case{1330922}{13125}\beta^{10}  
\nonumber \\ &-& 
   \case{12052904}{84375}\beta^{11} - \case{38292976}{118125}\beta^{12} + 
   \case{509694772}{590625}\beta^{13} + 
   \case{35255323868}{20671875}\beta^{14} - 
   \case{201482297164}{45478125}\beta^{15}  
\nonumber \\ &-& 
   \case{1185431950796}{136434375}\beta^{16} + 
   \case{33950985399928}{1461796875}\beta^{17} + 
   \case{2298488652590486}{51162890625}\beta^{18} - 
   \case{409115403009898784}{3325587890625}\beta^{19}  
\nonumber \\ &-& 
   \case{781623089480829248}{3325587890625}\beta^{20} + 
   \case{2184027254831875996}{3325587890625}\beta^{21} + 
   \case{185758308159909008476}{149651455078125}\beta^{22}  
\nonumber \\ &-& 
   \case{35281258231914454156624}{9999438134765625}\beta^{23} -
   \case{726617483139164562921284}{109993819482421875}\beta^{24} + 
   \case{2395898361931558337852}{125707222265625}\beta^{25}  
\nonumber \\ &+& 
   \case{556037939109294945044734}{15713402783203125}\beta^{26} - 
   \case{4835166268399659267621472288}{46747373280029296875}\beta^{27} - 
   \case{763751109137093877154435748}{4006917709716796875}\beta^{28}  
\nonumber \\ &+& 
   \case{711299012158799734199117996536}{1262179078560791015625}\beta^{29} + 
   \case{100262522489076708186582427306586}{97187789049180908203125}
      \beta^{30} +
    O(\beta^{31}) \\
M_G^2 &=&
   \case{4}{3}\beta^{-1} -4 + \case{56}{15}\beta - 
   \case{608}{225}\beta^{3} + \case{8544}{875}\beta^{5} - 
   \case{56}{25}\beta^{6} - \case{690104}{16875}\beta^{7} + 
   \case{30632}{1875}\beta^{8} + \case{895240544}{5315625}\beta^{9} - 
\nonumber \\ &-& 
   \case{2698048}{28125}\beta^{10} - 
   \case{15337773008}{20671875}\beta^{11} + 
   \case{78233152}{140625}\beta^{12} + 
   \case{15033061154264}{4385390625}\beta^{13} - 
   \case{46357786976}{14765625}\beta^{14}  
\nonumber \\ &-& 
   \case{1502465010829864}{92093203125}\beta^{15} + 
   \case{11492930412128}{664453125}\beta^{16} + 
   \case{1111624153887072856}{13967469140625}\beta^{17} - 
   \case{3450172284172424}{36544921875}\beta^{18}  
\nonumber \\ &-& 
   \case{3734860553852716409584}{9428041669921875}\beta^{19} + 
   \case{8295415887259675624}{16116310546875}\beta^{20} + 
   \case{1982675233173138163625048}{989944375341796875}\beta^{21}  
\nonumber \\ &-& 
   \case{309125247582087653324456}{109993819482421875}\beta^{22} - 
   \case{16910028827683018859169992}{1649907292236328125}\beta^{23} + 
   \case{1691096970202616702325224}{109993819482421875}\beta^{24}  
\nonumber \\ &+& 
   \case{66840015065786580503988075712}{1262179078560791015625}\beta^{25} - 
   \case{46345645264745356443888008}{549969097412109375}\beta^{26}  
\nonumber \\ &-& 
   \case{1206164175984894430511591451154792}{4373450507213140869140625}
      \beta^{27} + 
   \case{57262390753050551283591565888}{123743046917724609375}\beta^{28} +
    O(\beta^{29}) \\
M_{\rm v}^2 &=&
   \case{2}{9}\beta^{-2} -\case{34}{45} + \case{568}{675}\beta^{2} + 
   \case{13784}{23625}\beta^{4} - \case{69604}{16875}\beta^{6} + 
   \case{241041191}{15946875}\beta^{8} - 
   \case{185964672997}{2790703125}\beta^{10}  
\nonumber \\ &+& 
   \case{1392084587186}{4385390625}\beta^{12} - 
   \case{10666554935618557}{6906990234375}\beta^{14} + 
   \case{120272997266688239927}{15713402783203125}\beta^{16} +
    O(\beta^{18}) \\
M_{\rm h}^2 &=&
   \case{4}{3}\beta^{-1} -4 + \case{56}{15}\beta - 
   \case{608}{225}\beta^{3} + \case{24652}{2625}\beta^{5} - 
   \case{224}{375}\beta^{6} - \case{673136}{16875}\beta^{7} + 
   \case{10976}{5625}\beta^{8} + \case{899887928}{5315625}\beta^{9}  
\nonumber \\ &-& 
   \case{480088}{84375}\beta^{10} - \case{5216665892}{6890625}\beta^{11} + 
   \case{53139224}{2109375}\beta^{12}  +
   \case{15464225123456}{4385390625}\beta^{13} - 
   \case{8888980504}{73828125}\beta^{14}  
\nonumber \\ &-& 
   \case{38922396533024104}{2302330078125}\beta^{15} + 
   \case{5789142673744}{9966796875}\beta^{16} + 
   \case{32467756338289108}{391904296875}\beta^{17} - 
   \case{3641291779914748}{1279072265625}\beta^{18}  
\nonumber \\ &-& 
   \case{19467809813152568040512}{47140208349609375}\beta^{19} + 
   \case{85327706507419012316}{6043616455078125}\beta^{20} + 
   \case{10330548206236154611036876}{4949721876708984375}\beta^{21}  
\nonumber \\ &-& 
   \case{117133256340256981616728}{1649907292236328125}\beta^{22} - 
   \case{8000075417927232351906676}{749957860107421875}\beta^{23} + 
   \case{1656430217477736394276208}{4583075811767578125}\beta^{24} 
\nonumber \\ &+& 
    O(\beta^{25}) \\
u &=&
   3\beta - 27\beta^{2} + \case{1173}{5}\beta^{3} - \case{10179}{5}\beta^{4} + 
   \case{88357}{5}\beta^{5} - \case{3834927}{25}\beta^{6} + 
   \case{9320831}{7}\beta^{7} - \case{404542143}{35}\beta^{8}  
\nonumber \\ &+& 
   \case{87789657841}{875}\beta^{9} - \case{3810245382897}{4375}\beta^{10} + 
   \case{4465049407354747}{590625}\beta^{11} - 
   \case{4306488725191523}{65625}\beta^{12}  
\nonumber \\ &+& 
   \case{336438262155902647}{590625}\beta^{13} - 
   \case{3785726352066159219}{765625}\beta^{14} + 
   \case{9759896773518266783449}{227390625}\beta^{15}  
\nonumber \\ &-& 
   \case{28239924187021013712337}{75796875}\beta^{16} + 
   \case{33093054294196142797506511}{10232578125}\beta^{17} - 
   \case{1612013627927391682265393}{57421875}\beta^{18}  
\nonumber \\ &+& 
   \case{162079960512817503825568623029}{665117578125}\beta^{19} - 
   \case{14211295974908998829041938103}{6718359375}\beta^{20}  
\nonumber \\ &+& 
   \case{427441298299484525828261030140363}{23279115234375}\beta^{21} -
   \case{25240543508916505408741986975703}{158361328125}\beta^{22}  
\nonumber \\ &+& 
   \case{152179939416162537931621662918853279261}{109993819482421875}
      \beta^{23} -
   \case{146775799365247069031550755820556182991}{12221535498046875}
      \beta^{24} 
\nonumber \\ &+& 
   \case{11466649049267256355352323832429180274931}{109993819482421875}
      \beta^{25} -
   \case{55297254904420195051495336139745677825041}{61107677490234375}
      \beta^{26} 
\nonumber \\ &+& 
   \case{367201599583278031923778483847484819042846311}{46747373280029296875}
      \beta^{27} 
\nonumber \\ &-& 
   \case{1062485144503393173493863633401017982717619013}%
      {15582457760009765625}\beta^{28} 
\nonumber \\ &+& 
   \case{57465089834508188027065801793670434250632397031}%
      {97090698350830078125}\beta^{29} 
\nonumber \\ &-& 
   \case{277399209545717425078049048151477437693717560565633}%
      {53993216138433837890625}\beta^{30} +
    O(\beta^{31})
\end{eqnarray}
\end{mathletters}



\input psfig
\pssilent
\def\centerline#1{\hbox to \hsize {\hss #1\hss}}

\newpage
\begin{figure}
\centerline{\psfig{figure=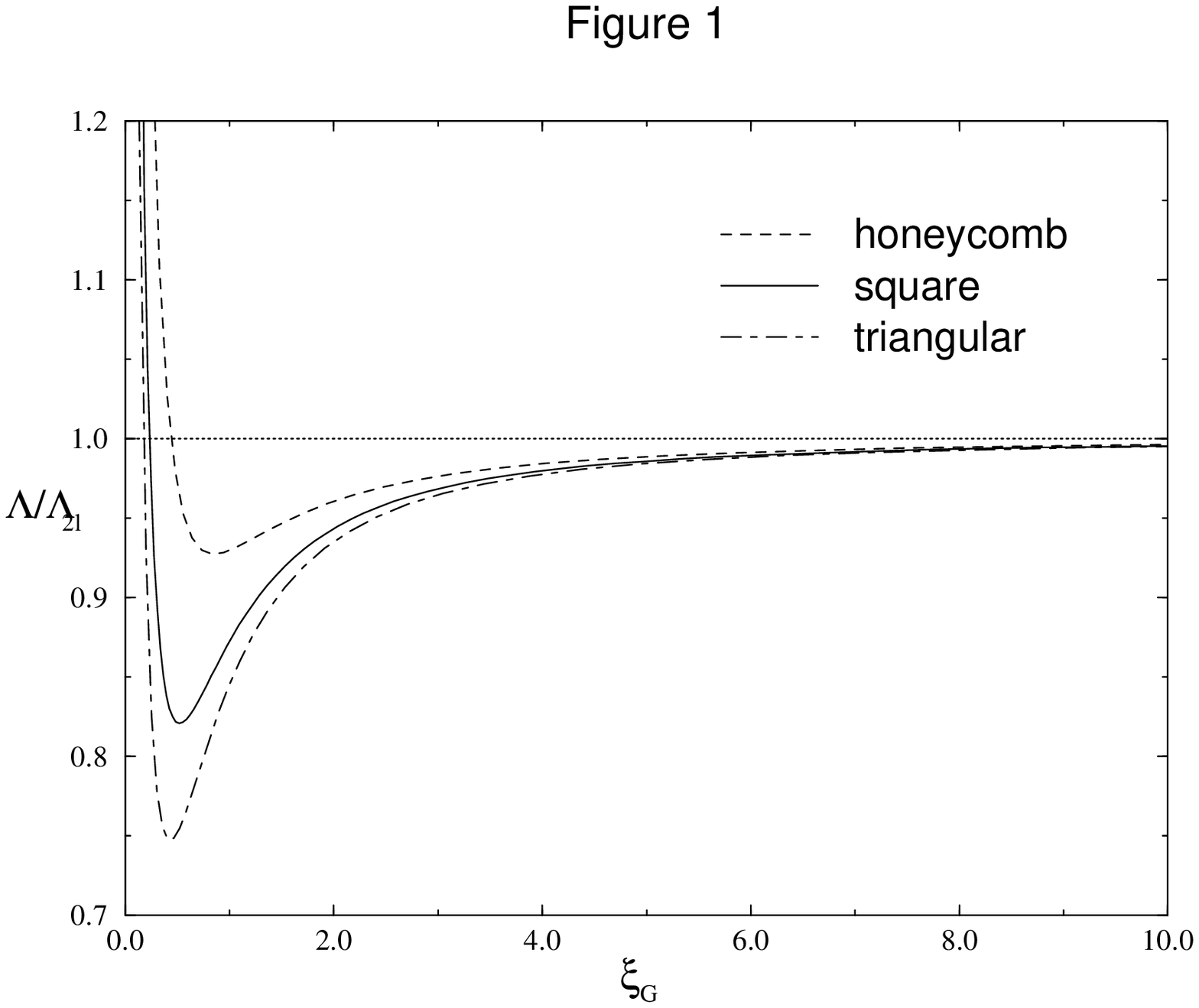}}
\caption{The large-$N$ limit of the ratio between
$M_{G}$ and the corresponding weak coupling asymptotic
formula for the square, triangular and honeycomb lattices.}
\label{asyiN}
\end{figure}

\newpage
\begin{figure}
\centerline{\psfig{figure=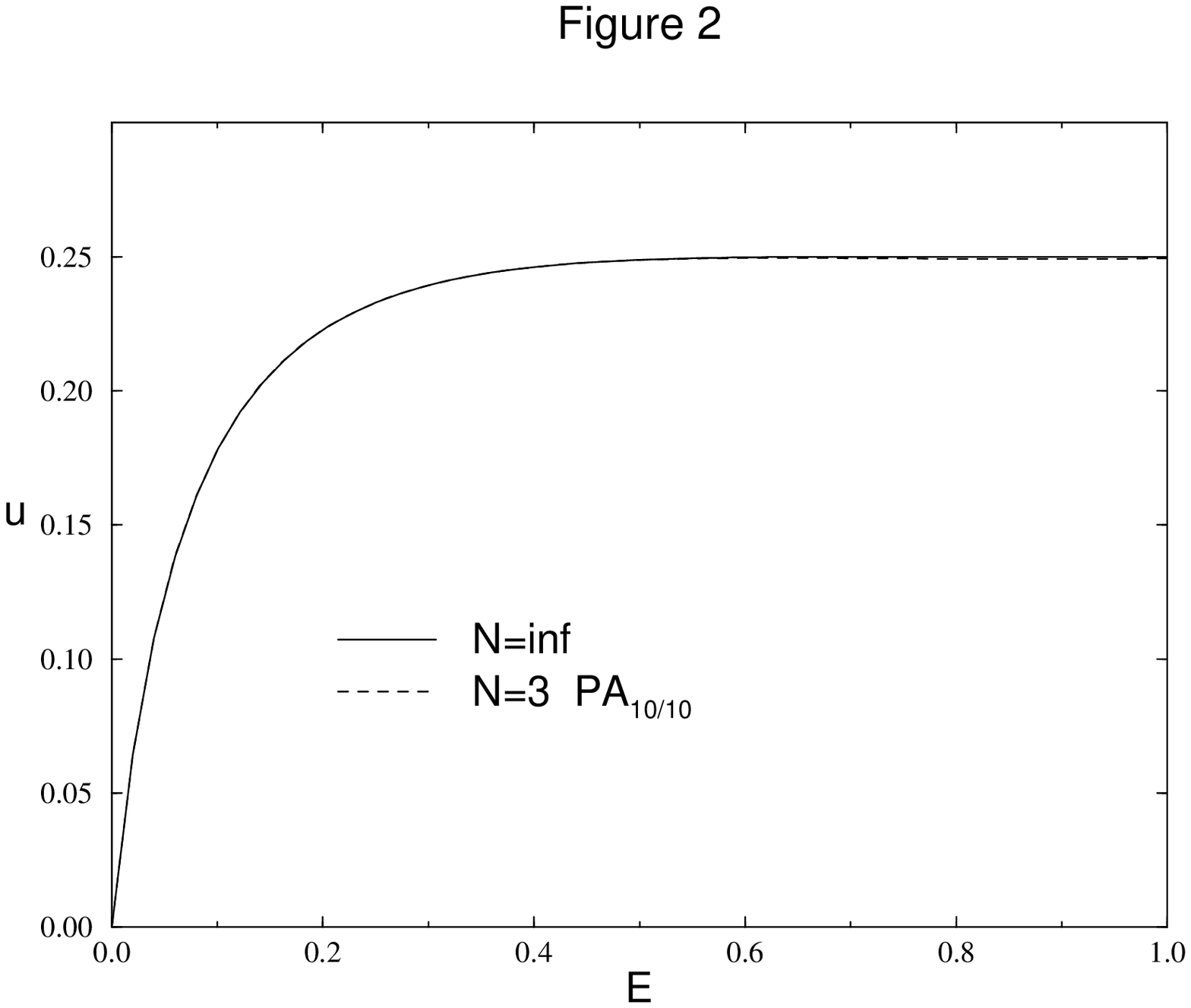}}
\caption{$u \equiv m_2^2 / (\chi m_4)$ versus $E$ for 
$N=3$ (as obtained by the $[10/10]$ PA) and $N=\infty$ (exact),
on the square lattice.}
\label{figomsq}
\end{figure}

\newpage
\begin{figure}
\centerline{\psfig{figure=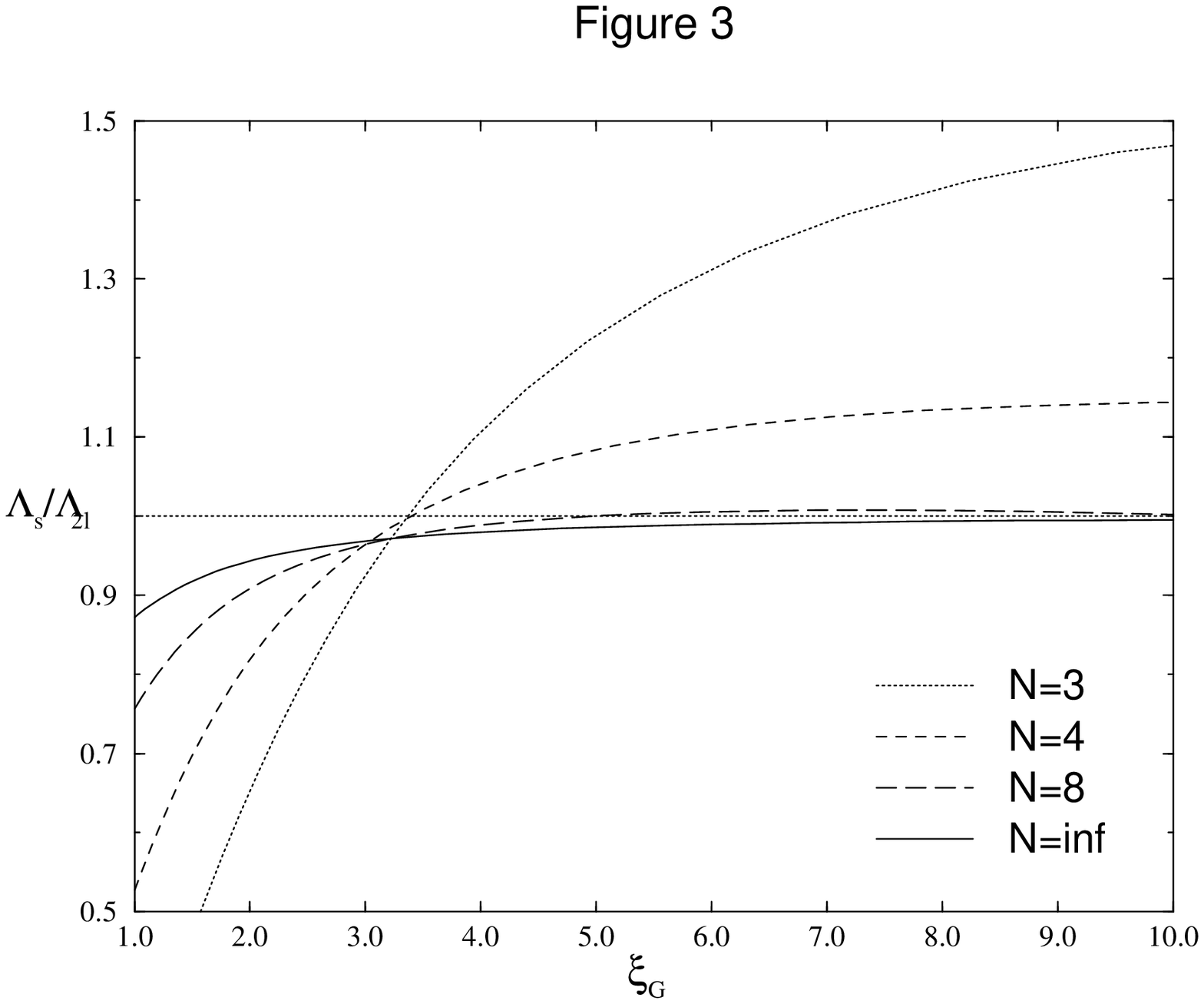}}
\caption{
Asymptotic scaling test from the strong-coupling
determinations of $\xi_{G}^2$ on the square lattice. 
We show curves of $\Lambda_{\rm s}/\Lambda_{2l}$, defined by  
Eqs.(\protect\ref{RL}--\protect\ref{twoloopla}), for $N=3,4,8$ and for 
$N=\infty$ (exact).}
\label{asysc}
\end{figure} 
  
\newpage
\begin{figure}
\centerline{\psfig{figure=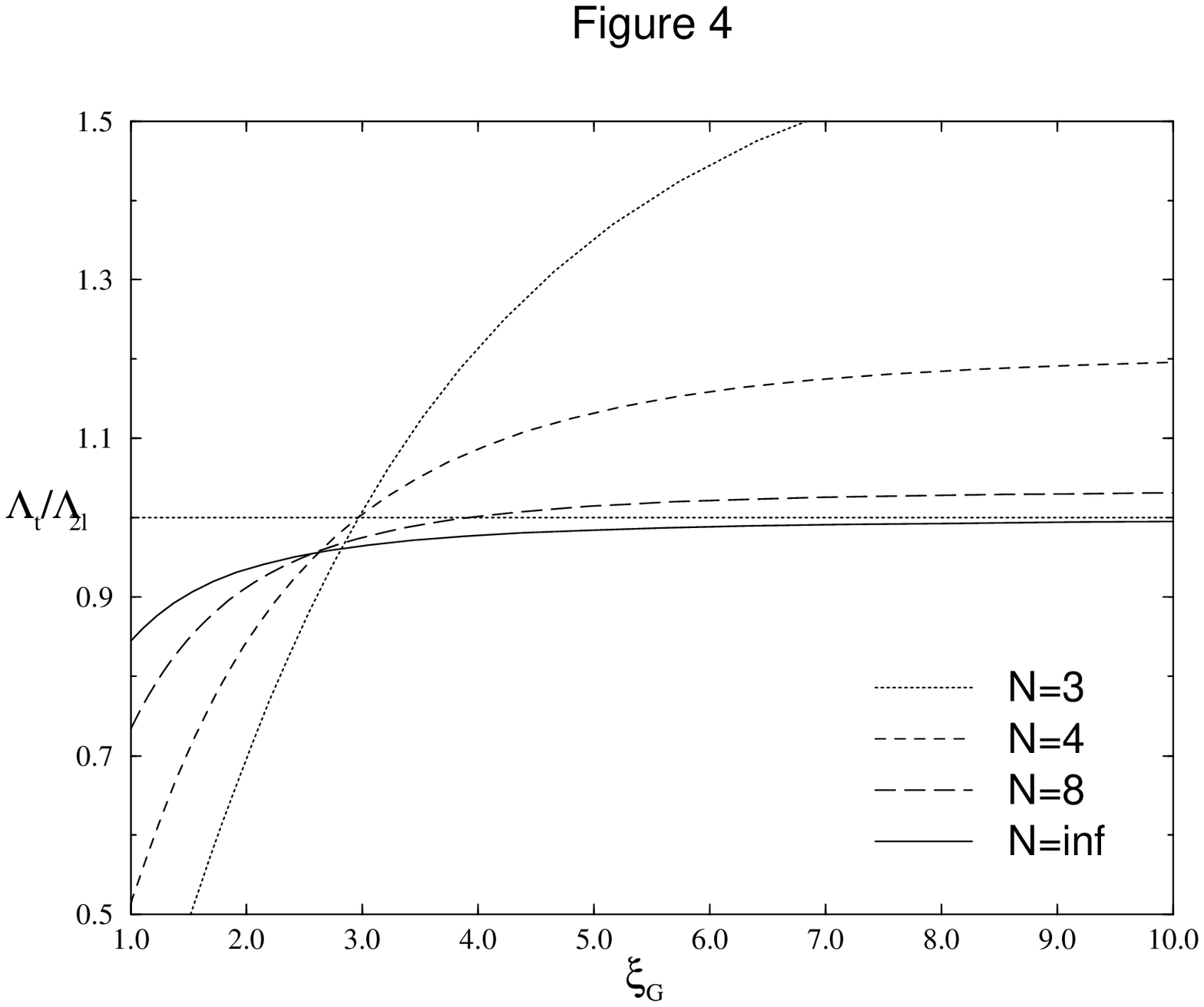}}
\caption{Asymptotic scaling test from the strong-coupling
determinations of $\xi_{G}^2$ on the triangular lattice. 
Curves of $\Lambda_{\rm t}/\Lambda_{2l}$ for $N=3,4,8$ 
and for $N=\infty$ (exact) are shown vs. $\xi_G$.}
\label{asysctr}
\end{figure}

\newpage
\begin{figure}
\centerline{\psfig{figure=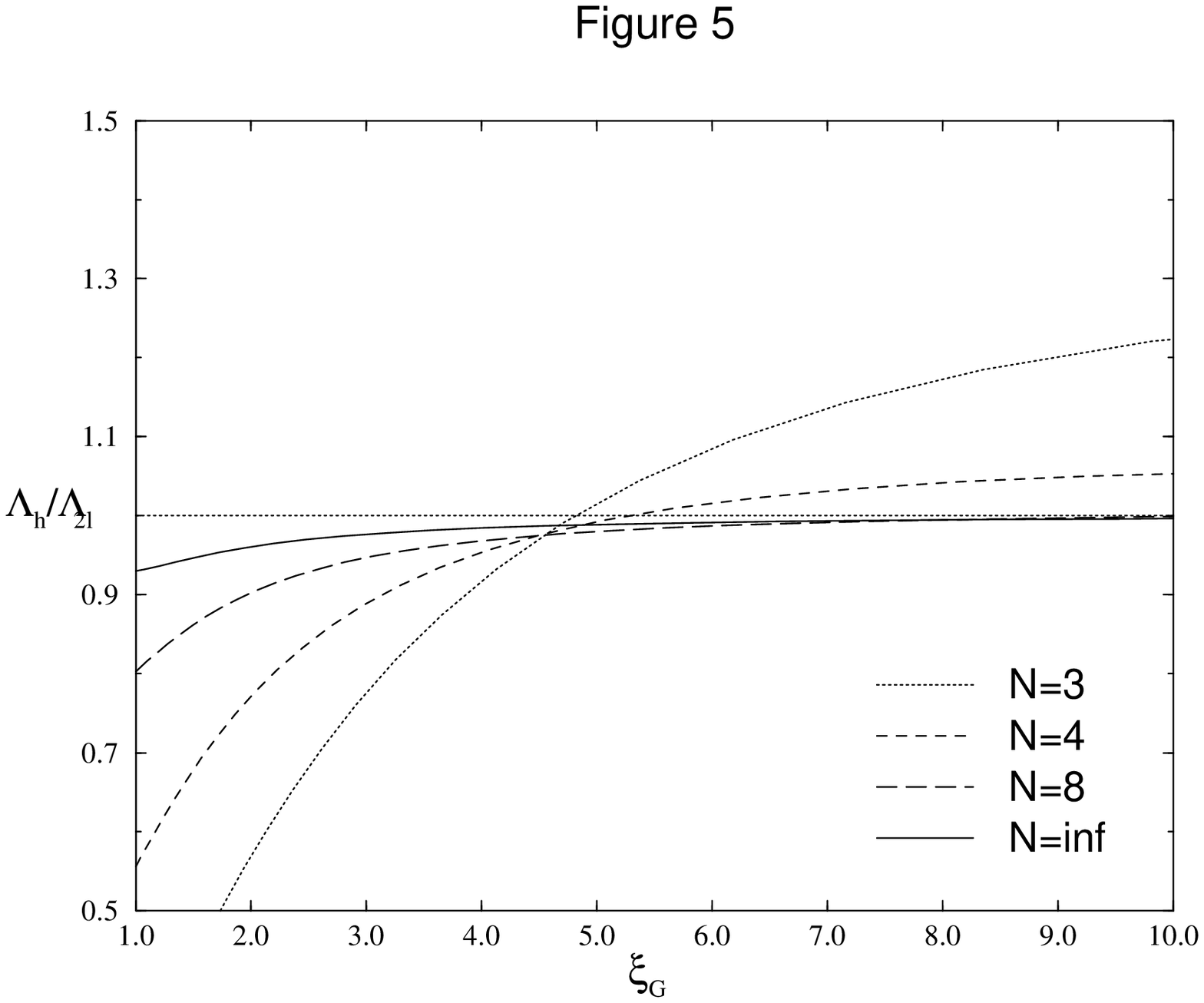}}
\caption{Asymptotic scaling test from the strong-coupling
determinations of $\xi_{G}^2$ on the honeycomb  lattice. 
Curves of $\Lambda_{\rm h}/\Lambda_{2l}$ for $N=3,4,8$ 
and for $N=\infty$ (exact) are shown vs. $\xi_G$.}
\label{asyscho}
\end{figure} 
  

\begin{table}
\squeezetable
\caption{
Summary of the large-$N$ calculations in ${\rm O}(N)$ $\sigma$ models
on the square, triangular and honeycomb lattices. All quantities 
appearing in this table have been defined in Sec.~\protect\ref{secNi}.
\label{sumiN}} 
\renewcommand\arraystretch{1.3}
\begin{tabular}{cccc}
\multicolumn{1}{c}{$$}&
\multicolumn{1}{c}{square}&
\multicolumn{1}{c}{triangular}&
\multicolumn{1}{c}{honeycomb}\\
\tableline \hline
$c$ & 4 & 6 & 3 \\\hline
$v_s$ & 1 & ${\sqrt{3}/ 2}$ & ${3\sqrt{3}/ 4}$ \\\hline  
$t$ & ${4v_s/c\beta}$ &  ${4v_s/ c\beta}$ 
& ${4v_s/ c\beta}$ \\ \hline
$\chi$ & ${4/c\beta z}$ & ${4/ c\beta z}$ & 
${4/c\beta  z}$ \\\hline
$M_{G}^2$ & $z$ & $z$ & $z$ \\\hline
$u$ & ${1\over 4}\left( 1 + {1\over 16}z\right)^{-1}$ & 
${1\over 4}\left( 1 + {1\over 16}z\right)^{-1}$ & 
${1\over 4}\left( 1 + {1\over 16}z\right)^{-1}$ \\\hline
$E$ & $1-1/(c\beta)+{1\over 4}z$  & 
$1-1/(c\beta)+{1\over 4}z$ 
& $1-1/(c\beta)+{1\over 4}z$ \\\hline
$M^2$ &  
$M_{\rm s}^2\equiv 2\bigl( {\rm cosh}\mu_{\rm s}-1\bigr)=z$ &
$M_{\rm t}^2\equiv {8\over 3}\bigl( {\rm cosh}
(\sqrt{3}\mu_{\rm t}/2) - 1\bigr)=z$ &
$M_{\rm h}^2\equiv {8\over 3}\bigl( {\rm cosh}
(\sqrt{3}\mu_{\rm h}/2) - 1\bigr)=z$ \\
&
$M_{\rm d}^2\equiv 
4\bigl( {\rm cosh}(\mu_{\rm d}/\sqrt{2}) - 1\bigr)=z$ &
&
$M_{\rm v}^2\equiv {8\over 9}\bigl( {\rm cosh}
({3\over2} \mu_{\rm v}) - 1\bigr) = z \bigl( 1+{1\over 8}z\bigr)$ \\
\hline
$\Lambda_{\overline{\rm MS}}/\Lambda_{\rm L}$&
$4\sqrt{2}$& $4\sqrt{3}$& 4 \\
\end{tabular}
\end{table}

\begin{table}
\caption{For various values of $N$ we report the singularity closest
to the origin on the square, triangular, and honeycomb lattices, as
obtained by a Dlog-PA analsysis of the strong-coupling series of
$\chi$ ($\bar{\beta}_\chi$) and $\xi_{G}^2$ ($\bar{\beta}_\xi$), and
the corresponding convergence radius of the strong-coupling expansion
$\beta_r$. The errors we display are related to the spread of the
results coming from different quasi-diagonal $[l/m]$ Dlog-PA's using
all available terms of the series, or a few less, while the difference
between $\bar{\beta}_\chi$ and $\bar{\beta}_\xi$ should give an idea
of the systematic error in the procedure.
\label{zeroes}} 
\begin{tabular}{ccccc}
\multicolumn{1}{c}{lattice}&
\multicolumn{1}{c}{$N$}&
\multicolumn{1}{c}{$\bar{\beta}_\chi$}&
\multicolumn{1}{c}{$\bar{\beta}_\xi$}&
\multicolumn{1}{c}{$\beta_r$}\\
\tableline \hline
square & 3  &  $0.590(1)\pm  \,i\,0.156(1)$  &  
               $0.586(1)\pm \,i\,0.157(1)$   & 0.61 \\
       & 4  &  $0.557(10)\pm \,i\,0.226(4)$  &  
               $0.555(5)\pm \,i\,0.225(5)$   & 0.60 \\
       & 8  &  $0.467(4)\pm \,i\,0.298(3)$   &  
               $0.467(3)\pm \,i\,0.293(1)$   & 0.55 \\
       & $\infty$ & $0.321621..(1\pm i)$     &
               $0.321621...(1\pm i)$         & 0.454841...\\
\hline
triangular & 3 & $0.3582(2)\pm \,i\,0.085(1)$ & 
                 $0.357(1)\pm \,i\,0.089(4)$    & 0.37     \\
           & 4 & $0.343(1)\pm \,i\,0.121(1)$  & 
                 $0.341(2)\pm \,i\,0.124(2)$    & 0.36     \\
           & 8 & $0.2901(1)\pm \,i\,0.1654(1)$  & 
                 $0.283(4)\pm \,i\,0.163(4)$    & 0.33     \\
           & $\infty$ & $0.206711...\pm \,i\,0.181627...$ & 
                 $0.206711...\pm \,i\,0.181627...$ & 0.275169...\\
\hline

honeycomb  & 3        & $\pm \,i\,0.459(1)$      & 
                        $\pm \,i\,0.461(1)$  & 0.46 \\
           & 4        & $\pm \,i\,0.4444(1)$     & 
                        $\pm \,i\,0.445(2)$  & 0.44  \\
           & 8        & $\pm \,i\,0.4161(1)$     & 
                        $\pm \,i\,0.4169(2)$ & 0.42  \\
           & $\infty$ & $\pm \,i\,0.362095...$ & 
                        $\pm \,i\,0.362095...$ & 0.362095...\\
\end{tabular}
\end{table}

\begin{table}
\squeezetable
\caption{
Analysis of the 14th order strong-coupling series of $r\equiv
M_{\rm s}^2/M_{\rm d}^2$ for $N=3$ on the square lattice, expressed in
powers of $E$ and $\beta$.  The first two lines report the values of
the $[l/m]$ PA's and Dlog-PA's at $E=1$.  The last two lines report
the values of $[l/m]$ PA's and Dlog-PA's at $\beta=0.55$ corresponding
to $\xi\simeq 25$.  We show data for PA's and Dlog-PA's with $l+m\geq
11$ and $m\geq l\geq 5$.  Asterisks mark defective PA's, i.e., PA's
with spurious singularities close to the real axis for 
$E\protect\lesssim 1$ in the energy series case, or for
$\beta\protect\lesssim 0.55$ in the $\beta$-series case.
\label{sqr}}
\begin{tabular}{llr@{}lr@{}lr@{}lr@{}lr@{}lr@{}lr@{}lr@{}l}
\multicolumn{1}{c}{$$}&
\multicolumn{1}{c}{$$}&
\multicolumn{2}{c}{$5/6$}&
\multicolumn{2}{c}{$6/6$}&
\multicolumn{2}{c}{$5/7$}&
\multicolumn{2}{c}{$6/7$}&
\multicolumn{2}{c}{$5/8$}&
\multicolumn{2}{c}{$7/7$}&
\multicolumn{2}{c}{$6/8$}&
\multicolumn{2}{c}{$5/9$}\\
\tableline \hline
$E=1$& PA & *& & 0&.9965 & 0&.9967 & 0&.9955 & 1&.0126 & 0&.9980 &
                 1&.0007 & 1&.0120\\
     &DLPA & 1&.0002 & 1&.0011 & 1&.0023 & 1&.0005 & 
0&.9995 &&&&&&\\\hline
$\beta=0.55$&PA & 0&.9986 & 0&.9996 & 1&.0015 & 1&.0007 & 1&.0010 
& 1&.0007 & 1&.0007 &1&.0012\\
&DLPA & 0&.9996 & 1&.0007 & 0&.9993 & 1&.0006 & 0&.9999 &&&&&&\\
\end{tabular}
\end{table}

\begin{table}
\squeezetable
\caption{
Analysis of the 16th order strong-coupling series of $s\equiv
M_{\rm s}^2/M_{G}^2$ for $N=3$ on the square lattice.  The first two
lines report the values of the $[l/m]$ PA's and Dlog-PA's at $E=1$.
The last two lines report the values of $[l/m]$ PA's and Dlog-PA's at
$\beta=0.55$.  We show data for PA's and Dlog-PA's with $l+m\geq 13$
and $m\geq l\geq 5$.  Asterisks mark defective PA's.
\label{sqs}} 
\begin{tabular}{llr@{}lr@{}lr@{}lr@{}lr@{}lr@{}lr@{}lr@{}l
r@{}lr@{}lr@{}lr@{}l}
\multicolumn{1}{c}{$$}&
\multicolumn{1}{c}{$$}&
\multicolumn{2}{c}{$6/7$}&
\multicolumn{2}{c}{$5/8$}&
\multicolumn{2}{c}{$7/7$}&
\multicolumn{2}{c}{$6/8$}&
\multicolumn{2}{c}{$5/9$}&
\multicolumn{2}{c}{$7/8$}&
\multicolumn{2}{c}{$6/9$}&
\multicolumn{2}{c}{$5/10$}&
\multicolumn{2}{c}{$8/8$}&
\multicolumn{2}{c}{$7/9$}&
\multicolumn{2}{c}{$6/10$}&
\multicolumn{2}{c}{$5/11$}\\
\tableline \hline
$E=1$ & PA &0&.9947 &0&.9938 &0&.9941 &0&.9942 &*& &0&.9944 &1&.0020
      &*& &0&.9961& 1&.0028 & 1&.0028& *& \\ 
 & DLPA &0&.9941 &0&.9971 &*& &0&.9992 &0&.9978 &0&.9951 &0&.9973 
      &0&.9984 && && && && \\\hline
$\beta=0.55$ &PA&*& &0&.9971 &0&.9972 &0&.9974 &*& &0&.9976 &0&.9988
&0&.9998 &0&.9980 &1&.0023 &0&.9996 &0&.9995 \\
&DLPA&0&.9971 &0&.9980 &0&.9974 &0&.9992 &0&.9985 &0&.9977 &0&.9982
&0&.9976 && && && && \\
\end{tabular}
\end{table}

\begin{table}
\squeezetable
\caption{
Analysis of the 20st order strong-coupling series of
$E^{-1}u(E)$ and $\beta^{-1}u(\beta)$, where 
$u\equiv m_2^2/(\chi m_4)$, for $N=3$ on the square lattice.  The
first two lines report the values of $u$ as obtained from the $[l/m]$
PA's and Dlog-PA's at $E=1$.  The last two lines report the values of
$u$ from $[l/m]$ PA's and Dlog-PA's at $\beta=0.55$.  The analysis
detected a pole at $E_0=-0.086418$ in the energy series, and at
$\beta_0=-0.085545$ in the $\beta$ series, corresponding to
$M_{G}^2=-16.000$.  We show data for PA's and Dlog-PA's with $l+m\geq
16$ and $m\geq l\geq 8$.  Asterisks mark defective approximants.
\label{sqom}} 
\begin{tabular}{llr@{}lr@{}lr@{}lr@{}lr@{}lr@{}lr@{}lr@{}l}
\multicolumn{1}{c}{$$}&
\multicolumn{1}{c}{$$}&
\multicolumn{2}{c}{$8/8$}&
\multicolumn{2}{c}{$8/9$}&
\multicolumn{2}{c}{$9/9$}&
\multicolumn{2}{c}{$9/10$}&
\multicolumn{2}{c}{$8/11$}&
\multicolumn{2}{c}{$10/10$}&
\multicolumn{2}{c}{$9/11$}&
\multicolumn{2}{c}{$8/12$}\\
\tableline \hline
$E=1$&PA&    0&.2491 &0&.2502 &0&.2495 &0&.2488 &0&.2491 &0&.2495 
&0&.2504 &0&.2496 \\
     &DLPA&   0&.2497 &0&.2524 &0&.2510 &0&.2486 &0&.2492 & & && && \\
\hline
$\beta=0.55$&PA &0&.2493 &0&.2496 &0&.2488 &0&.2496 &0&.2500 
&*&      &0&.2495 &0&.2503 \\
            &DLPA&0&.2493 &0&.2495 &0&.2491 &0&.2498 &0&.2500 
& & && && \\
\end{tabular}
\end{table}

\begin{table}
\caption{
In this Table we summarize our strong-coupling results for $N=3$,
giving the estimates of $r^*$, $s^*$ and $u^*$ 
from the PA and Dlog-PA analysis of both
the energy and $\beta$-series of $r$, $s$, and $u$.
For all lattices considered the values of $\beta$ where
the $\beta$-approximants have been evaluated correspond to a
correlation length $\xi\protect\gtrsim 20$.
\label{sum}} 
\begin{tabular}{lllr@{}lr@{}lr@{}l}
\multicolumn{1}{c}{$$}&
\multicolumn{1}{c}{$$}&
\multicolumn{1}{c}{$$}&
\multicolumn{2}{c}{$r^*$}&
\multicolumn{2}{c}{$s^*$}&
\multicolumn{2}{c}{$u^*$}\\
\tableline \hline
square &  $E=1$      & PA  & 1&.004(8)  & 1&.000(5)  & 0&.2498(6) \\
       &             & DLPA& 1&.0000(12)& 0&.997(2)  & 0&.249(2)  \\
       &$\beta=0.55$ & PA  & 1&.0009(6) & 1&.000(2)  & 0&.2499(6) \\
       &             & DLPA& 1&.0002(6) & 0&.9978(8) & 0&.2499(5) \\
\hline
triangular &$E=1$    & PA  &  &         & 1&.000(4)  & 0&.2497(15) \\
       &             & DLPA&  &         & 0&.997(3)  & 0&.248(2) \\
       &$\beta=0.33$ & PA  &  &         & 0&.9985(13)& 0&.2504(3) \\
       &             & DLPA&  &         & 0&.9980(9) & 0&.2499(4) \\
\hline
honeycomb  &$E=1$    & PA  & 1&.01(4)   & 0&.999(4)  & 0&.250(2) \\
       &             & DLPA& 0&.991(13) & 0&.999(3)  & 0&.247(2) \\
       &$\beta=0.85$ & PA  & 1&.001(2)  & 0&.9987(5) & 0&.2490(3) \\
       &             & DLPA& 1&.0009(8) & 0&.9987(5) & 0&.2491(3) \\
\end{tabular}
\end{table}

\begin{table}
\caption{
The strong-coupling estimates of $\xi_G$ are compared with the
available Monte Carlo results on the square lattice
for various values of $N$.
The strong-coupling estimates of $\xi_G$ come from the plain series of 
$\xi_G^2$, and from ([9/10],[8/11],[9/9],[8/10]) Dlog-PA's of
$\beta^{-1} \xi_G^2$.  The $N=3,4,8$ M.C. data are taken respectively
from Refs.~\protect\cite{xiO3,Sokal},
\protect\cite{Wolff}. The asterisk indicates that the number concerns
$\xi_{\rm exp}$, and not $\xi_G$.
\label{mc}} 
\begin{tabular}{ccr@{}lr@{}lr@{}l}
\multicolumn{1}{c}{$N$}&
\multicolumn{1}{c}{$\beta$}&
\multicolumn{2}{c}{plain series}&
\multicolumn{2}{c}{Dlog-PA's}&
\multicolumn{2}{c}{M.C.}\\
\tableline \hline
3  & ${1.4/3}$ & 6&.567   & 6&.869(1) &  6&.90(1)$^*$\\
   & 0.5            & 9&.939   &11&.036(4) & 11&.05(1)$^*$ \\
   & ${1.6/ 3}$ &15&.429   &18&.90(2)  & 19&.00(2)$^*$ \\
   & ${1.7/ 3}$ &24&.300   &33&.9(1)   & 34&.44(6)$^*$ \\
   & 0.6            &38&.459   &61&.0(4)   & 64&.7(3)$^*$\\\hline
4  & 0.45  & 4&.665   & 4&.672(1) &  4&.67(1)\\
   & 0.5   & 7&.845   & 7&.87(1)  &  7&.83(1) \\
   & 0.55  &13&.879   &13&.88(5)  & 13&.99(3) \\
   & 0.575 &18&.701   &18&.6(2)   & 18&.91(5) \\
   & 0.6   &25&.329   &24&.8(4)   & 25&.5(2) \\\hline
8  & 0.5   & 5&.432   & 5&.459(1) &  5&.461(5)$^*$\\
   & 0.525 & 6&.584   & 6&.651(1)  &  & \\
   & 0.55  & 7&.981   & 8&.139(2)  &  & \\
   & 0.575 & 9&.659   &10&.01(1)  &  9&.884(13)$^*$ \\
\end{tabular}
\end{table}

\begin{table}
\squeezetable
\caption{
Analysis of the 11th order strong-coupling series of $s\equiv
M_{\rm t}^2/M_{G}^2$ for $N=3$ on the triangular lattice.  The first
two lines report the values of the $[l/m]$ PA's and Dlog-PA's at
$E=1$.  The last two lines report the values of $[l/m]$ PA's and
Dlog-PA's at $\beta=0.33$ corresponding to $\xi\simeq 22$.  Asterisks
mark defective PA's.
\label{trs}} 
\begin{tabular}{llr@{}lr@{}lr@{}lr@{}lr@{}lr@{}l}
\multicolumn{1}{c}{$$}&
\multicolumn{1}{c}{$$}&
\multicolumn{2}{c}{$4/4$}&
\multicolumn{2}{c}{$4/5$}&
\multicolumn{2}{c}{$5/5$}&
\multicolumn{2}{c}{$4/6$}&
\multicolumn{2}{c}{$5/6$}&
\multicolumn{2}{c}{$4/7$}\\
\tableline \hline
$E=1$
&PA   & 0&.9993 & 0&.9972 & 1&.0005 & 0&.9927  & 0&.9954 & 1.&0039 \\
&DLPA & 0&.9963 & 1&.0014 & 0&.9948 &  *& &  &      &   & \\\hline
$\beta=0.33$
&PA   & 0&.9993 & 0&.9989 & 1&.0005 & 0&.9969 & 0&.9974 & 0&.9995   \\
&DLPA & 0&.9987 & *&      & 0&.9972 & 0&.9975 &  &      & &\\\hline
\end{tabular}
\end{table}

\begin{table}
\squeezetable
\caption{
Analysis of the 14st order
strong-coupling series of $E^{-1}u(E)$ and $\beta^{-1}u(\beta)$,
where $u\equiv m_2^2/(\chi m_4)$,
for $N=3$ on the triangular lattice.
The first two lines report the values
of $u$ as obtained from the $[l/m]$ PA's and Dlog-PA's 
at $E=1$. 
The last two lines report the values
of $u$ from $[l/m]$ PA's and Dlog-PA's 
at $\beta=0.33$ corresponding to $\xi\simeq 22$.
A pole has been detected at $E_0=-0.050655$,
corresponding to $M_{G}^2=-16.000$.
Asterisks mark defective PA's.
\label{trom}}
\begin{tabular}{llr@{}lr@{}lr@{}lr@{}lr@{}lr@{}lr@{}lr@{}l}
\multicolumn{1}{c}{$$}&
\multicolumn{1}{c}{$$}&
\multicolumn{2}{c}{$5/6$}&
\multicolumn{2}{c}{$6/6$}&
\multicolumn{2}{c}{$5/7$}&
\multicolumn{2}{c}{$6/7$}&
\multicolumn{2}{c}{$5/8$}&
\multicolumn{2}{c}{$7/7$}&
\multicolumn{2}{c}{$6/8$}&
\multicolumn{2}{c}{$5/9$}\\
\tableline \hline
$E=1$
&PA  &0&.2442 &0&.2502 & 0&.2533 & 0&.2483 & 0&.2492 & 0&.2494 & 
      0&.2500 & 0&.2497 \\
&DLPA&0&.2433 &0&.2521 & 0&.2502 & 0&.2477 & 0&.2491 &  & && && \\
\hline
$\beta=0.33$
&PA  &0&.2521 &0&.2500 & 0&.2500 & 0&.2502 & 0&.2502 & 0&.2504 & 
      0&.2505 & 0&.2502 \\
&DLPA&0&.2496 &0&.2494 & 0&.2494 & 0&.2496 & 0&.2502 && && && \\
\hline
\end{tabular}
\end{table}

\begin{table}
\squeezetable
\caption{
Analysis of the 19th order strong-coupling series of $r/\beta$, where
$r\equiv M_{\rm h}^2/M_{\rm v}^2$, for $N=3$ on the honeycomb lattice,
expressed in powers of $E$ and $\beta$.  The first two lines report
the values of the $[l/m]$ PA's and Dlog-PA's at $E=1$.  The last two
lines report the values of $[l/m]$ PA's and Dlog-PA's at $\beta=0.85$
corresponding to $\xi\simeq 22$.  Asterisks mark defective PA's, i.e.,
PA's with spurious singularities close to the real axis for
$E\protect\lesssim 1$ in the energy series case, or for
$\beta\protect\lesssim 0.85$ in the $\beta$-series case.
\label{hor}} 
\begin{tabular}{llr@{}lr@{}lr@{}lr@{}lr@{}lr@{}lr@{}lr@{}lr@{}lr@{}l}
\multicolumn{1}{c}{$$}&
\multicolumn{1}{c}{$$}&
\multicolumn{2}{c}{$8/8$}&
\multicolumn{2}{c}{$7/9$}&
\multicolumn{2}{c}{$8/9$}&
\multicolumn{2}{c}{$7/10$}&
\multicolumn{2}{c}{$9/9$}&
\multicolumn{2}{c}{$8/10$}&
\multicolumn{2}{c}{$7/11$}&
\multicolumn{2}{c}{$9/10$}&
\multicolumn{2}{c}{$8/11$}&
\multicolumn{2}{c}{$7/12$}\\
\tableline \hline
$E=1$            
&PA & 1&.070 & 0&.963 & 1&.006 & 1&.035 &*&     &0&.981 &1&.035 
    &0&.980 &0&.980 &1&.080\\
&DLPA& 1&.007 & 0&.996 & 0&.977 & 1&.010 &0&.993 &0&.989 &*&     
    && && &&\\ \hline
$\beta=0.85$
&PA   & 1&.0039 & *&      & 1&.0020 & 1&.0031 &*&      &0&.9988 
      &*&      &0&.9996 &0&.9997 &1&.0024\\
&DLPA & 1&.0006 & 0&.9992 & 1&.0019 & 1&.0009 &1&.0009 &1&.0014 &
       1&.0005 && && &&\\
\end{tabular}
\end{table}

\begin{table}
\squeezetable
\caption{
Analysis of the 25th order strong-coupling series of $s\equiv
M_{\rm h}^2/M_{G}^2$ for $N=3$ on the honeycomb lattice.  The first
two lines report the values of the $[l/m]$ PA's and Dlog-PA's at
$E=1$.  The last two lines report the values of $[l/m]$ PA's and
Dlog-PA's at $\beta=0.85$ corresponding to $\xi\simeq 22$.  Asterisks
mark defective PA's.
\label{hos}} 
\begin{tabular}{llr@{}lr@{}lr@{}lr@{}lr@{}lr@{}lr@{}lr@{}lr@{}lr@{}l}
\multicolumn{1}{c}{$$}&
\multicolumn{1}{c}{$$}&
\multicolumn{2}{c}{$11/11$}&
\multicolumn{2}{c}{$10/12$}&
\multicolumn{2}{c}{$11/12$}&
\multicolumn{2}{c}{$10/13$}&
\multicolumn{2}{c}{$12/12$}&
\multicolumn{2}{c}{$11/13$}&
\multicolumn{2}{c}{$10/14$}&
\multicolumn{2}{c}{$12/13$}&
\multicolumn{2}{c}{$11/14$}&
\multicolumn{2}{c}{$10/15$}\\
\tableline \hline
$E=1$ 
&  PA& 0&.9956& 1&.0001& 1&.0052& 0&.9983& *&     & *&  & *&     & 
       *& &*& &*& \\ 
&DLPA& *&     & 0&.9964& 0&.9994& 0&.9963& 1&.0023& *&  & 0&.9972&  
        & & & & & \\ \hline
$\beta=0.85$
&  PA& 0&.9978& 0&.9984& 0&.9983& 0&.9982& *&     & 0&.9983& 0&.9991& 
       0&.9982& 0&.9989& 0&.9989 \\ 
&DLPA& 0&.9989& 0&.9982& *&     & 0&.9982& *&     & 0&.9992& 0&.9983&

        & & & & & \\ 
\end{tabular}
\end{table}

\begin{table}
\squeezetable
\caption{
Analysis of the 29st order
strong-coupling series of $E^{-1}u(E)$ and $\beta^{-1}u(\beta)$,
where $u \equiv m_2^2/(\chi m_4)$,
for $N=3$ on the honeycomb lattice.
The first two lines report the values
of $u$ as obtained from the $[l/m]$ PA's and Dlog-PA's 
at $E=1$. 
The last two lines report the values
of $u$ from $[l/m]$ PA's and Dlog-PA's 
at $\beta=0.85$ corresponding to $\xi\simeq 22$.
A pole has been detected at $E_0=-0.11404$,
corresponding to $M_{G}^2=-16.000$.
Asterisks mark defective PA's.
\label{hoom}} 
\begin{tabular}{llr@{}lr@{}lr@{}lr@{}lr@{}lr@{}lr@{}lr@{}lr@{}l}
\multicolumn{1}{c}{$$}&
\multicolumn{1}{c}{$$}&
\multicolumn{2}{c}{$13/13$}&
\multicolumn{2}{c}{$13/14$}&
\multicolumn{2}{c}{$12/15$}&
\multicolumn{2}{c}{$14/14$}&
\multicolumn{2}{c}{$13/15$}&
\multicolumn{2}{c}{$12/16$}&
\multicolumn{2}{c}{$14/15$}&
\multicolumn{2}{c}{$13/16$}&
\multicolumn{2}{c}{$12/17$}\\
\tableline \hline
$E=1$ 
&PA   &0&.2511 &0&.2484 &0&.2498 &*& &*& &*& &0&.2482 &0&.2495 & 
       0&.2526\\
&DLPA &0&.2465 &0&.2469 &0&.2485 &0&.2467 & 0&.2467 & *& && && & 
&\\\hline
$\beta=0.85$ 
&PA   &0&.2490& 0&.2490& 0&.2490& 0&.2489& 0&.2482& 0&.2488&0&.2489 &
       0&.2490& 0&.2490\\
&DLPA &0&.2493& 0&.2494& 0&.2488& *&     & *&     & *& && && & &\\
\end{tabular}
\end{table}

\begin{table}
\caption{
Singularities in the complex $\beta$-plane for the triangular and
honeycomb lattice at $N=\infty$ with positive real and imaginary part
for the lowest values of $(d,c)$.  The singularity on the the real
axis for the honeycomb lattice ($\beta = 0.627168$) does not appear on
the principal sheet of $w(\beta)$ as the corresponding $w$-value is
$0.962998 i$.
\label{tabellazeribeta}}
\begin{tabular}{ccc}
\multicolumn{1}{c}{$(d,c)$}&
\multicolumn{1}{c}{triangular}&
\multicolumn{1}{c}{honeycomb}\\
\tableline\hline
$(1, 0)$ &  --- & 0.3620955333 $ i $  \\ \hline
$(1, \pm 2)$ & 0.206711 + 0.181628 $ i $& 0.482696 + 0.628020 $ i $  \\
&  0.685669 + 0.749077 $ i $& 0.449772 + 0.583632 $ i $ \\
& --- & 0.627168  \\ \hline
$(3, \pm 4)$ & 0.240692 + 0.486530 $ i $& 0.566020 + 1.476842 $ i $\\
& 0.564118 + 0.203430 $ i $& 0.946032 + 1.663513 $ i $\\
& 1.469137 + 2.118380 $ i $& 1.237526 + 0.266631 $ i $\\  \hline
$(5, \pm 6)$ & 0.260362 + 0.780732 $ i $ & 0.627495 + 2.353352 $ i $\\
&  0.920774 + 0.210433 $ i $& 1.413547 + 2.691806 $ i $\\
&  2.244964 + 3.482094 $ i $& 1.836347 + 0.535158 $ i $\\  \hline
$(5, \pm 8)$ & 0.662428 + 0.858005 $ i $ &  1.502252 + 2.655737 $ i $\\
&  0.980284 + 0.576976 $ i $ & 1.907210 + 2.926524 $ i $\\
&  2.847075 + 3.621622 $ i $ & 2.022490 + 1.997953 $ i $\\
&  ---                    & 2.500324 + 0.265750 $ i $\\  \hline
$(7, \pm 8)$ & 0.274172 + 1.072281 $ i $ &  0.669817 + 3.225087 $ i $\\
&  1.763190 + 0.911108 $ i $& 1.881646 + 3.718975 $ i $\\
&  3.019312 + 4.844833 $ i $& 2.432118 + 0.804519 $ i $\\
\end{tabular}
\end{table}

\begin{table}
\caption{
For $N=\infty$ we report the zeroes closest to the origin as obtained
by an analysis of the strong-coupling series of $\chi$ by Dlog-PA's
and IA's. We consider the series at 15th and 30th
order on the triangular lattice, and 30th and 60th
order on the honeycomb lattice. $\gamma_s$ is the exponent
corresponding to the singularity $\beta_s$ in the IA
analysis~\protect\cite{IA} (its exact value is $\gamma_s=-1/2$). The
Dlog-PA analysis does not provide stable estimates of $\gamma_s$.
The values we quote are 
average and maximum spread of the Dlog-PA's $[m/n]$ with 
$5\le m \le 9$ (series with 15 terms), $12\le m \le 17$ (30 terms)
and $27 \le m \le 32$ (60 terms); for the IA we use in all
cases 6 quasi-diagonal approximants.
\label{HTzeroes}} 
\begin{tabular}{ccccc}
\multicolumn{1}{c}{lattice}&
\multicolumn{1}{c}{$n$}&
\multicolumn{1}{c}{$\beta_s$ (Dlog-PA)}&
\multicolumn{1}{c}{$\beta_s$ (IA)}&
\multicolumn{1}{c}{$\gamma_s$ (IA)}\\
\tableline \hline
triangular & 15 & $0.214(4) \pm i 0.1838(2)$   & 
                  $0.206(1) \pm i\, 0.1811(6)$ &  
                  $0.54(7) \mp i\,0.07(6)$ \\
& 30 & $0.2084(2) \pm i 0.1821(2)$  & 
       $0.206712(1) \pm i\, 0.181628(1)$ & 
       $-0.4997(4) \mp i\,0.0001(2)$ \\\hline 
honeycomb  & 30 & $\pm i\, 0.3648(8)$          & 
                  $\pm i\, 0.36211(3)$ & 
                  $-0.50(2)\mp i\,0.001(6)$\\
& 60 & $\pm i\, 0.36270(4)$         & 
       $\pm i\, 0.3620955327(5)$ &
       $-0.500001(1) \mp i \, 0.000001(1)$ \\
\end{tabular}
\end{table}


\begin{references}

\bibitem{lattice95} M.~Campostrini, A.~Pelissetto,
P.~Rossi and E.~Vicari, HEP-LAT/9509025,
proceedings of the conference
``Lattice 95'', Melbourne 1995, Nucl. Phys. {\bf B}
(Proc. Suppl.) in press;
M.~Campostrini, A.~Cucchieri, T.~Mendes, A.~Pelissetto,
P.~Rossi, A.~D.~Sokal, and E.~Vicari, HEP-LAT/9509034,
{\em ibid.}.

\bibitem{Nm2} M.~Campostrini, A.~Pelissetto,
 P.~Rossi and E.~Vicari,
``A strong-coupling analysis of two-dimensional
$O(N)$ $\sigma$ models with
$N\leq 2$ on square, triangular and honeycomb
lattices.'', IFUP-TH 6/96.

\bibitem{sqNi} H.~E.~Stanley, Phys.\ Rev.\ {\bf 176},
718 (1968).

\bibitem{SCUN2}M.~Campostrini, P.~Rossi and E.~Vicari,
Phys. Rev. {\bf D52}, 386 (1995).

\bibitem{CRcond} M.~Campostrini and P.~Rossi,
Phys.\ Lett.\ {\bf 242B}, 81 (1990).

\bibitem{gr} M.~Campostrini, A.~Pelissetto,
P.~Rossi and E.~Vicari,
``Four-point renormalized coupling constant in ${\rm O}(N)$ models'',
Pisa preprint IFUP-TH 24/95, hep-lat 9506002, Nucl. Phys.
{\bf B}, in press.

\bibitem{BCMO} 
P. Butera, M. Comi, G. Marchesini and E. Onofri, Nucl. Phys. 
{\bf B236} 758 (1989). 

\bibitem{Gradshteyn} 
I.S. Grad\v stein and I.M. Ry\v zik, {\it Table of Integrals,
  Series and Products\/} (Academic Press, Orlando, 1980).

\bibitem{Guttmann} A.~J.~Guttmann, ``Phase Transitions and Critical
Phenomena'', vol. 13, C.~Domb and J.~Lebowitz eds.  (Academic Press,
New York).

\bibitem{Luscher} M.~L\"{u}scher and P.~Weisz,
Nucl.\ Phys.\ {\bf B300}, 325 (1988).

\bibitem{Butera} P.~Butera, M.~Comi, and G.~Marchesini,
Phys.\ Rev.\ {\bf B 41}, 11494 (1990). 

\bibitem{Reisz} T.~Reisz, Nucl.\ Phys.\ {\bf B450}, 569 (1995).

\bibitem{Butera2} P.~Butera and  M.~Comi, 
``New extended high-temperature series for the
N-vector spin models on three-dimensional bipartite lattices'',
IFUM-TH 498, HEP-LAT/9505027 (1995). 

\bibitem{SCUN1}M.~Campostrini, P.~Rossi and E.~Vicari,
Phys. Rev. {\bf D52}, 358 (1995).

\bibitem{RV}
P.~Rossi and E.~Vicari,
Phys. Rev. {\bf D49}, 6072 (1994); {\bf D50}, 4718 (1994) (E).

\bibitem{Flyv} H.~Flyvbjerg,
Nucl.\ Phys.\ {\bf B348}, 714 (1991).

\bibitem{CR} P.~Biscari, M.~Campostrini and P.~Rossi,
Phys.\ Lett.\ {\bf 242B}, 225 (1990).

\bibitem{Meyer} S.~Meyer, unpublished.

\bibitem{CRselfenergy} M.~Campostrini and P.~Rossi,
Int.\ J.\ Mod.\ Phys.\ {\bf A 7}, 3265 (1992). 

\bibitem{IA} A.~J.~Guttmann and G.~S.~Joyce, J.  Phys. {\bf A5},
 L81 (1972); D.~L.~Hunter and G.~A.~Baker Jr., Phys. Rev. {\bf B 49},
 3808 (1979).

\bibitem{CEPS} S.~Caracciolo, R.~Edwards, A.~Pelissetto,
and A.~D.~Sokal, Phys. Rev. Lett. {\bf 75}, 1891, (1995).

\bibitem{Bonnier} B.~Bonnier, M.~Hontebeyrie,
Phys.\ Lett.\ {\bf 226B}, 361 (1989).

\bibitem{Hasenfratz} 
P.~Hasenfratz, M.~Maggiore, and
F.~Niedermayer, Phys.\ Lett.\ {\bf 245B}, 522 (1990);
P.~Hasenfratz and
F.~Niedermayer, Phys.\ Lett.\ {\bf 245B}, 529 (1990).

\bibitem{Wolff} U.~Wolff, 
Phys.\ Lett.\ {\bf 242B}, 335 (1990).

\bibitem{Falcioni}M.~Falcioni and A.~Treves,
Nucl.\ Phys.\ {\bf B265}, 671 (1986).

\bibitem{Chandra}  
K. Chandrasekharan, Elliptic Functions, Grundlehren der mathematischen
Wissenschaften, vol. 281 (Springer Verlag, Berlin-Heidelberg, 1985).

\bibitem{Akhiezer}
N. I. Akhiezer, Elements of the Theory of Elliptic Functions,
Transl. Math. Monographs vol. 79 (American Mathematical Society,
Providence, 1990).

\bibitem{xiO3} J.~Apostolakis, C.~F.~Baillie, and
G.~C.~Fox, Phys.\ Rev.\ {\bf D 43}, 2687 (1991). 

\bibitem{Sokal} R.~G.Edwards, S.~J.~Ferreira, J.~Goodman and 
A.~D.~Sokal, Nucl.\ Phys.\ {\bf B380}, 621 (1991).

\end{references}
\end{document}